\documentclass[
superscriptaddress,
reprint,
prx,
amsmath,
footinbib,
twocolumn,
amssymb,
bibnotes
]{revtex4-2}

\usepackage{graphicx}
\usepackage{soul}      
\usepackage{color}
\usepackage{txfonts}
\usepackage[colorlinks, citecolor=blue,linkcolor=blue,urlcolor=blue]{hyperref}
\usepackage{verbatim}
\usepackage{multirow}
\graphicspath{{TmpFigs/}}

\usepackage[inline,final]{showlabels}
\usepackage{relsize}
\relscale{1.5}

 \usepackage{mathtools}
 \usepackage{wrapfig} 
 \usepackage{mathrsfs}
 \usepackage{amssymb} 
 \usepackage[x11names]{xcolor}
 \usepackage{amsbsy}
 \usepackage{marginnote}
 \usepackage{slashed}
 \usepackage{marvosym}
 \usepackage{pifont}
 \usepackage{mathbbol}
 \usepackage{wrapfig}
 \usepackage{upgreek}
 \usepackage{bbm}
 \usepackage{yfonts}
 \usepackage{xspace}
 \usepackage{lineno}
 \usepackage{tikz}
\usepackage[most]{tcolorbox}
\usepackage{pgfplots}
\pgfplotsset{width=10cm,compat=1.9}
 \newcommand{\bcen}{\begin{center}}
 \newcommand{\ecen}{\end{center}}
 \newcommand{\btab}{\begin{tabular}}
 \newcommand{\etab}{\end{tabular}}
 \newcommand{\bdes}{\begin{description}}
 \newcommand{\edes}{\end{description}}

 \newcommand{\beq}{\begin{equation}}
 \newcommand{\eeq}{\end{equation}}
 \newcommand{\bea}{\begin{eqnarray}}
 \newcommand{\eea}{\end{eqnarray}}

 \newcommand{\half}{\frac{1}{2}}
 \newcommand{\bary}{\begin{array}}
 \newcommand{\eary}{\end{array}}
 \newcommand{\benum}{\begin{enumerate}}
 \newcommand{\eenum}{\end{enumerate}}
 \newcommand{\bitem}{\begin{itemize}}
 \newcommand{\eitem}{\end{itemize}}

 \newcommand{\dou}{\partial}

 \newcommand{\mean}[1]{\langle #1 \rangle}
 
 \newcommand{\bra}[1]{{{\langle #1 |}}}
 \newcommand{\ket}[1]{{| #1 \rangle}}
 \newcommand{\braket}[2]{\langle #1 | #2 \rangle}

 \newcommand{\eqn}[1]{eqn.~(\ref{#1})}

 \newcommand{\Fig}[1]{Fig.~\ref{#1}}
 \newcommand{\Integers}{{\mathbb{Z}}}

 \newcommand{\ind}{\alpha}
 
\newcommand{\oibook}[1]{}

\newcommand{\term}[1]{\left( #1 \right)}

\renewcommand{\half}{\frac{1}{2}}

\newcommand{\eqnref}[1]{Eq.~(\ref{#1})}
\newcommand{\figref}[1]{Fig.~\ref{#1}}
\newcommand{\sfigref}[2]{Fig.~\hyperref[#1]{\ref{#1}#2}}
\newcommand{\tabref}[1]{Table~\ref{#1}}

\newcommand{\appref}[1]{Appendix~\ref{#1}}

\newcommand{\B}[1]{{{\cal B}(#1)}}
\newcommand{\Bd}[2]{{{\cal B}^{#1}(#2)}}
\newcommand{\HB}[2]{{{\cal HB}(#1,#2)}}
\newcommand{\SC}[2]{{{\cal SC}(#1)_{#2}}}
\newcommand{\RC}[2]{{{\cal RC}(#1)_{#2}}}
\newcommand{\SHC}[3]{{{\cal SHC}(#1,#2)_{#3}}}
\newcommand{\RHC}[3]{{{\cal RHC}(#1,#2)_{#3}}}
\newcommand{\tSHC}[3]{{\widetilde{{\cal SHC}}(#1,#2)_{#3}}}
\newcommand{\tRHC}[3]{{\widetilde{{\cal RHC}}(#1,#2)_{#3}}}

\newcommand{\T}{{\cal T}}
\newcommand{\HT}{{\cal HT}}
\newcommand{\cart}{{\Box}} %we can redefine this later, if needed

\newcommand{\GS}{{\textup{GS}}}
\newcommand{\GQIM}{{\textup{GQIM}}}
\newcommand{\dontshowthis}[1]{{ }}
\newcommand{\Ztwo}{{\Integers_2}}

\newcommand{\Xcube}{{\textup{X-cube}}}

\pgfkeys{/pgf/declare function={arctanh(\x) = (1/2)*ln((1+\x)/(1-\x));}}
\newcommand{\IISc}{Centre for Condensed Matter Theory, Department of Physics, Indian Institute of Science, Bangalore 560012, India}

\newcommand{\mytitle}{Arboreal Topological and Fracton Phases}

\begin{document}
%\linenumbers 

\title{\mytitle}

\author{Nandagopal Manoj}\email{nandagopalm@iisc.ac.in}\affiliation{\IISc}
\author{Vijay B.~Shenoy}\email{shenoy@iisc.ac.in}\affiliation{\IISc}

\date{\today{}}
\begin{abstract} 
We describe topologically ordered and fracton ordered quantum systems on novel geometries which do not have an underlying manifold structure. Using tree graphs such as the $k$-coordinated Bethe lattice 
%$\B{k}$
${\cal B}(k)$
and a hypertree called the $(k,n)$-hyper-Bethe lattice 
%$\HB{k}{n}$
${\cal HB}(k,n)$ consisting of $k$-coordinated hyperlinks (links defined by $n$ sites), we construct  multidimensional arboreal arenas such as 
%$\B{k_1} \cart \B{k_2}$ 
${\cal B}(k_1) \square {\cal B}(k_2)$ 
by a generalized notion of a graph Cartesian product 
%$\cart$
$\square$.
We study various 
%$\Ztwo$ 
quantum systems such as the  
%$\Ztwo$ 
${\mathbb Z}_2$
gauge theory, generalized quantum Ising models (GQIM), the fractonic 
%\Xcube~model
X-cube
model, and related 
%\Xcube
X-cube
gauge theory defined on these arboreal arenas, finding several fascinating results.   Even the simplest 
%$\Ztwo$ 
${\mathbb Z}_2$
gauge theory on a two-dimensional is found to be fractonic -- an isolated monopole excitation is rendered fully immobile on an arboreal arena. The 
%\Xcube
X-cube
model on a generic three-dimensional arboreal arena is found to be fully fractonic in the magnetic sector, all multipoles of magnetic excitations are rendered immobile on the arboreal arena. We obtain variational ground state phase diagrams of the gauge theories (both 
%$\Ztwo$
${\mathbb Z}_2$
~and 
%\Xcube
X-cube
~gauge theories) which are shown to have deconfined and confined phases. These phases are usually separated by a first-order transition, while continuous transitions are obtained in some cases. 
Further, we find an intriguing class of dualities in  arboreal arenas as illustrated by the 
%$\Ztwo$ 
${\mathbb Z}_2$
gauge theory defined on 
%$\B{k_1}\cart\B{k_2}$
${\cal B}(k_1) \square {\cal B}(k_2)$
being dual to a 
%\GQIM
GQIM
~defined on 
%$\HB(2,k_1)\cart\HB(2,k_2)$
${\cal HB}(2,k_1) \square {\cal HB}(2,k_2)$. 
Finally, we discuss different classes of topological and fracton orders that appear on arboreal arenas.  We find three distinct classes of arboreal toric code orders on  two-dimensional arboreal arenas, those that occur on %$\B{2}\cart\B{2}$, $\B{k}\cart\B{2}, k>2$, 
${\cal B}(2) \square {\cal B}(2)$, ${\cal B}(k) \square {\cal B}(2), k >2$,
and 
%$\B{k_1} \cart \B{k_2}, k_1,k_2 >2$
${\cal B}(k_1) \square {\cal B}(k_2)$, $k_1,k_2>2$.
 Likewise,  four classes of 
%\Xcube
X-cube
~fracton orders are found on three-dimensional arboreal arenas which  correspond to those on
%$\B{2}\cart\B{2}\cart\B{2}$, $\B{k}\cart\B{2}\cart\B{2}, k>2$
${\cal B}(2)\square{\cal B}(2)\square {\cal B}(2)$, 
${\cal B}(k) \square {\cal B}(2)\square {\cal B}(2), k>2$, 
%$\B{k_1} \cart \B{k_2}\cart\B{2}, k_1,k_2 >2$,
${\cal B}(k_1) \square {\cal B}(k_2) \square {\cal B}(2), k_1,k_2 >2$,
and 
%$B{k_1} \cart \B{k_2}\cart\B{k_3}, k_1,k_2,k_3 >2$.
${\cal B}(k_1) \square {\cal B}(k_2) \square {\cal B}(k_3), k_1,k_2,k_3 >2$.

\end{abstract}

\pacs{}

\maketitle 

%\tableofcontents{}

%\input{maintext_tenfold}

%\tableofcontents

\section{Introduction}
\label{sec:Intro}

A central problem of condensed matter physics is the description and classification of phases of systems with many microscopic quantum degrees of freedom. Research over the last two decades has revealed that, in addition to the ideas of symmetry, notions of entanglement and topology have a preeminent role in the description and classification of phases~\cite{Ryu2010, Kitaev2009, Senthil2015, Chiu2016, Wen2017}. Broadly, quantum phases realized by many degrees of freedom can be classified as short-ranged entangled or those with topological order with more complex patterns of long-range entanglement.  Examples of novel phases already realized in the laboratory with short-range entanglement are topological insulators~\cite{Hasan2010,Qi2011,Chiu2016}. Similarly, fractional hall states~\cite{Wen1990} realize states with topological order and associated long-range entanglement.

Topologically ordered phases have enjoyed particular attention owing to their exotic properties that can be exploited for topologically protected quantum information processing~\cite{Nayak2008}. A particularly interesting example of a topologically ordered system useful as a quantum memory is the toric code~\cite{Kitaev2003}. The toric code has a ground state degeneracy that depends only on the topology of the manifold on which it is defined, and not on the microscopic details such as lattice structure. For example, the ground state degeneracy of a toric code defined on a torus is 4 irrespective of whether the underlying lattice is a square lattice or a triangular lattice. Excitations of the toric code also have exotic physics - the two types of excitations, magnetic monopoles (plaquette excitations) and electric charges (vertex excitations) are mutually semionic, and bound states of these excitations are fermions (see \cite{Kapustin2018} that explores interesting consequences and implications of this). From the perspective of quantum information processing/computing, the ground state degeneracy which is topologically protected is an attractive platform to store and manipulate quantum information~\cite{Terhal2015,Brown2016}.  A deeper understanding of topologically ordered phases was effected with the formulations of many exactly solvable, for example, string-net models~\cite{Levin2005,Wen2017}. Connections to topological field theories and modular tensor categories have been exploited for the classification (see recent work~\cite{Burnell2021} and references therein) of topologically ordered phases.

Although topologically ordered phases have degenerate ground states that are protected, there are issues associated with the stability of these phases at finite temperatures~\cite{Dennis2002,Castelnovo2007,Nussinov2008,Terhal2015,Brown2016}. This arises from fact that the monopole excitations discussed above, when thermally excited in pairs, can proliferate owing to the entropy gain. This, and related issues, have prompted researchers to formulate and explore models~\cite{Chamon2005,Bravyi2011,Castelnovo2012,Castelnovo2012,Haah2011,Yoshida2013,Bravyi2013,Vijay2015,Vijay2016,Williamson2016,Hsieh2017} where the excitations have limited mobility, possibly offering avenues to address the finite-temperature issues faced by topological codes, giving way to the discovery of fracton phases (see reviews~\cite{Nandkishore2019,PretkoChenYou2020}). Fracton phases are so named owing to the fact that the excitations of in this system have fractional mobility. As an example, the excitations of the exactly solvable X-cube model, the “toric-code of fracton physics”, has point magnetic monopoles which are fully immobile, even as, the dipoles of these magnetic monopoles are mobile in a plane. The analogs of electric charges can be moved along a line by local operations.  While the presence of immobile excitations is a generic feature of all fracton phases, some like the X-cube model discussed above, have bound states of excitations that are mobile. On the other hand models such as Haah’s code \cite{Haah2011} have only immobile excitations, leading to further classification of fracton phases as Type I and Type II fractons (see \cite{PretkoChenYou2020}). Another intriguing aspect of fracton phases is that they have a sub-extensive ground state degeneracy, i.~e., the degeneracy scales as, for example in the X-cube model,  $e^{a L}$, where $L$ is the length of the side of the cuboidal box with periodic boundary conditions used to define the model, and $a$ is a numerical constant. In addition, this ground state degeneracy, although sub-extensive, depends on the details of the lattice leading them to be termed as ``geometrically ordered’’~\cite{Slagle2018}. Based on these cues, the notion of foliated fracton order has been formulated and elaborated~\cite{Shirley2018,Shirley2019}. Yet another notable point is that discrete fracton models (such as the X-cube model) which are typically theories with a local gauge structure have been shown~\cite{Vijay2016} to be dual to discrete models with global subsystem symmetries where the symmetry operations act on a sub-extensive set of degrees of freedom.

The discrete models discussed above have motivated long-wavelength (field-theoretic descriptions) of topologically ordered and fracton phases. As alluded to above, topological field theories are used to describe topologically ordered phases. For example, Chern-Simons theories with a suitably chosen $K$-matrix, a topological field theory, can describe topologically ordered phases of the toric code~\cite{Kitaev2003} and the double-semion model~\cite{Levin2005}. In such theories, the gauge fields are assigned to every space-time point of the manifold of interest, and the action depends on these gauge fields only via a topological invariant associated with these fields.  The field theoretical description of fracton phases have more interesting underpinnings. Anticipated by work aimed at providing infrared descriptions of quantum spin liquids~\cite{Xu2006,RYX2016}, refs.~\cite{Pretko2017a,Pretko2017b} elaborated that higher rank tensor gauge theories offer themselves as natural candidates for the long-wavelength descriptions of fracton phases. Generalizations into include extended charges with different types mobility restrictions have also been explored~\cite{Pai2018,ShenoyMoessner2020}, and general fracton gauge principles have been formulated and explored~\cite{Pretko2018FGP,Seiberg2020FG}. Remarkably, some of these higher rank tensor gauge theories have been shown to be dual~\cite{PretkoRadzihovsky2018,Gromov2017,Gromov2019,Manoj2021} to many well-known and well-studied~\cite{Kleinert1989,DietelKleinert2006,Zaanen2004,Beekman2017} systems.  Field theories related to the discrete fracton models discussed above have also been postulated~\cite{Slagle2017, You2020, Slagle2021,ShaoSeiberg2021}.

As is clear from the discussion above, key advances have been made possible by the formulation of discrete models. These discrete models are typically defined on lattices that ``tessellate a manifold’’. For example, the square lattice with periodic boundary conditions tessellates the two-torus. The Chern-Simons theory~\cite{Wen1990} that describes the toric code uses topological data associated with U(1) gauge fields that are attached at points of the two-torus (and, of course, the time coordinate), i.~e., the manifold that the discrete square lattice tessellates. Our point here is that most of the discrete models defined and studied are lattices that tessellate a manifold, and such models often enjoy the advantage of a universal long-wavelength field theoretical description using fields attached to points on a manifold.  In other words, much of the current focus has been on systems defined on a manifold arena.

The creation, manipulation, and study of many-body quantum systems have seen spectacular advances in the last decade (see, for example, \cite{Cirac2012} and references in that issue). A number of platforms offer many opportunities to engineer a variety of discrete models, for example, circuit quantum electrodynamics~\cite{Blais2020} can be used to create topological codes and its excitations (see, for example, \cite{Song2018}). In this context, the unprecedented ability to control individual quantum degrees of freedom  raises the possibility of exploring discrete quantum systems that are built on {\em non-manifold arenas}.  In other words, systems whose degrees of freedom are placed in a fashion different from the tessellation of a manifold. The main thrust of this paper is to explore this direction, i.~e.,  to study many-body quantum systems that are defined on an arena that is not a discretization of a manifold. 

At a first glance, the problem statement seems unwieldy as there is a myriad of non-manifold structures on which quantum models can be defined. Here we focus on those discrete arenas that are templated on tree graphs~\cite{Voloshin2009}, which we dub as the {\em arboreal arena}. Further, we focus on arenas that allow us to keep a notion of ``translational invariance’’, and this restriction gives us a well-defined class of tree graphs that provide us the templates for the construction of the arboreal arena studied here. The key tree graph that we consider in our work is the $k$-Bethe lattice, denoted as $\B{k}$, which is an infinite graph where every vertex (or site) has $k$ links/edges (each link, as usual, is defined by two graph vertices/sites) attached to it. A generalized tree graph, called a hypertree graph~\cite{Voloshin2009}, also plays a crucial role in our discussion. A hypertree is a tree graph whose links, called hyperlinks, have an arbitrary number of sites attached to it. This leads naturally to a $(k,n)$-hyper-Bethe lattice denoted as $\HB{k}{n}$, where each vertex(site) has $k$ hyperlinks attached, and each hyperlink has $n$ vertices(sites) that define it. Next, we use a generalized notion of graph cartesian product, denoted by $\cart$, to construct ``higher dimensional’’ arboreal arenas, for example, $\B{3} \cart \B{2}$ is a two-dimensional arboreal arena. To explore the physics of models defined on these arboreal arenas, we also define finite arboreal arena made from the ones defined above by the introduction of ``surfaces’’.

Physics in arenas based on tree graphs have been studied in the past in various contexts, including classical statistical mechanics~\cite{Thompson1982,Baxter2013}, physics of interacting fermions~\cite{Mahan2001,Georges1996} etc.~to state a few, where often times, exact solutions are possible in the limit of large coordination number (for example, $\B{k}$, for large $k$). Our motivation is to use tree-graphs from a different perspective -- to create ``connectivities'' of underlying microscopic degrees of freedom (qubits) that produce different and varied patterns of quantum entanglement which could possibly offer new opportunities, for example, in quantum information processing.

Following the definition of the arboreal arenas, we define and study a variety of models on them. In particular, we define $\Ztwo$-gauge theory, generalized quantum Ising models (\GQIM) where interactions are defined on hyperlinks,  the \Xcube~model and related \Xcube~gauge theory, on the arboreal arena in a natural fashion. We summarize here are what we believe are intriguing and exciting results. First, we find that even the simple $\Ztwo$ gauge theory on a two-dimensional arboreal arena is fractonic -- isolated monopole excitations are fully immobile. Second, the \Xcube~model on a generic three-dimensional arboreal arena is ``fully fractonic’’ in the magnetic sector, no multipole of magnetic excitation is mobile. Third, the gauge theories (both $\Ztwo$ and \Xcube~gauge theories) have deconfined and confined phases at zero temperature, that are generically separated by a first-order transition upon tuning of the electric coupling constant (second-order transitions are also possible).  Fourth, we demonstrate a rich class of dualities between the models defined on arboreal arena. As an example, we show that the $\Ztwo$ gauge theory defined on $\B{k_1}\cart\B{k_2}$ is dual to a \GQIM~defined on $\HB{2}{k_1}\cart\HB{2}{k_2}$ where the latter model possesses subsystem symmetries! These dualities are natural generalizations of the dualities formulated in \cite{Vijay2016} to the arboreal arena. Finally, we show a rather intriguing and aesthetically appealing result. We show that there are only three distinct classes of arboreal toric code orders on the two-dimensional arboreal arenas, those that correspond to $\B{2}\cart\B{2}$, $\B{k}\cart\B{2}, k>2$, and $\B{k_1} \cart \B{k_2}, k_1,k_2 >2$. This implies, for example, that the ground state of $\Ztwo$ gauge theory defined on $\B{3} \cart \B{3}$  in its deconfined phase, can be transformed to that of $\B{k_1} \cart \B{k_2}$, for {\em any} $ k_1>3,k_2 >3$  by a finite depth unitary quantum circuit! Similarly, there are four classes of $\Xcube$ fracton orders on  three-dimensional arboreal arenas viz., those that correspond to $\B{2}\cart\B{2}\cart\B{2}$, $\B{k}\cart\B{2}\cart\B{2}, k>2$, and $\B{k_1} \cart \B{k_2}\cart\B{2}, k_1,k_2 >2$, and $\B{k_1} \cart \B{k_2}\cart\B{k_3}, k_1,k_2,k_3 >2$.

In the next section, we present definitions of the various arboreal arena. This is followed by a discussion of arboreal $\Ztwo$ gauge theory in section.~\ref{sec:AGT}. The \Xcube~model and realted gauge theory in the arboreal setting are discussed in section.~\ref{sec:FMArboreal}. Arboreal topological and fracton orders are discussed in section \ref{sec:ATFO}. Section.\ref{sec:CR}  concludes the paper with a discussion of arboreal quantum phases and possible applications, and directions of future research. 

\section{The Arboreal Arena}
\label{sec:ArborealArena}
%\NM{A suggestion -- can we call the higher dimensional objects as $d$-dimensional arboreal lattice/arena, for example? This avoids the confusion that $d$-dimensional trees are not trees, and we can clearly state that a 1d arboreal lattice is a simple hypertree.}\VBS{Yes, lets do this in the next round. I will try to put down a logical scaffold without worry too much about clear and correct terminology.} 

We adopt and adapt notions from graph theory to define arboreal arenas. A graph ${\cal G} = \{ {\cal S}, {\cal L} \}$ is a set of vertices ${\cal S}$ and a set of edges ${\cal L}$ that define connections between the vertices. In our discussion, we use the term ``sites'' synonymous with vertices and ``links'' also mean edges. Two sites are called ``adjacent'' if there is a link between them; we sometimes  call adjacent sites as neighbours. We consider only graphs whose links connect distinct sites, i.~e, there are no links that start and end at the same site. However, links that  connect to a single site are allowed. A path on the graph $\mathcal{G}$ is defined as an ordered (possibly repeating) list of sites such that adjacent elements in the ordered list are neighbours in $\mathcal{G}$. A loop is a path that starts and ends at the same site such that no other site is present more than once. A {\em tree}, denoted by $\T$ is a graph that does not have any loops.

\begin{figure}
    \centering
    \includegraphics[width=0.95\columnwidth]{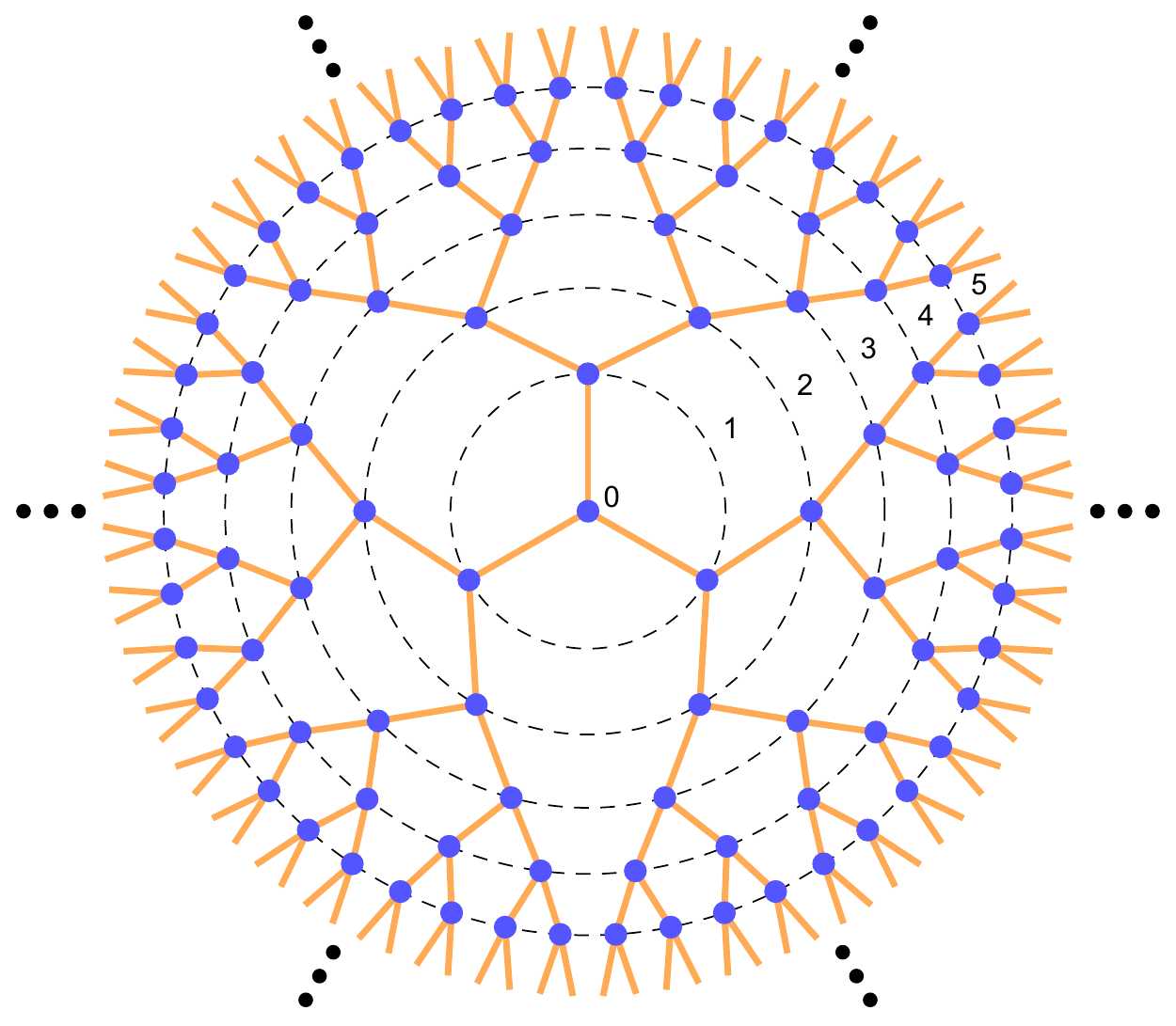}
    \caption{Bethe lattice $\B{k}$ with $k=3$. The tree is infinite, as indicated by the radial black ellipses. }
    \label{fig:BetheLattice}
\end{figure}

\begin{figure}
    \centering
    \includegraphics[width=\columnwidth]{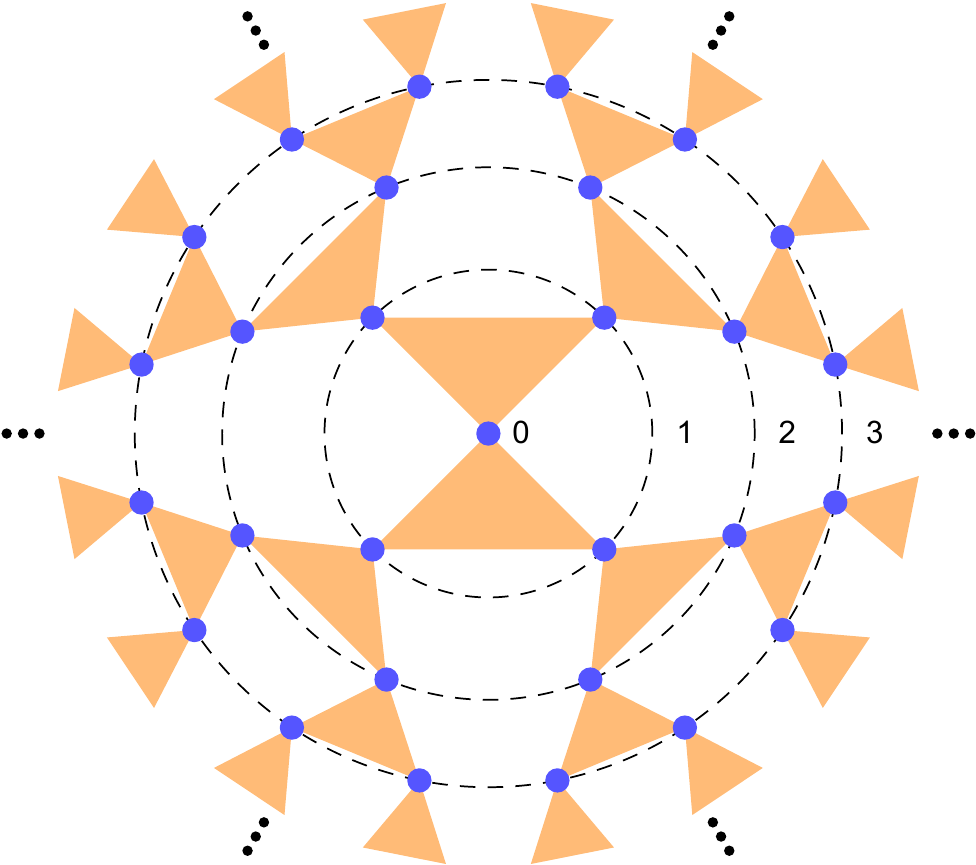}
    \includegraphics[width=\columnwidth]{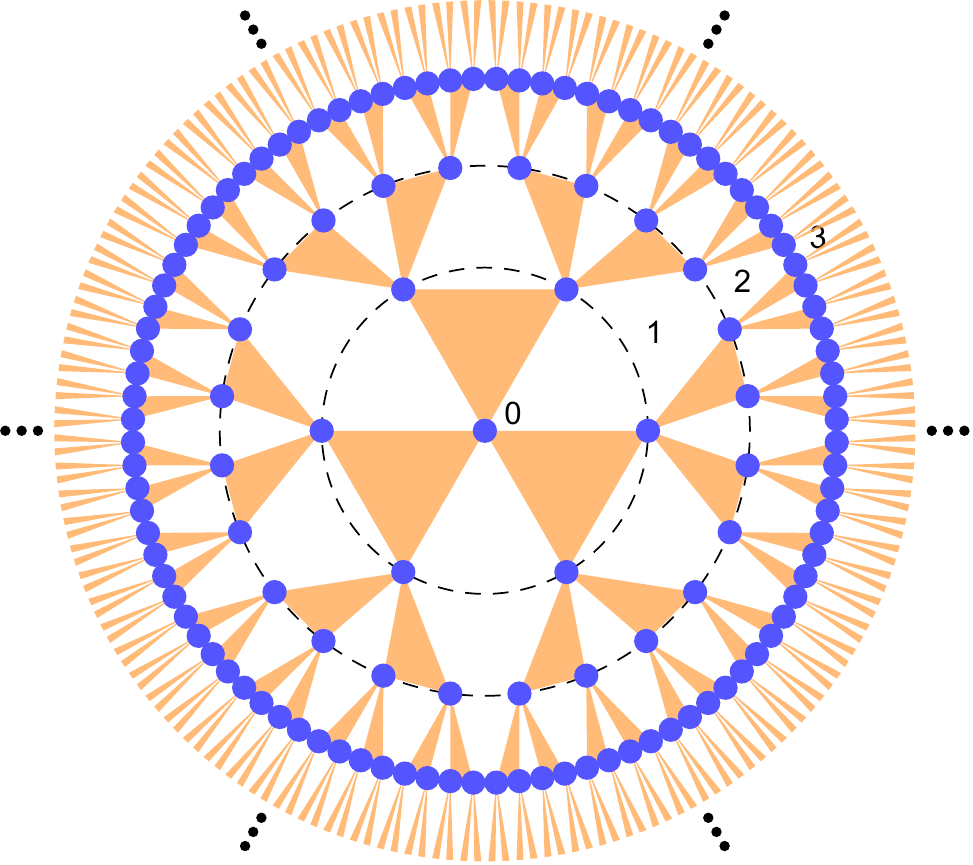}
    \caption{Hyper-Bethe lattices $\HB{2}{3}$, and $\HB{3}{3}$. Hyper-trees have generalized links (indicated by shaded triangles) which touch more than two sites (indicated by blue dots).}
    \label{fig:HyperBethe}
\end{figure}

We focus on tree graphs that posses additional properties, in particular, a notion of ``translational symmetry'' (we avoid a formal definition of this). Roughly, this means that every site on in the graph has an identical neighbourhood. A natural example of such a tree is a Bethe lattice (referred to as a ``regular tree'' in the mathematics literature). We introduce the notion of a $k$-Bethe lattice denoted as $\B{k}$ where each site is attached to $k$ links. Translational symmetry then necessitates that this graph is infinite (see \Fig{fig:BetheLattice}). We can impose a coordinate system on  $\B{k}$  by introducing the notion of ``generations''. Pick any site and declare it to be the generation $g=0$. All sites linked to the site at generation $g=0$ are said to be in generation $g=1$, and similarly for any other generation $g$. For any generation $g>0$, there are $k(k-1)^{g-1}$ sites which can be suitably numbered using $m$ (see \Fig{fig:BetheLattice}) where $m=0,\ldots,k(k-1)^{g-1}-1$. The cooridinate of any site in $\B{k}$ is then given by $(g,m)$.

An important extension of the tree graph that plays an important role in our work is the notion of a {\em hypertree}. Hypergraphs \cite{Voloshin2009} are graphs where the links can contain any number of vertices.  The desideratum of ``translation symmetry'' again requires that all the links contain the same number of vertices (say $n$). These ideas allow us to generalize $\B{k}$ to a  hypertree $\HB{k}{n}$, called a $(k,n)$-hyper Bethe lattice. A general way of stating the no loop condition for hypertrees is that there exists no finite subset of links such that any vertex that touches any link of this subset touches an even number of distinct links of the subset.   \figref{fig:HyperBethe} shows $\HB{2}{3}$ and $\HB{3}{3}$. Again, we can set up a coordinate system on this hyper-Bethe lattice using the notion of generations (see \figref{fig:HyperBethe}). Note that $\HB{k}{2}$ is same as $\B{k}$.

In order to understand the phases that appear on these arboreal arena, we also find it useful to ``break translational symmetry'' by introducing ``surfaces''. A convenient way to achieve this goal is by truncating a $\HB{k}{n}$ at some generation $M$ (typically a large number), i.~e., all sites up to an including generation $M$ are kept.  We call such hypertrees as {\em hyper-Cayley trees}. A site of a hyper-Cayley tree is called an interior site if all the $k$-links connected to it are present in the graph, and is termed a boundary site if the number of links connected to the site is less than $k$. Similary, a hyper-link is called called an interior link if there are $n$ sites connected to it, and a boundary hyperlink otherwise.
It is useful to define two types of hyper-Cayley trees. If the links connecting the sites of the $M$-th generation sites to the $M+1$-generation sites are kept in the graph, then  these trees are termed {\em rough hyper-Cayley trees} and denoted by $\RHC{k}{n}{M}$. In $\RHC{k}{n}{M}$ all the sites are interior sites, while the last set of links (coonecting $M$-th generation sites to $M+1$-th generation sites) are boundary links (other links are interior links). If these boundary links are not included, then they are called {\em smooth hyper-Cayley trees} and denoted as $\SHC{k}{n}{M}$. In $\SHC{k}{n}{M}$, all links are interior links, while the sites of generation $M$ are boundary sites (other sites are interior sites).  For $n=2$, $\RHC{k}{2}{M}$ and $\SHC{k}{2}{M}$ are denoted, respectively as, $\RC{k}{M}$ and $\SC{k}{M}$, i~e., rough and smooth Cayley trees.

Another important class of finite trees are useful for the discussion of arboreal quantum phases (especially in the construction of dual models). Choose a link in the infinite hyper-Bethe lattice $\HB{k}{n}$ to be the central link, and declare its centre as the origin. The $n$ vertices touching this link are said to be at generation 1. The vertices adjacent to these links are of generation 2, and so on. This defines a new coordinate system for the hyper-Bethe lattice. Now, as above, we can truncate the lattice at some generation $M$. For reasons that will become apparent in section \ref{sec:Duality}, we will call this lattice a \emph{dual} smooth hyper-Cayley tree ($\tSHC{k}{n}{M}$),  We also define dual rough hyper-Cayley trees ($\tRHC{k}{n}{M}$) as the lattice containing the sites up to generation $M$ and keeping the boundary links that connect generation $M$ to generation $M+1$ in this new coordinate system. 
\begin{table*}
	\centering
	 \begin{tabular}{| c |c | c |}
	 	\hline \rule{0pt}{3ex}
	 	Notation & Name & Description \\
	 	 \hline \rule{0pt}{4ex}    
	 	$\B{k}$ & $k$-Bethe lattice & \parbox{8cm}{Infinite ``translation invariant'' lattice with  $k$ links to each site. (see \figref{fig:BetheLattice})} \\
%	 	 \rule{0pt}{7ex}    
%	 	$\C{k}{g}$ & $k_g$-Cayley tree & \parbox{8cm}{Cayley tree with $g$ generations and $k$ links to each site except at boundary sites, which have only one link to it.} \\
	 	\rule{0pt}{7ex}    
	 	$\HB{k}{n}$ & \parbox{3.5cm}{\centering $(k,n)$ hyper-Bethe lattice} & \parbox{8cm}{Infinite ``translation invariant'' lattice with  $k$ generalized links to each site, and each generalized link containing $n$ sites. (see \figref{fig:HyperBethe})} \\
	 	\rule{0pt}{8ex}    
	 	$\SHC{k}{n}{M}$ & \parbox{3.5cm}{\centering $(k,n)_M$ hyper-Cayley tree (smooth boundaries)} & \parbox{8cm}{Hyper-Cayley tree with $M$ generations and $k$ generalized links to each site except at boundary sites, which have only one link to it. All links contain $n$ sites. } \\
	 	\rule{0pt}{10ex}    
	 	$\RHC{k}{n}{M}$ & \parbox{4.5cm}{\centering $(k,n)_M$ hyper-Cayley tree (rough boundaries)} & \parbox{8cm}{Hyper-Cayley tree with $M$ generations and $k$ generalized links to every site. All links except the ones emanating out of the boundary sites (boundary links) contain $n$ sites, and the boundary links contain only one site.} \\
	 	\rule{0pt}{8ex}
	 	$\tSHC{k}{n}{M}$ & \parbox{3.5cm}{\centering $(k,n)_M$ dual hyper-Cayley tree (smooth boundaries)} & \parbox{8cm}{Hyper-Cayley tree with an edge at the origin and $M$ generations and $k$ generalized links to each site except at boundary sites, which have only one link to it. All links contain $n$ sites. } \\
	 	\rule{0pt}{10ex}    
	 	$\tRHC{k}{n}{M}$ & \parbox{4.5cm}{\centering $(k,n)_M$ dual hyper-Cayley tree (rough boundaries)} & \parbox{8cm}{Hyper-Cayley tree with a generalized edge at the origin and $M$ generations and $k$ generalized links to every site. All (non-boundary) links contain $n$ sites, and the boundary links contain only one site.} \\
	 	\hline
	 \end{tabular}
	\caption{Notation, nomenclature, and description of various tree graphs used to construct arboreal arenas.}
	\label{tab:Catalogue}
\end{table*}

The (hyper)tree structures considered in this work summarized in \tabref{tab:Catalogue}, act as the building blocking for erecting arboreal arenas. ``Higher dimensional'' arboreal arenas can also be constructed in a natural fashion by using a generalization of a Cartesian product, called $\cart$ for any two hypergraphs\cite{Voloshin2009}. Let $\HT_1$ and $\HT_2$ be two hypertrees. Then, $\HT_1 \cart \HT_2$ is also a hypergraph with sites that are the collection of ordered pairs of sites $(u,v)$ where $u$ is in $\HT_1$ and $v$ is in $\HT_2$. A collection of $l$ vertices $\{(u_1,v_1), (u_2,v_2), \ldots, (u_l,v_l)\}$ represent a hyperedge of  $\HT_1 \cart \HT_2$
if and only if, {\em one of the two conditions} are satisfied. Either, $u_1=u_2=\ldots=u_l$ and $(v_1,v_2,\ldots,v_l)$ is a hyperlink in $\HT_2$, or $v_1=v_2=\ldots=v_l$ and $(v_1,v_2,\ldots,v_l)$ is a hyperlink in $\HT_2$, or $v_1=v_2=\ldots=v_l$ and $(v_1,v_2,\ldots,v_l)$ is a hyperlink in $\HT_2$, or $v_1=v_2=\ldots=v_l$ and $(u_1,u_2,\ldots,u_l)$ is a hyperlink in $\HT_1$. Indeed, the $\cart$-product  $\HB{k_1}{n_1} \cart \HB{k_2}{n_2}$ will result in ``two-dimensional'' arboreal arenas with a hyperpalquette defined by $n_1 \times n_2$ points with $n_1$+$n_2$ hyperedges. The $\cart$ product naturally generalizes to artbitrary ``dimensions'' $d$, for $\HT_1 \cart \HT_2 \ldots 
\cart \HT_d$ is naturally defined recurseively obtaining a {\em higher dimensional hypertree}.  For $\cart^d \HB{k}{n}$ has $d$-hypercubes with $n^d$ sites and $d n^{d-1}$-hyperedges. We shall term $\HB{k}{2} \cart \HB{2}{2} =\B{k}\cart\B{2}$ as an {\em extruded} $k$-tree.

A key point to be noted is that aside from whenever both $k$ and $n$ are not equal to $2$ (we exclusively apply this condition to when we refer to an arboreal arena), the $\HB{k}{n}$ does not represent a tessellation of a manifold, in the sense of a manifold is covered, for example, by simplices/cells\cite{Nakahara2003}. These $\HB{k}{n}$ with $k\ne2$ or $n\ne2$ offer a natural scaffold to construct ``non-manifold'' arenas to explore possible new physics in them. Additionaly, our construction has the advantage that $\cart^d \B{2}$ indeed is the $d$-dimensional cubic lattice and so the physics of models defined on manifold arenas can acessed  in the same framework.

\section{Arboreal Gauge Theories}
\label{sec:AGT}

\subsection{$\Ztwo$ Gauge Theories}

In this section we will construct and study gauge theories on multi-dimensional arenas. We will focus on $\Ztwo$ gauge structure  We begin with the two-dimensional arboreal arena -- $  \B{k_1} \cart \B{k_2}$ with $k_1 > 2$. The coordinate of any site $s$ on this arena is given by $((g_1,m_1),(g_2,m_2))$ where $g_i \in \{ 0,1,2,\ldots \}$, $m_i= 0, g_i = 0$ or $m_i \in \{0,1,\ldots,k_i (k_i-1)^{g_i-1} \}$. The links are defined by unordered pairs of sites such as $[((g_1,m_1),(g_2,m_2)),((g_1,m_1),(g_2+1,m_2'))]$, or  $[((g_1,m_1),(g_2,m_2)),((g_1+1,m_1'),(g_2+1,m_2))]$. On each of these links, labelled by $I$, a qubit (spanned by $\ket{\uparrow}_I$ and $\ket{\downarrow}_I$) is placed, and the tensor product of all of these two dimensional vector spaces define the full Hilbert space of our system (which could be modified by Gauss' law, see below).
The Hamiltonian of the system is
\beq\label{eqn:Z2GT}
H = -J \sum_p B_p - h \sum_I X_I
\eeq
where $X_I$ is the Pauli operator (other operator of interest is $Z_I$) on the link $I$, $p$ stands for a plaquette which consists of four links denoted by $I/p$  $[((g_1,m_1),(g_2,m_2)),((g_1+1,m_1'),(g_2,m_2))]$ (link along the ``$1$-direction''), $[((g_1+1,m_1'),(g_2,m_2)),((g_1+1,m_1'),(g_2+1,m_2'))]$ (link along the ``$2$-direction''), $[((g_1,m_1),(g_2+1,m_2')),((g_1+1,m_1'),(g_2+1,m_2'))]$, and 
$[((g_1,m_1),(g_2,m_2)),((g_1,m_1),(g_2+1,m_2'))]$,  such that
\beq
B_p =\prod_{I/p} Z_{I/p}.
\eeq
The system is invariant to a local (gauge) transformation defined at any site $s$ by
\beq
A_s = \prod_{I/s} X_{I/s}
\eeq
where $I/s$ are the links that touch the site $s$. Of course, $A_s$ commutes with $B_p$, for all $s,p$. The question we pose is nature of the ground state as a function of $h/J$.

\subsection{Ground State and Excitations}

Taking $J>0$, we see that that a ground state  of the system is  $\ket{\Uparrow} =\prod_I \ket{\uparrow}_I$ such that $B_p = 1$ for all $p$. Excitations above the ground state are gapped; indeed, flipping the spin on the link along the ``1-direction'' e.~g., $[((g_1,m_1),(g_2,m_2)),((g_1+1,m_1'),(g_2,m_2))]$, flips $k_2$ plaquettes with an energy penalty of $2 k_2 J$. Similarly, flipping a spin on a link along the $2$-direction produces $k_1$ plaquettes with $B_p = -1$ and an energy cost of $2 k_1 J$. 

The ground state for small $h$, $h \ll J$ can be obtained by noting the gapped nature of the system at $h=0$. We see, using standard perturbation theory,  that the effective toric code Hamiltonian is 
\beq\label{eqn:HToricCode}
H_{\textup{eff}} = - J \sum_p B_p - K \sum_s A_s
\eeq
where
\beq
K = C(k_1,k_2) \frac{h^{k_1 + k_2}}{J^{k_1+k_2 -1}}
\eeq
and $C(k_1,k_2)$ is a positive number. We thus see that the ground state for $h \ll J$ requires $A_s = 1$ leading to
\beq\label{eqn:TCGS}
\ket{GS, h \ll J} = \prod_s (1+A_s) \ket{\Uparrow}.
\eeq
quite similar to what is found in the usual toric code\cite{Kitaev2003}. 
Although the ground state bears a strong resemblance to that found in the usual toric code, there is more interesting physics in the arboreal arena. 

Consider the excitations in the system. First, we have the ``electric charges'' of the gauge theory, described by states where $A_s = -1$. Just as in the toric code, a pair of these charges appear at the sites connected by the link $I$ when we operate $Z_I$ on the ground state, with an energy cost of $4K$. Remarkably, these two charges can be ``moved away'' from each other arbitrarily ``far away'' by successive operation of $Z$ operators, while keeping the energy fixed. Just as the toric code, the arboreal $\Integers_2$ gauge theory is thus electrically deconfined when $h \ll J$. 

\begin{figure}
    \centering
     \includegraphics[width=0.95\columnwidth]{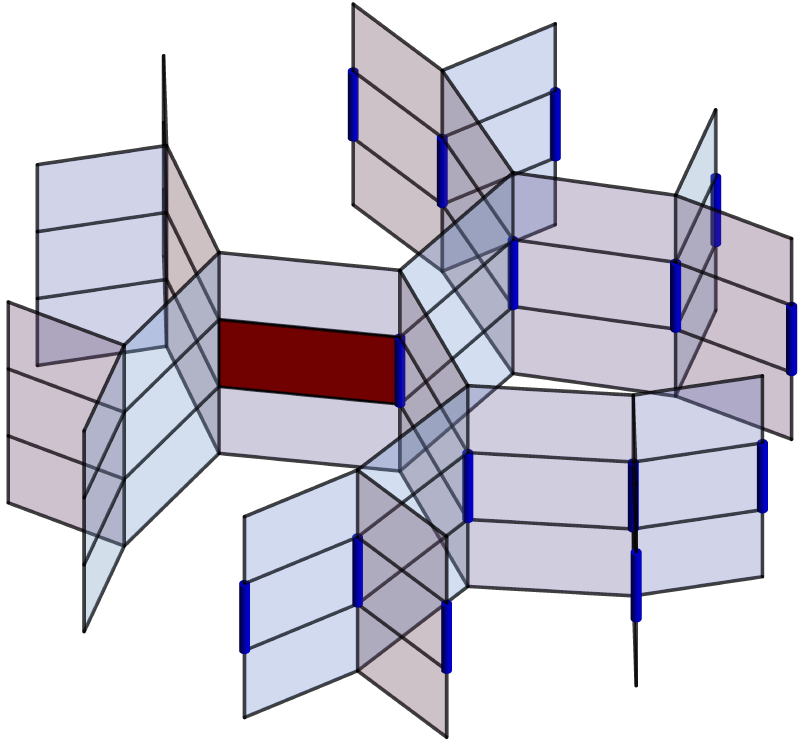}
    \caption{An isolated monopole excitation with energy $2J$ created in an arboreal arena $\B{3} \cart \B{2}$ created from the state $\ket{\Uparrow}$. The links with thick dark blue lines have flipped spins. The monopole in  this case is a lineon. In the general $\B{k_1} \cart \B{k_2}$ ($k_1,k_2>2$) arena for which an illustration is difficult, it is immobile, i.~e., a fracton. }
    \label{fig:fracton_monopole}
\end{figure}

The crucial difference with the usual toric code is in the magnetic plaquette excitations or {\em monopoles}. Action of $X_I$ on the link $I$ on the ground state produces $k_1$ flipped plaquettes with an energy cost of $2k_1 J$ if the link $I$ is in the ``2-direction'', and $k_2$ flipped plaquettes with energy cost $2 k_2 J$ if the link $I$ is in the ``$1$-direction''.  The difference with the toric code is  most easily seen when $k_2 \ge 2$. The set of $k_1 (k_2)$ monopoles in the arboreal arena cannot be separated from each other while being in the energy degenerate subspace, unlike in the usual 2 dimensional toric code where the monopoles can be freely separated arbitrarily apart without recourse to any further excited states. One might suspect that the situation in the arboreal arena is akin to a toric code on a three dimensional cubic lattice~\cite{Castelnovo2008}, but there is, again a crucial difference. One can produce {\em an isolated monopole } (plaquette excitation) with energy $2J$ on the infinite arboreal arena, unlike in the the three dimensional toric code where plaquette excitations are necessarily associated with loop like entities. Stated in other words, the plaquette excitations are point like excitations in the two dimensional arboreal arena (hence naturally called monopoles), while the plaquette excitations of the three dimensional toric code are loop-like  (not point) excitations. In particular,  an {\em isolated monopole} with energy $2J$ can be created in an arboreal arena. This is readily demonstrated by an explicit construction. The state (see \figref{fig:fracton_monopole})
\beq\label{eqn:Monopole}
\ket{M} = X_{I(0,0)} \left(\prod_{g=1}^{\infty} \prod_{m_g=0}^{k_1 (k_1-1)^{g-2} -1} X_{I(g,m_g)} \right) \ket{GS, h\ll J}
\eeq
where
\beq
I(g,m_g) = \begin{cases}
[((0,0),(0,0)),((0,0),(1,0))], & g=0,m_g=0\\
[((g,m_g),(0,0)),((g,m_g),(1,0))], & g \ne 0
\end{cases}
\eeq
contains an isolated monopole on the plaquette defined by the sites $((0,0),(0,0)), ((0,0),(1,0)), ((1,k_1-1),(1,0)), ((1,k_1-1),(0,0))$. A key observation that follows is that if $k_2 > 2$, then the isolated monopole is {\em completely immobile}, since attempts to move it by local spin flips necessarily produces additional monopoles. However, when $k_2=2$, the monopole is mobile solely along the 2-direction as shown in \figref{fig:fracton_monopole}. When $k_2>2$, the monopole is, thus, an immobile fracton, while for $k_2=2$ (extruded tree) the monopole is {\em lineon} as it can move freely along the 2-direction. We thus arrive at the inevitable conclusion that {\em  even  the simplest gauge theory on the arboreal arena is  ``magnetically fractonic''!}.

\begin{figure}
    \centering
     \includegraphics[width=0.95\columnwidth]{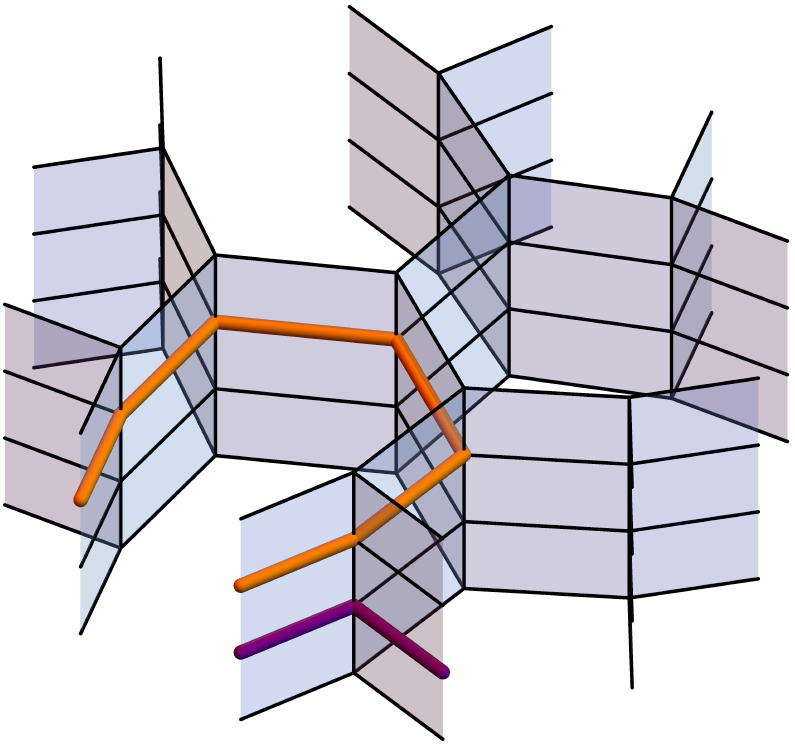}
    \caption{Boundary operators and global Wilson lines in a finite arboreal arena. The purple solid line runs over the links such that the product of $Z$ operators on these links define a boundary operator. The orange solid line shows a global Wilson line (again a product of $Z$ operators of all the links that make up the orange line).
    }
    \label{fig:BoundaryOperators}
\end{figure}

\subsection{Ground State Degeneracy}

It is natural to enquire if the novel aspects of gauge theory on arboreal arenas uncovered above manifest in other properties such as the ground state degeneracy. This is best studied focusing on two-dimensional Cayley trees. The ground state degeneracy, and associated topological order, can be studied using finite systems with smooth and rough boundaries\cite{Bravyi1998}. An important point to be noted in the construction of gauge theory on {\em finite} arboreal arenas is that, in addition to the $B_p$ terms that are associated  naturally to plaquettes that arise from the Cartesian product of graphs, there are additional gauge invariant local operators that arise at the boundary. By ``local operators'', we here mean the interactions that are defined on adjacent links (which share a site) such that the number of links are not more than four (the number of links that defines a plaquette). An instance of the boundary operator is illustrated in \figref{fig:BoundaryOperators}. Every such local boundary operators that arise in the case of the rough boundaries, collectively denoted by $H_\dou$, 
\beq\label{eqn:HwithDou}
H = -J \sum_p B_p -h \sum_I X_I + H_\dou,
\eeq
commutes with the Hamiltonian.

We will obtain the ground state degeneracies $D_G$ of two dimensional Cayley trees by considering four cases. First we consider completely smooth two dimensional trees of the type $\SC{k_1}{M_1} \cart \SC{2}{M_2}$, which is a smooth extruded Cayley tree. Here the number of qubits are
\beq\label{eqn:nqubits}
\begin{split}
&  N_{Q}\left(\SC{k_1}{M_1} \cart  \SC{2}{M_2}\right)  = \\ 
&  \frac{k_1 \left(\left(k_1-1\right){}^{M_1}+M_2 \left(4
   \left(k_1-1\right){}^{M_1}-2\right)-1\right)-4 M_2}{k_1-2}
   \end{split}  
\eeq
The number of conserved plaquette operators $B_p$ is
\beq
N_B\left(\SC{k_1}{M_1} \cart  \SC{2}{M_2}\right) = \frac{2 k_1 M_2 \left(\left(k_1-1\right){}^{M_1}-1\right)}{k_1-2}
\eeq
and the number of conserved charges are 
\beq\label{eqn:NA}
N_A\left(\SC{k_1}{M_1} \cart  \SC{2}{M_2}\right) = \frac{\left(2 M_2+1\right) \left(k_1 \left(k_1-1\right){}^{M_1}-2\right)}{k_1-2}.
\eeq
There are no constraints on the $B_p$, but  $\prod_s A_s =1$ when the boundaries are smooth. Further, there are no boundary operators, i.~e., $H_\dou = 0$.
In this smooth two dimensional Cayley tree, therefore, we see that the ground state  is non-degenerate.'
% i.~e.,
% \beq
% \ln{_2 \left[ D_G\left(\SC{k_1}{M_1} \cart  \SC{2}{M_2}\right) \right]} = N_Q - (N_B + N_A - 1) = 0.
% \eeq

Next, we consider two-dimensional Cayley trees with rough boundaries. The first case we consider is $\RC{k_1}{M_1} \cart \SC{2}{M_2}$, a rough extruded tree. The number of qubits in this case is
\beq
\begin{split}
& N_Q(\RC{k_1}{M_1} \cart \SC{2}{M_2}) = \\
&  \frac{k_1^2 \left(2 M_2+1\right) \left(k_1-1\right){}^{M_1}-k_1
   \left(\left(k_1-1\right){}^{M_1}+2 M_2+1\right)-4 M_2}{k_1-2}
\end{split}  
\eeq
and the number of conserved plaquettes $B_p$ is
\beq
N_Q(\RC{k_1}{M_1} \cart \SC{2}{M_2}) = \frac{2 k_1 M_2 \left(\left(k_1-1\right){}^{M_1+1}-1\right)}{k_1-2}
\eeq
while the number of conserved charges is same as in \eqnref{eqn:NA}. The new feature here, as mentioned above, are the boundary terms $H_\dou$; we a number of additional terms in the Hamiltonian that are gauge invariant and describe the interactions of four (or less) adjacent links. The number of such terms is in $H_\dou$ is
\beq\label{eqn:extrudeBO}
N_\dou(\RC{k_1}{M_1} \cart \SC{2}{M_2}) = \left(k_1-2\right) k_1^2 \left(k_1-1\right){}^{M_1-2},
\eeq
each of which commutes with the Hamiltonian \eqnref{eqn:HwithDou}. There are no global constraints on the plaquettes, charges or the boundary operators. We thus obtain the ground state degeneracy $D_G$
\beq\label{eqn:ExtrudeRSDegen}
\begin{split}
\ln{_2} D_G & = N_Q -(N_B + N_\dou + N_A) \\ 
&= k_1 \left(k_1-1\right){}^{M_1-2}-1
\end{split}
\eeq
which results in a degeneracy of the ground state whose logarithm scales as the exponential of the system size $M_1$! What is the origin of such large degeneracies? These degeneracies can be traced to the global Wilson line operators 
\beq
W(I_a,I_b) = \prod_{I=I_{b}}^{I_a} Z_I
\eeq
where the index $I$ runs over links that provide the ``shortest path'' from boundary link $I_b$ to $I_a$ (see \figref{fig:BoundaryOperators}). It must be noted that these  Wilson lines are not all of the same length. The shortest of them will be six links long, while the longest of them will contain links of the order $M_1$. Due to this reason, the degeneracy of the ground state is not ``topologically protected'' -- perturbations that span over six links can mix states with distinct values of the short Wilson lines. However, if the perturbations are short ranged, spanning at most over $L \ll M_1$ links, a large number  
of these degenerate states cannot be perturbed, and degeneracy survives. This is akin to the topological protection of the ground state degeneracy in the toric code giving rise to the notion of topological order~\cite{Wen2017}. Taking a cue from this we term the degeneracy induced by the large number of ``global Wilson lines'' to be {\em arboreal topological order}. This notion along with a fractonic monopole excitation and a fully mobile charge excitation provide a novel form of quantum matter in the arboreal arena. 

Most interestingly, the large degeneracy discussed above is absent if the a two dimensional Cayley tree of the type $\SC{k_1}{M_1} \cart \RC{2}{M_2}$ is considered (it be can show that the degeneracy in this case is 2). This observation further corroborates in the importance of the tree structure ($k_1 > 2$) in providing for the large number of global Wilson lines.

We now consider the last type of extruded tree $\RC{k_1}{M_1} \cart \RC{2}{M_2}$ where all boundaries are rough. Here the number of qubits are
\beq
\begin{split}
& N_Q( \RC{k_1}{M_1} \cart \RC{2}{M_2}) = \\
&\frac{k_1 \left(\left(k_1 \left(2 M_2+1\right)+1\right)
   \left(k_1-1\right){}^{M_1}-2 M_2-1\right)-4 \left(M_2+1\right)}{k_1-2}.
 \end{split}  
\eeq
The number of conserved plaquettes are
\beq
 N_B( \RC{k_1}{M_1} \cart \RC{2}{M_2}) = \frac{2 k_1 \left(M_2+1\right)
   \left(\left(k_1-1\right){}^{M_1+1}-1\right)}{k_1-2}
\eeq
along with number charges being  given \eqnref{eqn:NA}. There are no independent boundary terms when both Cayley trees are rough.
Notably, there are constraints on $B_p$. Indeed, we have
\beq
\prod_{p \in \text{WS}} B_p = 1
\eeq
where $\textup{WP}$ is a ``Wilson surface'' whose ``$1$-direction'' is the global Wilson line and the ``$2$-direction'' is along the 1-d chain forming the $\RC{2}{M_2}$ Cayley tree. The total number of such constraints are
\beq
\begin{split}
N_C( \RC{k_1}{M_1} \cart \RC{2}{M_2}) = k_1 \left(k_1-1\right){}^{M_1}-1
\end{split}
\eeq
All these considerations results in the ground state degeneracy given by
\beq
\begin{split}
& \ln{_2 D_G}   = N_Q -( N_B +  N_A - N_C) = 0\end{split}
\eeq
That the ground state of this system on the arboreal areana is non-degenerate, a feature that  it shares with a fully rough square lattice toric code~\cite{Bravyi1998} define by $\RC{2}{M_1} \cart \RC{2}{M_2}$ which is also non-degenerate.

We can extend these considerations to generic two-dimensional Cayley lattices, i.~e., with $k_1, k_2 > 2$.  For $\SC{k_1}{M_1}\cart \SC{k_2}{M_2}$, we find that the ground state is non degenerate. For $\RC{k_1}{M_1}\cart \SC{k_2}{M_2}$, the ground state state degeneracy $\ln{_2 D_G( \RC{k_1}{M_1}\cart \SC{k_2}{M_2})} \sim (k_1-1)^{M_1 - 2}$, with an analogous result for the smooth-rough case. For the rough-rough case, we, again, obtain a non-degenerate ground state.

\subsection{Properties of Excitations}
As noted above, the excitations above the gapped ground state in the deconfined phase ($h \ll J$) of the gauge theory are the deconfined electric charges and the fractonic magnetic monopoles. We now discuss the generalized braiding properties of these excitations. Consider an isolated monopole (as discussed near equation \eqnref{eqn:Monopole}). Consider any ``surface'' $S$ containing this monopole plaquette such that a site at the boundary of this surface has an electric charge.  We consider this surface to be simply connected in that it has ``no holes'' etc. This charge can be transported ``around'' the monopole plaquette by the operator
\beq
T_e = \prod_{I \in \dou S} Z_I
\eeq
such that the set of links from a closed loop. Presence of the monopole is detected by the condition that $T_e = -1$, i.~e., a deconfined electric charge will pick up an Aharonov-Bohm phase of $\pi$ when transported around a magnetic monopole.

Curiously, the process of transporting a monopole around an electric charge is not so straightforward. If, for example, the process used for the electric charge is applied with the role of the electric and magnetic charges reversed and a transport operator of the from $T_m = \prod_{I \perp \dou S} X_I$ (where $I$ are now links not in included in the surface $S$, but attached to the boundary vertices), $T_m \ket{\Psi}$ will not, in general, restore the state back to $\ket{\Psi}$. To alleviate this problem, we consider a generalized ``braiding process'' by the following construction. Consider an electric charge located at a site $s_e$ (i.~e., $A_{s_e} = -1$). Now consider a ``volume'' $V$ that consists this site. The volume $V$ consists of a set of sites and links (i.~e., is a subgraph of arboreal arena $\B{k_1} \cart \B{k_2}$) such that every two sites are path connected, i.e., the volume is ``simply connected''. Further, for any link present in $V$, both the sites connected to it are present in $V$. A site in the volume $V$ is called an interior point if all the links incident  on the site are present in the subgraph $V$. Thus the boundary of $V$ consists of vertices $v$ in $V$ such that some of the links of these vertices are not included in $V$. The links of the boundary points not included in $V$ are the the boundary links and denoted by ${\cal L}(\dou V)$. The volume $V$ also does not contain any ``holes''. This is ensured by the condition that there is a ``boundary path'' connecting any two boundary points which contain only sites that are boundary sites. Now consider the operator
\beq
T_m = \prod_{I \in {\cal L}(\dou V)} X_{I}
\eeq
which acts on the boundary links of $V$. Although this operator does not enjoy the direct interpretation as the transport operators of monopole charges, it detects the presence of electric charges in the volume $V$. Indeed for the state $\ket{\Psi}$ with the electric charge at $s_e$ discussed above, $T_{m} \ket{\Psi} = - \ket{\Psi}$. Further, this operator generalizes to any number of charges, as is immediately evident from the fact that $T_m = \prod_{s \in V} A_s$.  As a concrete example of this consider $\Integers_2$ gauge theory defined on $\B{k_1} \cart \B{k_2}$ with an electric charge present at the sites $((0,0),(0,0))$. Now consider the the subgraph $V =\SC{k_1}{G_1} \cart \SC{k_2}{G_2}$ which is made of Cayley trees with $G_1$ and $G_2$ generations. The boundary links of $V$ will now be those links that are present in $\RC{k_1}{G_1} \cart \RC{k_2}{G_2}$, but not present in $V$. Indeed, it is immediate that $T_m = -1$, i.~e., this operator detects the presence of the electric charge at $((0,0),(0,0))$.

\subsection{Ground State Phases}

Having established the state in the regime $h \ll J$, we investigate opposite regime where $h \gg J$. For $J=0$, the state is given by a product state $\ket{\Rightarrow} = \prod_I \ket{\rightarrow}_I$ where $X_I \ket{\rightarrow}_I = \ket{\rightarrow}_I$. The state is non-degenerate and gapped with a gap of order $h$. For finite $J$ (with $h/J \gg 1$) we see that plaquette terms only produce a dispersive change of the ground state energy of order $J^2/(8h)$ and the state $\ket{\Rightarrow}$ continues to be the ground state. Further, the electric charges confined -- two adjacent charges can be separated over $L$ links only via an energy penalty of order $2Lh$.

These observations raise the  natural question apropos the nature of  quantum transition from the deconfined arboreal ordered state to a confined state up on the tuning of $h$. We explore this question using a variational approach aimed at understanding the overall physics~\cite{DusuelVidal2015,ReissSchmidt2019,MuhlhauserSchmidt2020}. We work in the sector of the Hilbert space that imposes the Gauss' law, $A_s = 1$. The wave function we use,
\beq\label{eqn:GSb}
\ket{\GS(b)} = C(b) \left[\prod_{p} ( 1+ b B_p) \right] \ket{\Rightarrow}
\eeq
where $b$ is a real variational parameter, and $C(b)$ is a normalization constant, and $\ket{\Rightarrow} = \prod_I \frac{1}{\sqrt{2}} \left(\ket{\uparrow}_I + \ket{\downarrow}_I \right)$. The ground state occurs when $b=b_{\GS}$ at which $C(b)^2\bra{\GS(b)}H\ket{\GS(b)}$ is minimized.  This variational state recovers the exact ground state for $h \ll J$ when $b_{\GS}=1$, and the ground state $\ket{\Rightarrow}$ for $h \gg J$ when $b_{\GS}=0$.

\begin{figure}
    \centering
    \includegraphics[width=0.95\columnwidth]{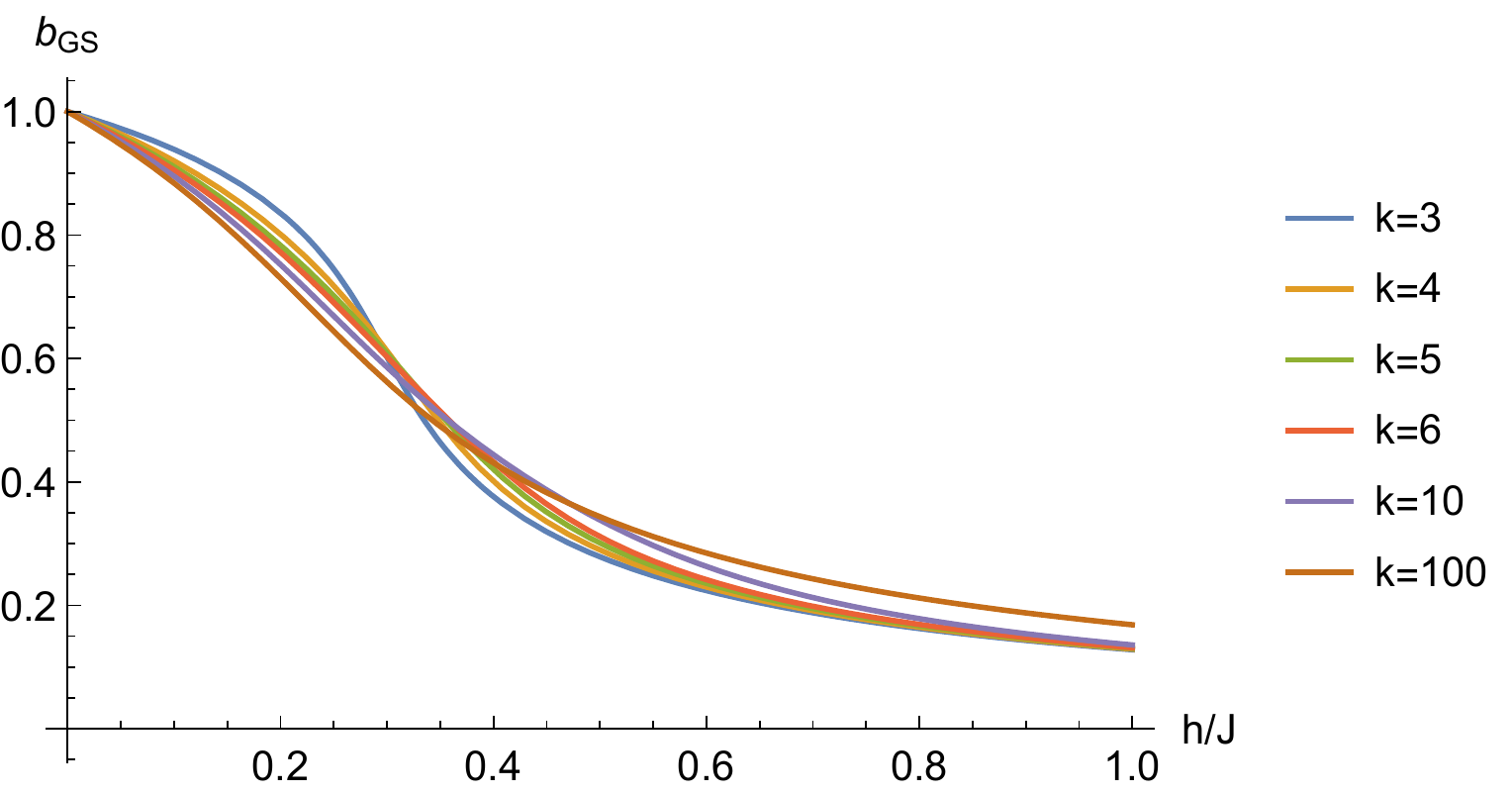}
    \caption{Plot of $b_{\GS}$ as a function of $h/J$ for a $\Integers_2$ gauge theory \eqnref{eqn:HwithDou} defined on $\SC{k}{M}\cart \SC{2}{M_2}$-- {\em extruded $k$-Cayley trees with smooth-smooth boundaries} with $M,M_2$ large. There is no quantum phase transition and topological order does not persist at finite $h$ for any $k \geq 3$.}
    \label{fig:Var_Manyp_S}
\end{figure}
\begin{figure}
    \centering
    \includegraphics[width=0.95\columnwidth]{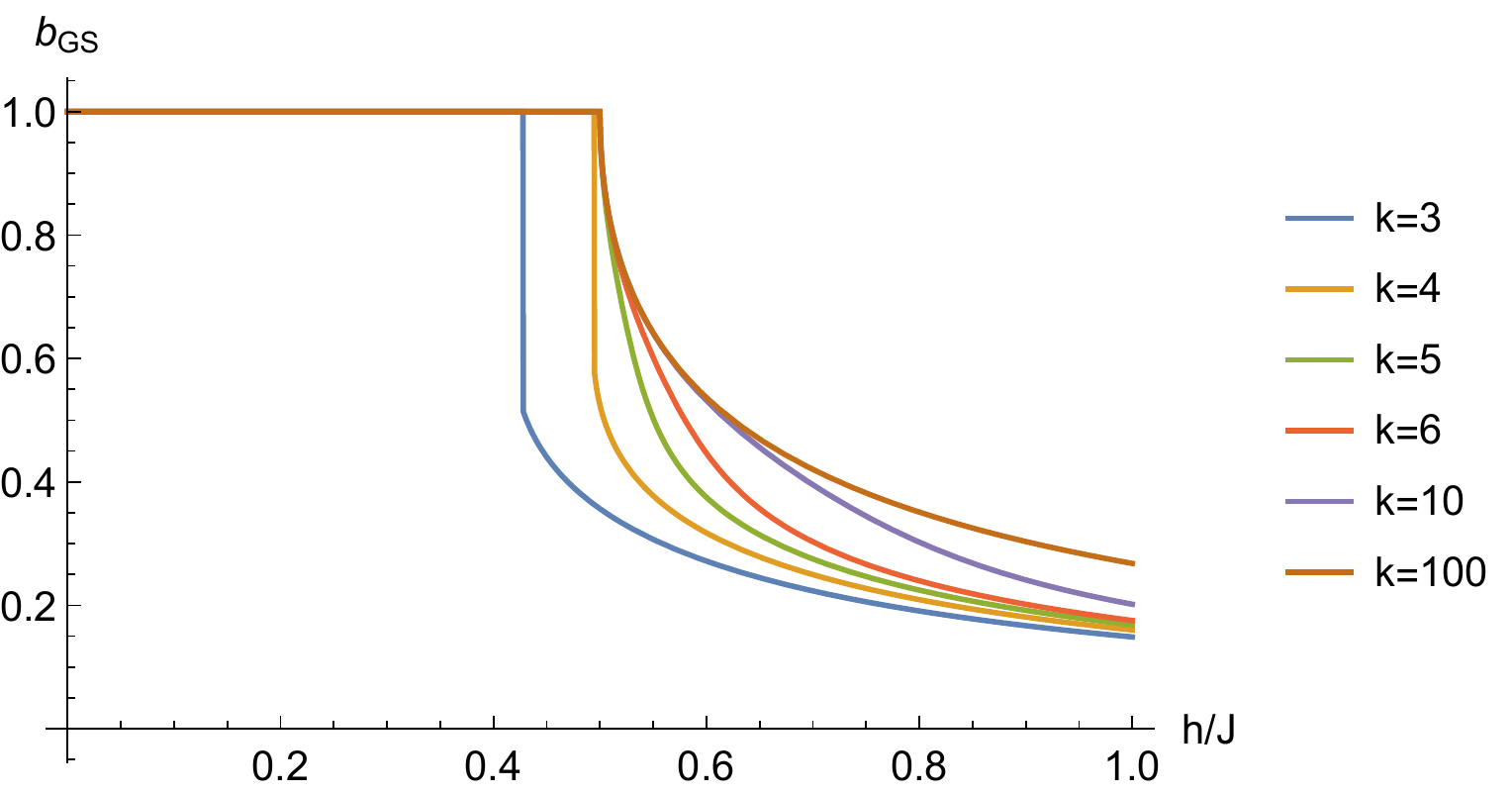}
    \caption{ Plot of $b_{\GS}$ as a function of $h/J$ for a $\Integers_2$ gauge theory \eqnref{eqn:HwithDou} defined on $\RC{k}{M}\cart \SC{2}{M_2}$-- extruded $k$-Cayley trees with rough-smooth boundaries ($M,M_2$ large). There is a deconfined phase for $h < h_c$ and a confined phase for $h>h_c$. The transition is first order for $k=3$ ($h_c \approx 0.42 J$) and $k=4$ ($h_c = 0.49 J$). For all $k \ge 5$, a continuous transition from the deconfined to the confined phase occurs at $h_c= J/2$. The same results are obtained for fully rough trees of the type $\RC{k}{M} \cart \RC{2}{M_2}$.}
    \label{fig:Var_Manyp_R}
\end{figure}

The details of the variational calculations are presented in \appref{sec:VarCalc}. Here we describe the key results. First, we note that the variational state \eqnref{eqn:GSb} recovers the continuous phase transition of the $\Integers_2$-gauge theory on the square lattice at an $h_c = 0.25J$ (to be compared with the exact result of 0.22$J$ (cf.~ \cite{TupitsynProkofev2010}), albeit with Landau critical exponents. 

The physics in the arboreal arena is richer. Starting with extruded trees, we see that boundary conditions of the tree play a crucial role in determining the phases of the gauge theory on the arboreal arena. For $\SC{k}{M} \cart \SC{2}{M_2}$, we find that in the limit of large $M,M_2$, the state changes smoothly up on the tuning of $h/J$ (see \figref{fig:Var_Manyp_S}). On the other hand for $\RC{k}{M_1} \cart \SC{2}{M_2}$, the system does encounter a phase transition (\figref{fig:Var_Manyp_R}). Quite interestingly, the nature phase transition found in this variational description depends on the value of $k$. For $k=3$ and $k =4$ the transition is {\em first order}, while for $k \ge 5$, the transition is continuous. For $k=3$, the critical value $h_c \approx 0.418625 J$ and for $k=4$, $h_c \approx 0.49295 J$. For $k \ge 5$, the continuous quantum phase transition occurs at $h_c = J/2$ as can be obtained from analytical considerations. The same results are obtained for fully rough extruded $k$-Cayley trees ($\RC{k}{M} \cart \RC{2}{M_2}$).

\begin{figure}
    \centering
    \includegraphics[width=0.99\columnwidth]{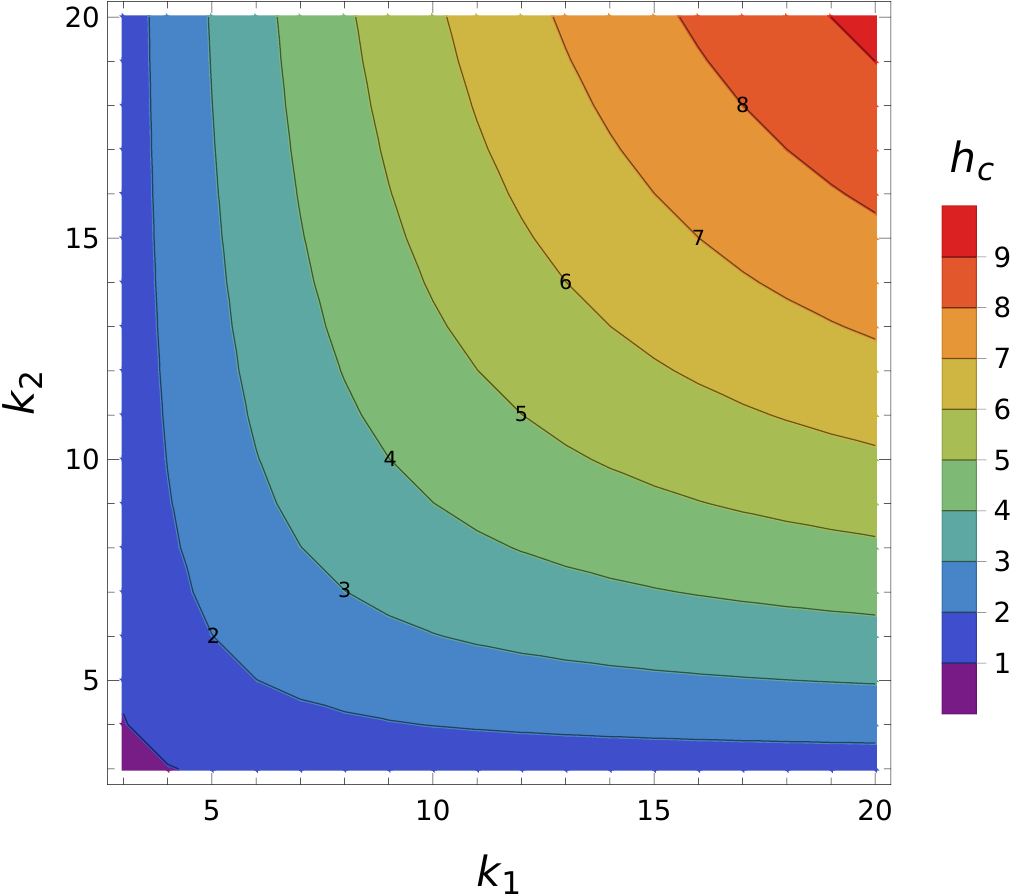}
    \caption{ Critical value $h_c$ of $h$ for general finite arboreal lattices $\RC{k_1}{M_1} \cart \RC{k_2}{M_2}$ ($M_1,M_2$ large) with rough boundaries. There is a deconfined phase for $h < h_c$ and a confined phase for $h>h_c$. }
    \label{fig:hcRR}
\end{figure}

Considering more general two dimensional arboreal lattices, we find that both $\SC{k_1}{M_1} \cart \SC{k_2}{M_2}$ and $\RC{k_1}{M_1} \cart \SC{k_2}{M_2}$ do not undergo a phase transition with increasing $h$, i.~e., there is no deconfined phase in these systems. The rough-rough Cayley trees $\RC{k_1}{M_1} \cart \RC{k_2}{M_2}$, on the other hand, have a first order transition up on increase of $h$. In these systems, there is a deconfined phase for $h < h_c$ and a confined phase for $h > h_c$. The dependence of $h_c$ on $k_1$ and $k_2$ is plotted in \figref{fig:hcRR}. For $k_1 = k_2$, $h_c$ increases monotonically and approximately linearly with increasing $k_1$ ($h_c \approx 0.5 k_1$ for large $k_1$). On the other hand keeping $k_1$ fixed, and increasing $k_2$, $h_c$ saturates to a value determined by the fixed value of $k_1$.

\subsection{Dual Models}
\label{sec:Duality}

The results discussed above raise a set intriguing questions pertaining to the crucial role played by the boundary conditions on the phases obtained in the arboreal arena. While this may not be unexpected considering the fact that the number of boundary degrees of freedom of the arboreal arena are of the same order as the number of ``bulk'' degrees of freedom, significant insights are obtained by constructing and studying dual models.
%that reveal precisely how the boundary degrees of freedom determine the physics in these systems. 

\begin{figure}
    \centering
    \includegraphics[width=\columnwidth]{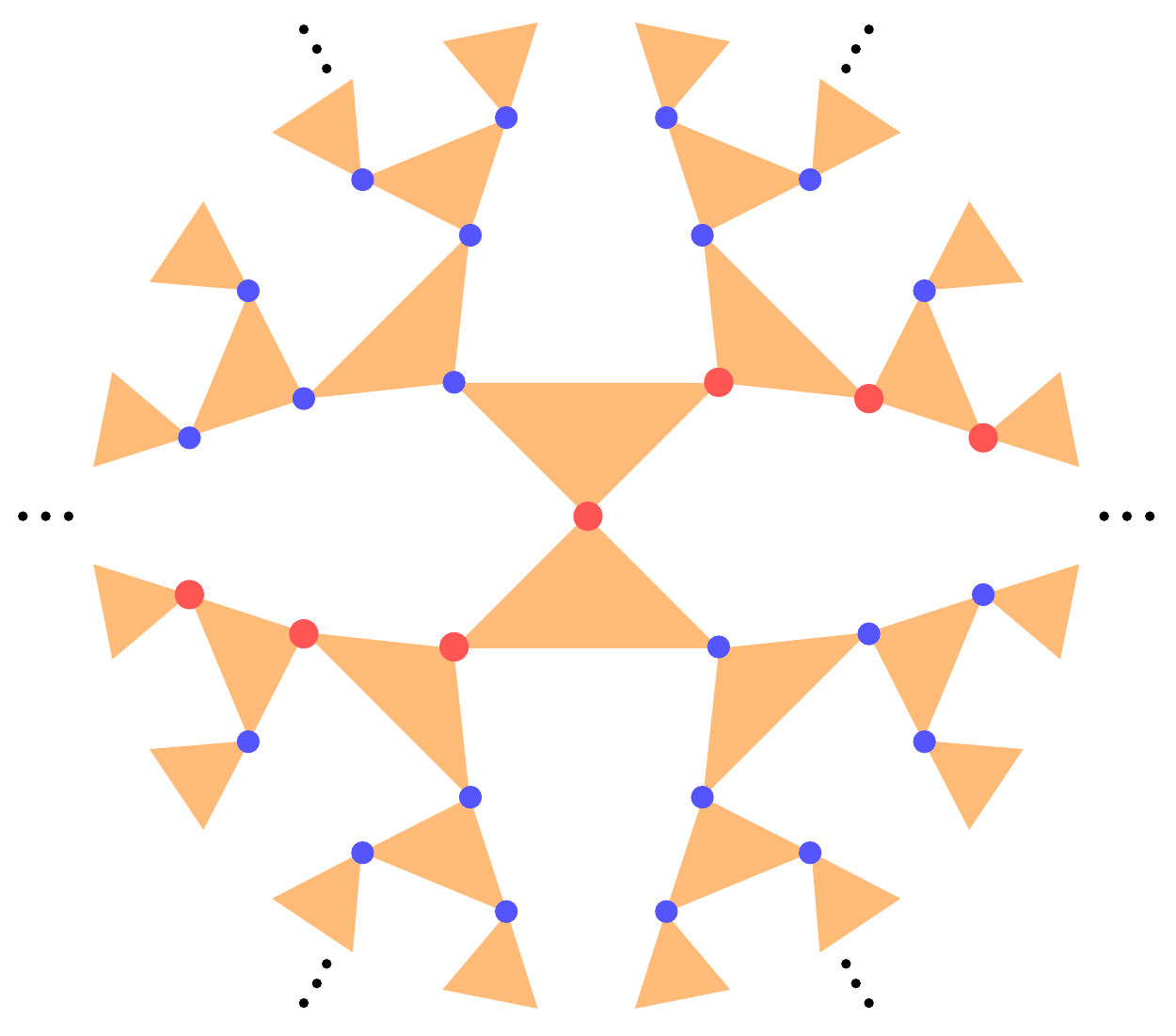}
    \includegraphics[width=\columnwidth]{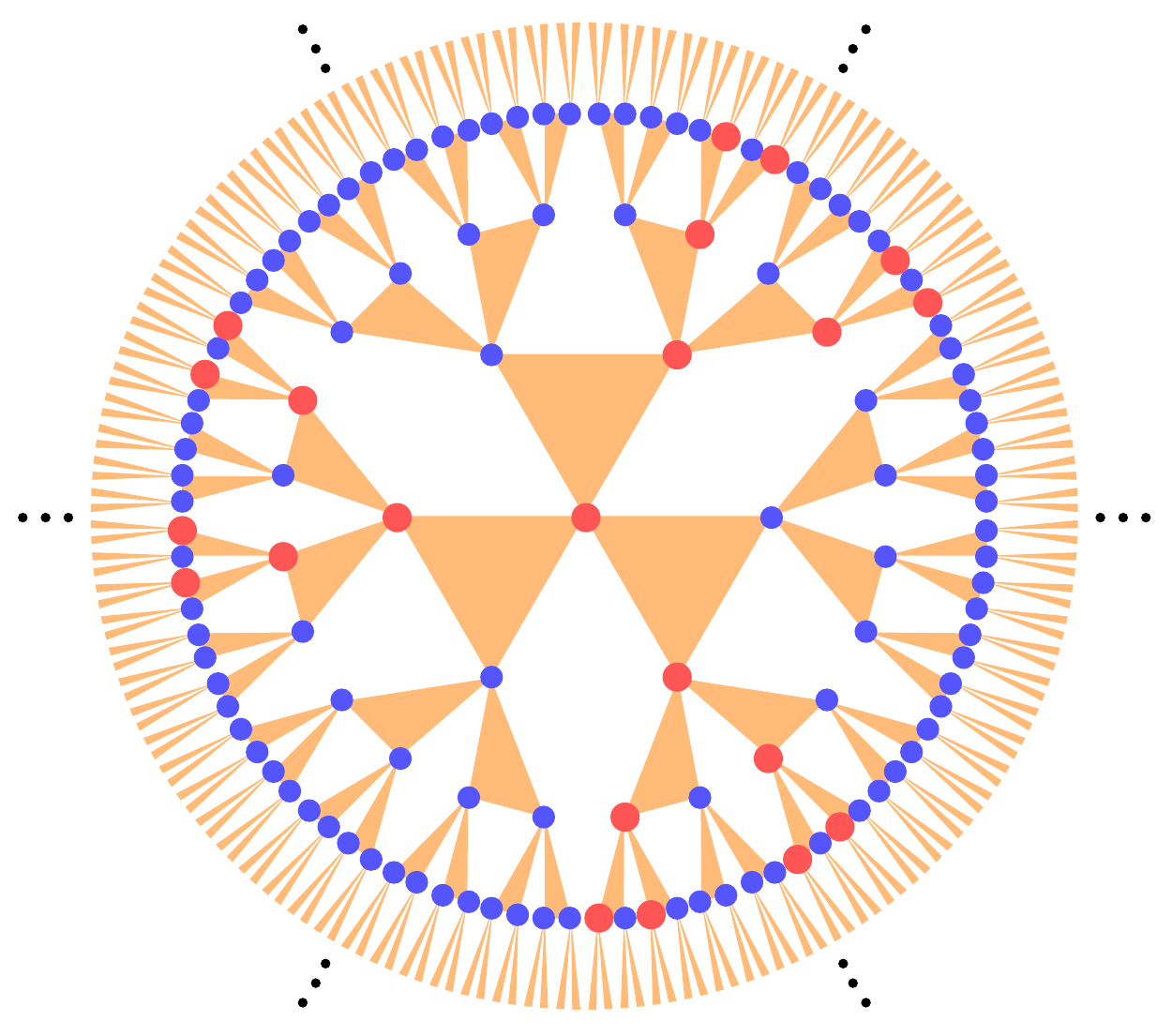}
    \caption{The subsystem symmetry operators for the GQIM are given by applying $X$ on the highlighted (red) spins as shown. Note that the symmetry operators for GQIM on $\HB{2}{3}$ and $\HB{3}{3}$ live on $\B{2}$ and $\B{3}$ respectively.}
    \label{fig:subsystem_symmetry}
\end{figure}

\begin{figure}
    \centering
    \includegraphics[width=0.9\columnwidth]{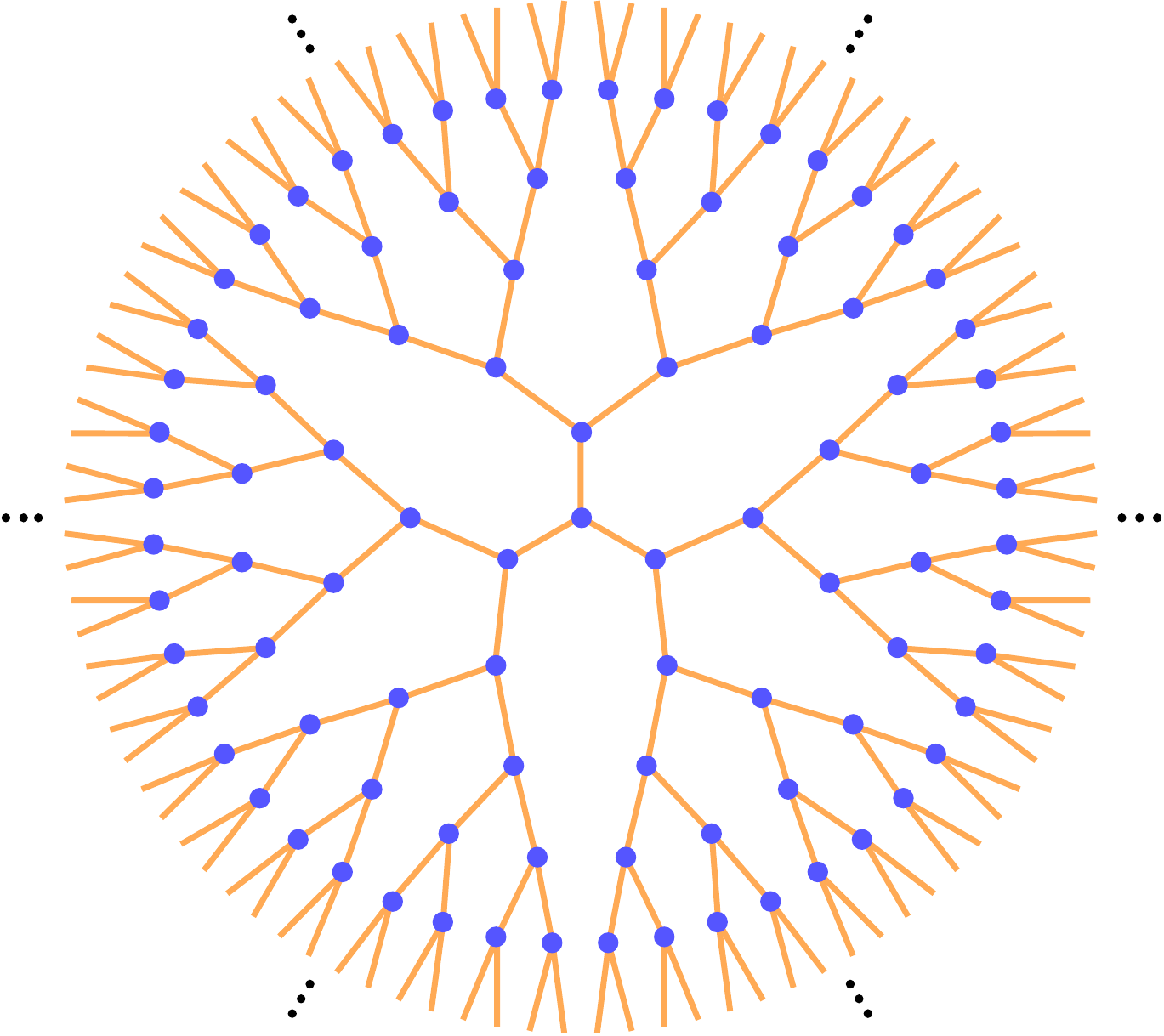}
    \medskip
    \includegraphics[width=0.9\columnwidth]{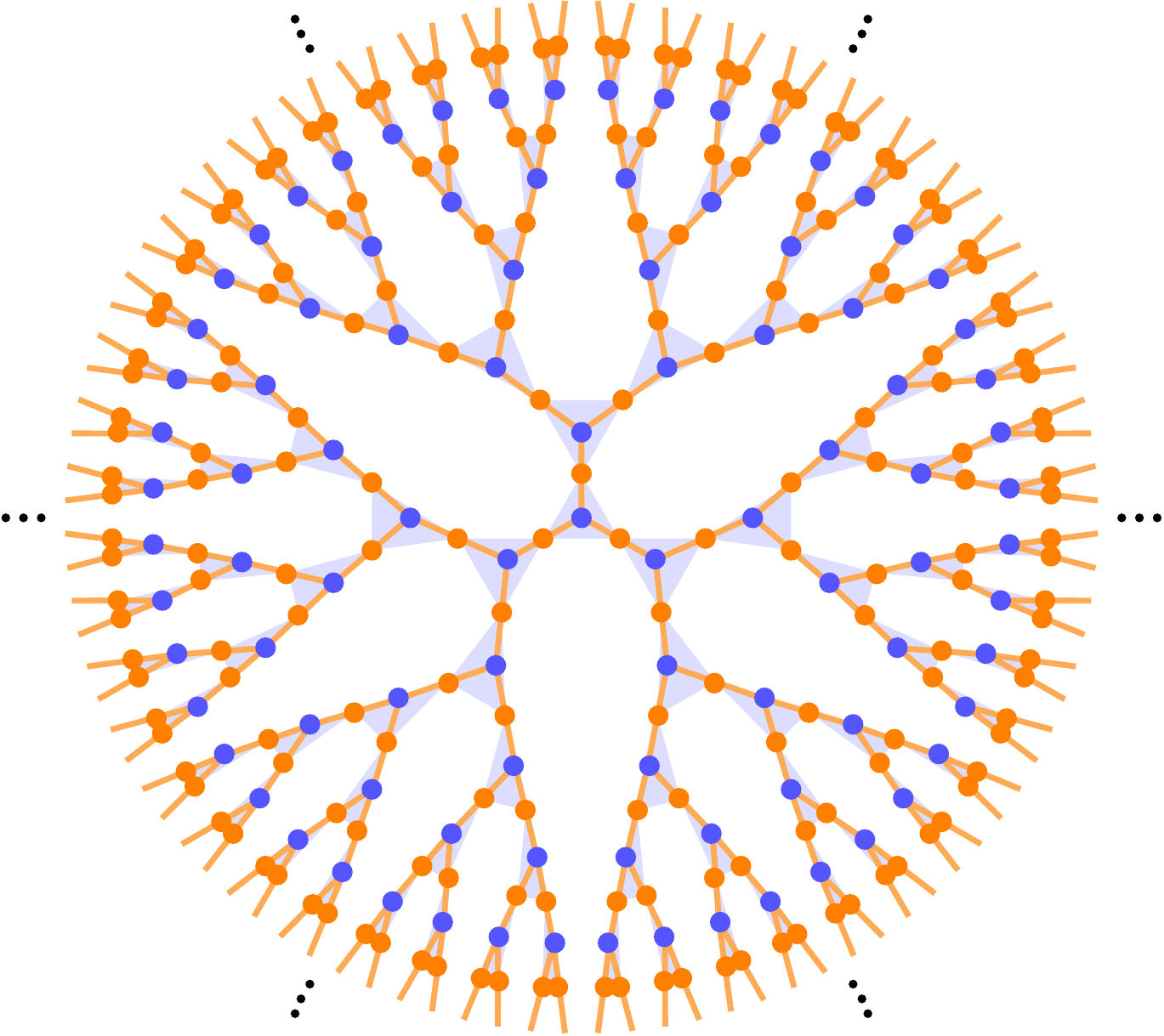}
    \medskip
    \includegraphics[width=0.9\columnwidth]{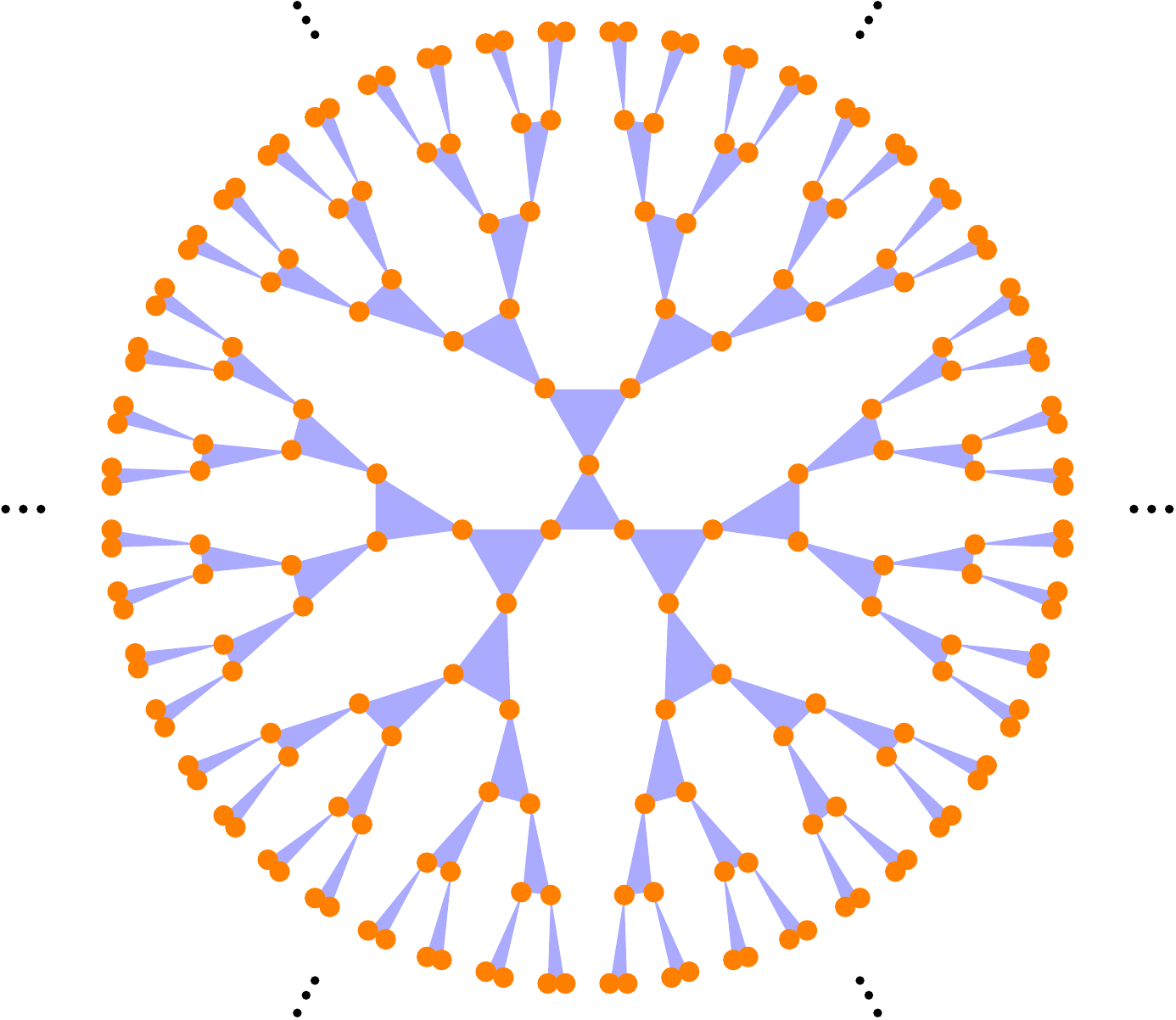}
    \caption{Illustrating the generalized Kramers-Wannier duality of generalized quantum Ising models on $\HB{k}{n}$ and $\HB{n}{k}$. Colour scheme shows the mapping between sites and edges. Dual qubits (orange dots) are placed at the centres of the links of the top panel as shown in the middle panel. Hyperlinks (light blue triangles) are defined using the dual qubits, to obtain the dual hypertree shown in the bottom panel.}
    \label{fig:KW}
\end{figure}

Before we discuss models dual to gauge theories defined in general arboreal arenas, we note that many of the dualities that we uncover can be elucidated using a basic duality in a hyper-Bethe lattice which we dub as the arboreal Kramers-Wannier duality. Consider a generalized quantum Ising model (\GQIM) defined on a Hilbert space of qubits placed on the sites of a hypertree $\HB{k}{n}$ with the Hamiltonian
\beq\label{eqn:GQIM}
H = - J \sum_{I} \left(\prod_{s/I} Z_s \right) - h \sum_s X_s.
\eeq
Here $I$ labels the links and $s$ the sites, respectively of $\HB{k}{n}$. The first term defines the generalized Ising interaction on the hyperlink $I$  that is a product of $Z$-operators on $n$ sites that belong to hyperlink $I$.  Interestingly, this \GQIM ~has a {\em subsystem symmetry}  for all $n > 2$, in that flipping spins on a subset of sites $S$ (such as those shown in \figref{fig:subsystem_symmetry}), described by the transformation operator $U_S = \prod_{s\in S} X_s $,  leaves the Hamiltonian $H$ (\eqnref{eqn:GQIM}) invariant. There are many such distinct subsets of sites, and these generate all the subsystem symmetries.

To find a dual, we introduce a second  set of qubits located at the {\em centres} of the links $I$, and define operators $\tilde{Z}_I$ and $\tilde{X}_I$ that act on them. We then make the following identifications
\begin{align}
\tilde{X}_I &\equiv \prod_{s/I} Z_s\label{eqn:tilX}\\
\prod_{I/s} \tilde{Z}_I &\equiv X_s \label{eqn:tilZ}
\end{align}
that preserve all the necessary algebraic relations between the operators $X_s$ and $\prod_{s/I} Z_s$. The dual Hamiltonian is 
\beq
\tilde{H} = -J \sum_{I} \tilde{X}_I - h \sum_s \left( \prod_{I/s} \tilde{Z}_I \right),
\eeq
where $I$ define {\em sites} and $s$ define the links of a  hyper-Bethe lattice $\HB{n}{k}$ (see \figref{fig:KW}). This dual model also has a subsystem symmetry analogous to the original model provided $k>2$. Since the lattice is infinite we have, from \eqnref{eqn:tilX} that $\prod_{I \in S'} \tilde{X}_I = 1$ where $S'$ the subset of sites as shown in \figref{fig:subsystem_symmetry}, and similarly $\prod_{s \in S} \tilde{X}_s = 1$ (following \eqnref{eqn:tilZ}, see \cite{Radicevic2018}). We thus conclude that the \GQIM~defined on $\HB{k}{n}$ is dual to \GQIM~defined on $\HB{n}{k}$,  when restricted to the sub-system symmetry singlet sectors of both models. This is the statement of the arboreal Kramers-Wannier duality.

\begin{figure}
    \centering
     \includegraphics[width=0.9\columnwidth]{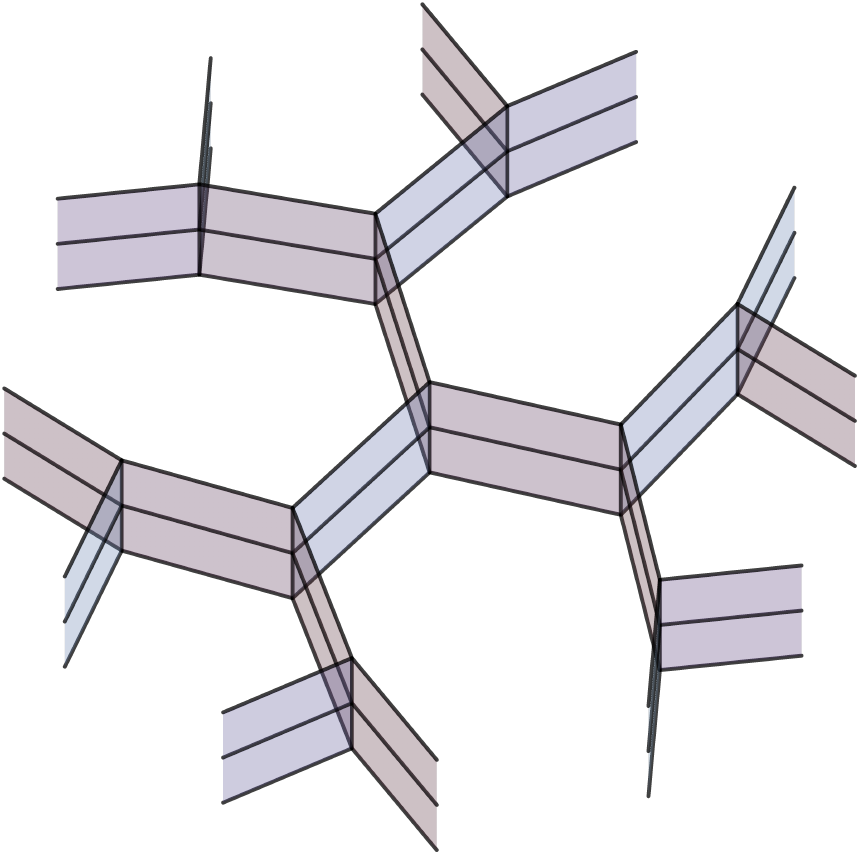}
     \includegraphics[width=0.9\columnwidth]{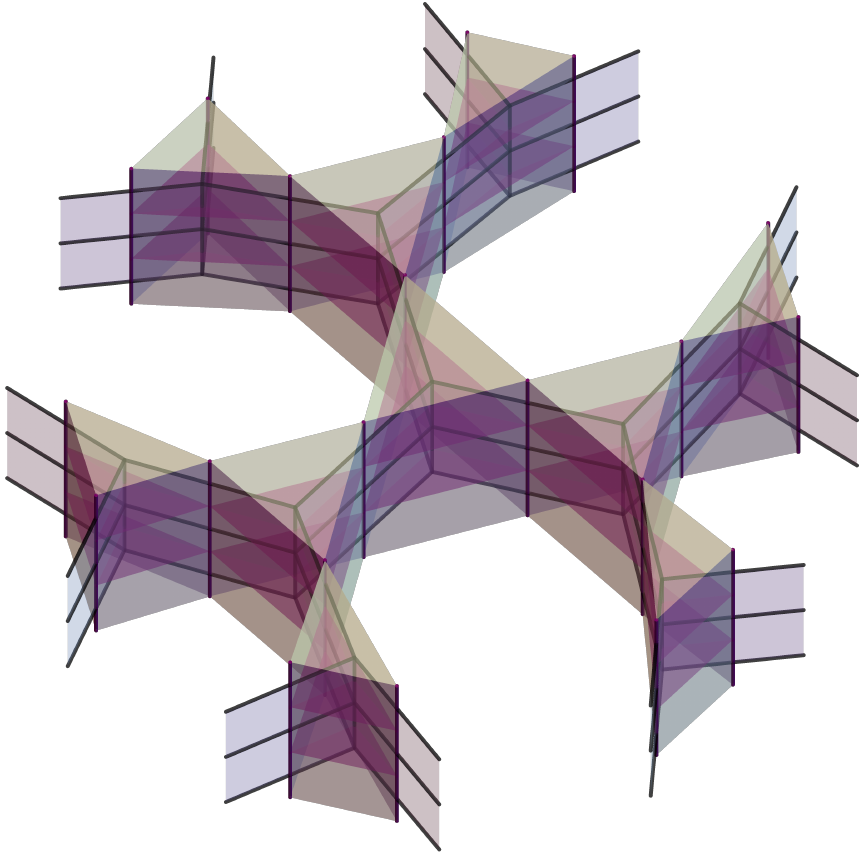}
     \includegraphics[width=0.85\columnwidth]{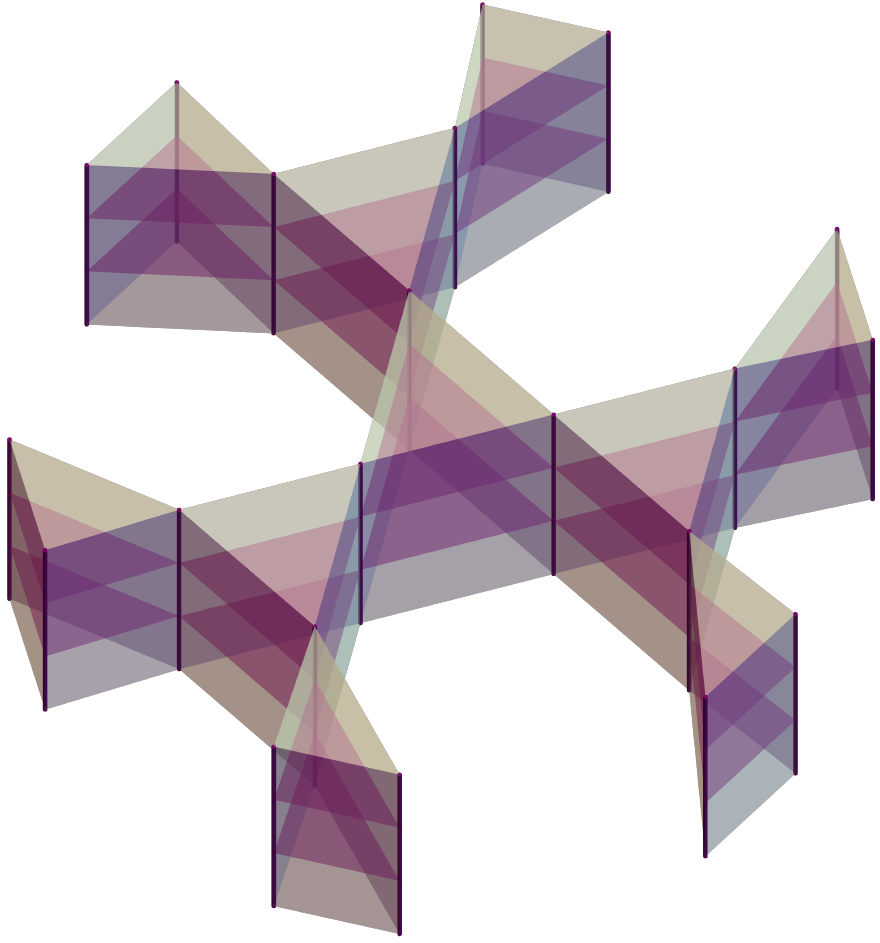}
    \caption{Theory dual to $\Integers_2$ gauge theory defined on $\B{3} \cart \B{2}$ (extruded tree). Construction of the dual is analogous to \figref{fig:KW}. Dual qubits are placed at the centres of the plaquettes of the top panel, and connected by hyperlinks as shown in the middle panel. The resulting dual model defined on $\HB{2}{3} \cart \HB{2}{2}$ is shown in the bottom panel. }
    \label{fig:ETDuality}
\end{figure}

Armed with the arboreal Kramers-Wannier duality, we next  construct a model dual  to the $\Integers_2$ gauge theory (\eqnref{eqn:Z2GT}) defined on $\B{k} \cart \B{2}$, i.~e., an infinite extruded tree. Recall that the Hilbert space of this theory is defined by a set of qubits $(Q)$ that reside on the links of this extruded tree, and satisfy the Gauss' law constraint $A_s = 1$, all sites $s$. The key sets of operators that act on this Hilbert space are $B_p$, plaquette operators where $p$ labels the plaquettes, and $X_I$ the transverse field operators on the links labelled by $I$. These operators satisfy the following relations
\beq\label{eqn:GTOperators}
\begin{split}
[B_p,B_{p'}] &= 0, \forall~\textup{plaquettes}\, p, p' \\
[X_I,X_{I'}] &= 0, \forall~\textup{links}\, I, I' \\
[B_p,X_I] &= 
0, \;\textup{if link} \; I \; \textup{is not a part of the plaquette $p$} \\ \{B_p, X_I \} &= 0, \; \textup{if link} \; I \; \textup{is part of the plaquette } \; p
\end{split}
\eeq
with $[~,~]$ and $\{~,~\}$ denoting, respectively, the commutator and anti-commutator.
Note that there are two types of links, labelled by $I_1$ which are ``along the 1-direction, or tree direction'', and by $I_2$ which are ``along the 2-direction, or extrusion direction''.
To obtain the dual model, we define a different Hilbert space made of qubits $(\tilde{Q})$ placed at the centers of each plaquettes and define operators $\tilde{Z}_p$ and $\tilde{X}_p$ that act on these new qubits which are naturally labelled by the plaquettes $p$. We  make the following dual identifications
\beq\label{eqn:Dualization}
\begin{split}
B_p &\equiv \tilde{X}_p \\
X_{I_1} &\equiv \prod_{p/I_1} \tilde{Z}_p \\
X_{I_2} &\equiv \prod_{p/I_2} \tilde{Z}_p
\end{split}
\eeq
where the links $I_1$ and $I_2$ are, as described above, along the $1$ and $2$ directions respectively. Note that the product $p/I_2$ runs over $k$ plaquettes and thus $X_{I_2}$ dualizes to a $k$-qubit generalized Ising interaction. Similarly, $X_{I_1}$ dualizes to a $2$-site Ising interaction. The resulting dual Hamiltonian is
\beq\label{eqn:GQIMgt}
\tilde{H} = -J \sum_p \tilde{X}_p - h \sum_{I_1} \left(\prod_{p/I_1} \tilde{Z}_p \right) - h \sum_{I_2}   \left(\prod_{p/I_2} \tilde{Z}_p \right).
\eeq
A study of \figref{fig:ETDuality} reveals that this is a generalized quantum Ising model (\GQIM) defined on a two-dimensional hypertree $\HB{2}{k}\cart\HB{2}{2}$.  Most interestingly, this dual model has an apparent ``subsystem'' Ising symmetry, where transformation $U_W = \prod_{p \in W} \tilde{X}_p$ leaves the system unchanged. Here, the plaquettes $p$ belong to $W$, the global Wilson surface of the $\Integers_2$ gauge theory on the extruded tree introduced earlier (see \figref{fig:BoundaryOperators} and \figref{fig:ETDuality}). Every distinct global Wilson surface $W$ produces a symmetry operation $U_W$ that acts only on a subset of qubits and in this sense is ``subsystem''. Of course, viewed from the the perspective of the original operators $U_W = \prod_p B_p$ which are constrained, i.~e., $\prod_p B_p = 1$ since each qubit in a plaquette is shared by one other plaquette in $W$. This forces, $U_W = 1$, which allows us to make a remarkable conclusion:  the theory dual to  $\Ztwo$ gauge theory $H$ (\eqnref{eqn:Z2GT}) with Gauss' law constraint ($A_s = 1, \forall s$) defined on $\B{k}\cart\B{2}$ is the generalized quantum Ising model (\GQIM, \eqnref{eqn:GQIMgt}) defined on $\HB{2}{k} \cart \HB{2}{2}$ in its singlet sector of its subsystem symmetry. This result also provides a nice connection to earlier work. Many known fractonic gauge theories defined on lattice (such as the X-cube model) are dual to models with subsystem symmetries. Indeed, as we discussed above the simple  $\Integers_2$ gauge theory $H$ (\eqnref{eqn:Z2GT}) defined on $\B{k}\cart\B{2}$ does support fractonic excitations, and its duality to the \GQIM~on $\HB{2}{k} \cart \HB{2}{2}$ with subsystem symmetries fits nicely into this picture.

Finally, we mention the model dual to an $\Integers_2$ gauge theory defined on a general two dimensional arboreal lattice $\B{k_1} \cart \B{k_2}$ with Gauss' law constraint $A_s = 1$ imposed. Using the procedure developed above, we see that the dual theory is the \GQIM~defined on two-dimensional arboreal lattice $\HB{2}{k_1} \cart \HB{2}{k_2}$. This $\GQIM$ has a set of subsystem symmetries constructed using the Wilson surfaces of the gauge theory, and duality holds in the singlet sector of all generators of the subsystem symmetries. These results can be generalized to higher dimensional arboreal lattices.

%Can the duality help us gain insights into the physics of system with boundaries? The answer is in the affirmative, and this is best motivated by the example of GQIM defined on smooth and rough hypertrees. 

We will now explore the dualities in finite arboreal lattices. First consider a \GQIM~defined on a smooth hyper-Cayley tree $\SHC{k}{n}{M}$. This tree consists sites $s$ that are bulk sites and boundary sites (sites of the $M$-th generation), while all links $I$ are bulk links  The Hamiltonian is same as \eqnref{eqn:GQIM} and this system possesses subsystem symmetries effected by simultaneous spin flips on the subset of sites $S$ such as shown in  \figref{fig:subsystem_symmetry}.

On the rough hyper-Cayley tree $\RHC{k}{n}{M}$, all the sites $s$ are bulk sites while there are both bulk links and boundary links (which are the last set of links). The Hamiltonian is defined as
\beq\label{eqn:GQIMonRHC}
H = -J \sum_{I \in \textup{bulk links}}\left(\prod_{s/I} Z_s \right) - J \sum_{I_\dou \in \textup{boundary links}} Z_{s/I_\dou} - h \sum_s X_s
\eeq
Note that this Hamiltonian does not have an global flip symmetries and the definition is motivated by the anticipation of a duality. %\ttd{Another point we can make is that the above Hamiltonian has a $Z$-magnetic field on the boundary spins, which are a large number. Therefore, we do not expect a magnetic phase transition as this is like a ferromagnet with a $Z$ magnetic field. This can later be used as an explanation as to why smooth boundary gauge theory does not have a phase transition.}

We can naturally extend the definition of $\GQIM$ to higher dimensional arboreal lattices with boundaries. Indeed a $\GQIM$ defined on $\SHC{k_1}{n_1}{M_1} \cart \SHC{k_2}{n_2}{M_2}$ has a Hamiltonian
\beq
H = -J \sum_{I_1} \underbrace{\left( \prod_{s/I_1} Z_s \right)}_{n_1 terms} - J \sum_{I_2} \underbrace{\left( \prod_{s/I_2} Z_s \right)}_{n_2 terms} - h \sum_s X_s
\eeq
where $I_{1,2}$ are hyperlinks along the 1 and 2 directions respectively. This model, again, has a large number of subsystem symmetries whenever $k_1,k_2 >2$. More general higher dimensional lattices with different boundary conditions can similarly be defined.

We now show that $\GQIM$ defined on $\SHC{k}{n}{M}$ is dual to a \GQIM~defined on %\footnote{This finite rough hyper-Cayley tree is slighly different from the one constructed in section.~\ref{SECTION}\ttd{section} in this hypertree is constructed with all vertices on the central link being treated as generation 1.}
$\tRHC{n}{k}{M}$ (see Table~\ref{tab:Catalogue} for definitions).  To this end, identify each link $I$ of $\SHC{k}{n}{M}$  with a {\em site} of $\tRHC{n}{k}{M}$ on which a dual qubit is placed. For each link $I$ of $\SHC{k}{n}{M}$ define,
\beq
\tilde{X}_I \equiv \prod_{s/I} Z_s.
\eeq
Further, associate with each bulk sites $s$ of  $\SHC{k}{n}{M}$ a hyperlink of $\tRHC{n}{k}{M}$ (this dual  hyperlink will touch all the dual sites that are the hyperlinks $\SHC{k}{n}{M}$ which touch the site $s$), and define
\beq
\prod_{I/s}\tilde{Z}_I \equiv X_s, \;\;\; s \in \textup{bulk sites of}~ \SHC{k}{n}{M}
\eeq
We now see that the boundary sites of $\SHC{k}{n}{M}$ will be identified with the dual {\em boundary links} of $\tRHC{n}{k}{M}$, such that 
\beq
\tilde{Z}_{I/s_\dou} \equiv X_{s_\dou}, \;\;\;\;  s_\dou \in \textup{boundary sites of}~ \SHC{k}{n}{M}
\eeq
We see the Hamiltonian \eqnref{eqn:GQIM} defined on $\SHC{k}{n}{M}$  dualizes to
\beq
\tilde{H} = -J \sum_{I} \tilde{X}_I - h \sum_{s \in \textup{bulk sites}} \left(\prod_{I/s}\tilde{Z}_I \right) -h \sum_{s_\dou \in  \textup{boundary sites}} \tilde{Z}_{I/s_\dou}
\eeq
which is exactly the \GQIM~defined on $\tRHC{n}{k}{M}$ with coupling constants $J$ and $h$ interchanged (see \eqnref{eqn:GQIMonRHC}). 
Finally,  consider $\prod_{s \in S} X_s$ (which defines a subsystem symmetry transformation of the \GQIM defined on $\SHC{k}{n}{M}$, see \figref{fig:subsystem_symmetry}, which maps under duality to
\beq \label{eqn:tilZcons}
\prod_{s \in S} \prod_{I} \tilde{Z}_{I/s} = 1.
\eeq
In other words, the duality is operates in the singlet sector of all the subsystem symmetries of \GQIM~defined on  $\SHC{k}{n}{M}$.

The discussion above provides a platform for us to discuss the theory dual to $\Integers_2$ gauge theory defined on $\SC{k_1}{M_1} \cart \SC{k_2}{M_2}$. By placing dual qubits on the faces of the plaquettes of the arboreal lattice, and performing identifications similar to \eqnref{eqn:Dualization} we see that dual theory is  a \GQIM~defined on $\tRHC{2}{k_1}{M_1} \cart \tRHC{2}{k_2}{M_2}$. The rough boundaries arise from the identification of $X$ operators at the boundary sites of $\SC{k_1}{M_1} \cart \SC{k_2}{M_2}$ with the $\tilde{Z}$ operators on the sites of $\tRHC{2}{k_1}{M_1} \cart \tRHC{2}{k_2}{M_2}$ that host its boundary links. The Gauss' law constraint of the gauge theory is identically satisfied in its dual description. Finally, the duality is valid in the singlet sector where certain 't-Hooft operators $\prod_{s \in T} X_s = \prod_{s \in T} \left( \prod_{I/s} \tilde{Z}_{I} \right) = 1$, which is analogous to the singlet condition discussed earlier.

Turning now to the $\Integers_2$ gauge theory defined on $\RC{k_1}{M_1} \cart \RC{k_2}{M_2}$, we see immediately that 
theory is dual to \GQIM~defined on $\tSHC{2}{k_1}{M_1+1} \cart \tSHC{2}{k_2}{M_2+1}$ with the dual qubits, placed again on the faces of the plaquettes of $\RC{k_1}{M_1} \cart \RC{k_2}{M_2}$, and the dualization effected via \eqnref{eqn:Dualization}. Now, there are restrictions on the subsystem symmetries of the dual \GQIM~defined on $\tSHC{2}{k_1}{M_1+1} \cart \tSHC{2}{k_2}{M_2+1}$ arising from the constraints imposed by ``surfaces'' such as the Wilson surfaces discussed above, for $\prod_{p \in W} \tilde{X}_p = \prod_{p \in W} B_p = 1$. Thus the duality operates in the singlet sector of the subsystem symmetries of the of the dual \GQIM~defined on $\tSHC{2}{k_1}{M_1+1} \cart \tSHC{2}{k_2}{M_2+1}$.

Finally, we note that the  $\Integers_2$ gauge theory defined on $\SC{k_1}{M_1} \cart \RC{k_2}{M_2}$ is dual to \GQIM~defined on $\tRHC{2}{k_1}{M_1} \cart \tSHC{2}{k_2}{M_2+1}$ using procedure outlined above. The duality operates in the singlet sector 't-Hooft operators along the 1-direction. 
The dualities discussed here are summarized in \tabref{tab:dualities}. 

\begin{table*}
    \centering
    \begingroup
\setlength{\tabcolsep}{12pt} % Default value: 6pt
\renewcommand{\arraystretch}{1.8} % Default value: 1
    \begin{tabular}{||c|c||c|c||}
    \hline \hline
       \multicolumn{2}{||c||}{Model} & \multicolumn{2}{c||}{Dual Model} \\
       \hline \hline
       Hamiltonian & Arena & Hamiltonian & Arena \\
       \hline \hline
        \GQIM & $\HB{k}{n}$ & \GQIM &  $\HB{n}{k}$  \\ \hline
        
        $\Ztwo$-GT & $\B{k_1} \cart \B{k_2}$ & \GQIM & $\HB{2}{k_1} \cart \HB{2}{k_2}$ \\ \hline
        \Xcube-GT & $\B{k_1} \cart \B{k_2} \cart \B{k_3}$ & GFIM & $\HB{2}{k_1} \cart \HB{2}{k_2} \cart \HB{2}{k_3}$ \\ \hline
        \GQIM & $\SHC{k}{n}{M}$ & \GQIM & $\tRHC{n}{k}{M}$ \\ \hline
        \GQIM & $\RHC{k}{n}{M}$ & \GQIM & $\tSHC{n}{k}{M+1}$ \\ \hline
        $\Ztwo$-GT & $\SHC{k_1}{n_1}{M_1} \cart \SHC{k_2}{n_2}{M_2}$ & \GQIM &  $\tRHC{n_1}{k_1}{M_1} \cart \tRHC{n_2}{k_2}{M_2}$\\ \hline
        $\Ztwo$-GT & $\RHC{k_1}{n_1}{M_1} \cart \RHC{k_2}{n_2}{M_2}$ & \GQIM &  $\tSHC{n_1}{k_1}{M_1+1} \cart \tSHC{n_2}{k_2}{M_2+1}$\\ \hline
        $\Ztwo$-GT & $\SHC{k_1}{n_1}{M_1} \cart \RHC{k_2}{n_2}{M_2}$ & \GQIM &  $\tRHC{n_1}{k_1}{M_1} \cart \tSHC{n_2}{k_2}{M_2+1}$\\
        \hline \hline
    \end{tabular}
    \endgroup
    \caption{ Summary of dualities. Most of the dualities operate in the singlet sectors of certain operators on either side of the duality. \GQIM: Generalized quantum Ising model (\eqnref{eqn:GQIM} and \eqnref{eqn:GQIMonRHC}), $\Ztwo$-GT: $\Ztwo$ gauge theory (\eqnref{eqn:Z2GT} and \eqnref{eqn:HwithDou}), \Xcube-GT : \Xcube~gauge theory (\eqnref{eqn:XGT}). GFIM: Generalized quantum face Ising model -- this is the generalization of \GQIM~to include only face interactions. The arena are as described in table.~\ref{tab:Catalogue}. The first three dualities are for infinite arboreal arenas, while the remainder are on finite systems.  }
    \label{tab:dualities}
\end{table*}

The dualities developed above allow us to obtain further insights into the phases of the gauge theory discussed earlier using the variational approach. We  exploit the duality between $\Integers_2$ gauge theory defined on $\B{k_1} \cart \B{k_2}$ and \GQIM~defined on $\HB{2}{k_1} \cart \HB{2}{k_2}$, by redefining the coupling constants of the $\Integers_2$ gauge theory \eqnref{eqn:Z2GT} via $J \to  J/\lambda$ and $h \to \lambda J$ where $J $  (on the r.h.s) is an energy scale. The dual $\GQIM$ on $\HB{2}{k_1} \cart \HB{2}{k_2}$ is obtained as
\beq
\tilde{H} = - \frac{J}{\lambda} \sum_p \tilde{X}_p - \lambda J \sum_{I_1} \underbrace{\left( \prod_{p/I_1} \tilde{Z}_{p/I_1} \right)}_{k_2 \;\; \textup{$\tilde{Z}$ operators}} - \lambda J \sum_{I_2} \underbrace{\left( \prod_{p/I_2} \tilde{Z}_{p/I_2}\right)}_{k_1 \;\; \textup{$\tilde{Z}$ operators}}
\eeq
where $p$ are plaquettes and $I_{1,2}$ are the links in the $1,2$-directions of $\B{k_1} \cart \B{k_2}$; $p$s label the sites and $I_1, I_2$ label the hyperlinks of $\HB{2}{k_1} \cart \HB{2}{k_2}$. The Trotterized finite temperature partition function of this model is 
\beq\label{eqn:TrotterH}
\beta\tilde{H} = -K_\tau \sum_{I_\tau} \left(\prod_{p/I_\tau} Z_{p/I_\tau}  \right) - K \sum_{I_1} \left( \prod_{p/I_1} \tilde{Z}_{p/I_1} \right) - K \sum_{I_2} \left( \prod_{p/I_2} \tilde{Z}_{p/I_2}\right)
\eeq
where $p$, $I_1,I_2,I_\tau$ are the sites and hyperlinks of the a {\em three} dimensional arboreal arena $\HB{2}{k_1} \cart \HB{2}{k_2} \cart \HB{2}{2}$. Here $K_\tau = - \half \ln(\Delta \tau J/\lambda)$ and $K = \Delta \tau \lambda J$. Taking $K_\tau = K$, we see that \eqnref{eqn:TrotterH} is a generalized {\em classical} Ising model defined on $\HB{2}{k_1} \cart \HB{2}{k_2} \cart \HB{2}{2}$. The key point here is the any thermal phase transition at finite $K$ of this model describes the quantum phase transition of the $\Integers_2$ gauge theory defined on $\B{k_1} \cart \B{k_2}$. 

We now study the phases of the theory \eqnref{eqn:TrotterH} using the Bragg-Williams mean-field ansatz. Defining $m = \mean{\tilde{Z}_p}$ (where $\mean{a}$ stands for the thermal average of the quantity $a$),  we get the self consistency relation:
\beq\label{eqn:BWSelfCons}
m = \tanh{\left(K \sum_{\alpha=1}^D 2 m^{k_\alpha -1} \right)}
\eeq
where $D=3$, $\alpha=1,2,3$ and $3$ is the $\tau$-direction \footnote{A more general result for the self consistency for a generalized classical Ising model on $d$ dimensional arboreal arena $\cart_{\alpha=1}^{D} \HB{k_\alpha}{n_\alpha}$ reads as $m = \tanh{\left(K \sum_{\alpha=1}^{d}  k_\alpha m^{n_\alpha -1}\right)}$}. An analysis of the \eqnref{eqn:BWSelfCons} reveals that whenever $k_1,k_2 > 2$, we obtain a first order transition, i.~e., there is a $K_c$ at which a finite $m$ non-trivial solution appears. On the other hand, if $k_1=2$, we obtain a first order transition for $k_2=3,4$, while for $k_2\ge 5$ continuous transition is obtained where a nonzero solution of $m$  begins to appear for $K \ge K_c$ with the solution vanishing at $K=K_c$. It is reassuring that the results obtained from this dual picture qualitatively matches the results obtained using the variational approach apropos the nature of the transition from the confined to the deconfined phase.  

Finally, the duality analysis also offers insight into why the $\Ztwo$ gauge theory defined on arboreal lattices with smooth boundaries do not have a phase transition (see \figref{fig:Var_Manyp_S}). As discussed the dual to this theory is a \GQIM~defined on an arboreal lattice with {\em rough boundaries}. This entails extra boundary terms (see \eqnref{eqn:GQIMonRHC}) which act like a ``boundary magnetic field'' along the $z$ direction on the boundary spins (dual qubits). It is natural that no phase transition occurs in the system due to large number boundary spins which experience this field.

\section{Fractonic Models on Arboreal Arenas}
\label{sec:FMArboreal}

\subsection{\Xcube~model}

In this section, we explore fracton models  defined on the arboreal arenas. We will focus particularly on the X-cube model~\cite{Vijay2016} defined on a three-dimensional arboreal arena. Consider a three dimensional arboreal lattice  $\Bd{3} {k_1,k_2,k_3} \equiv \B{k_1} \cart \B{k_2} \cart \B{k_3}$ where $k_1,k_2,k_3 > 2$ with sites denoted by $s$ and links denoted by $I$. The links of this arboreal lattice can be naturally classified as $1$-links, $2$-links and $3$-links, indicating their ``direction'' (or the parent tree to which they belong). To aid the discussion, we introduce an index $\alpha$ which can take values $1,2,3$. Further, $\alpha'=2,3,1$ and $\alpha''=3,1,2$, respectively, for $\alpha=1,2,3$.

 A set of links of this arboreal arena can act as the bounding links of ``cubes'' with twelve edges. For example, using the coordinate system defined on $\B{k_1},\B{k_2},\B{k_3}$(see~section~ \ref{sec:ArborealArena}), the following twelve links $(s_1,s_2)$, $(s_2,s_3)$, $(s_3,s_4)$, $(s_4,s_1)$, $(s_5,s_6)$, $(s_6,s_7)$, $(s_7,s_8)$, $(s_8,s_9)$, $(s_1,s_5)$, $(s_2,s_6)$, $(s_3,s_7)$, $(s_4,s_8)$ make up a cube. Here the sites $s_i$ are, for example, 
\beq
\begin{split}
s_1 & = ((g_1,m_1),(g_2,m_2),(g_3,m_3)) \\
s_2 & = ((g_1+1,m'_1),(g_2,m_2),(g_3,m_3)) \\
s_3 & = ((g_1+1,m'_1),(g_2+1,m'_2),(g_3,m_3)) \\
s_4 & = ((g_1,m_1),(g_2+1,m'_2),(g_3,m_3)) \\
s_5 & = ((g_1,m_1),(g_2,m_2),(g_3+1,m'_3)) \\
s_6 & = ((g_1+1,m'_1),(g_2,m_2),(g_3+1,m'_3)) \\
s_7 & = ((g_1+1,m'_1),(g_2+1,m'_2),(g_3+1,m'_3)) \\
s_8 & = ((g_1,m_1),(g_2+1,m'_2),(g_3+1,m'_3)) 
\end{split}
\eeq
with $m'_\alpha$ are suitably chosen coordinates such that $((g_\alpha, m_\alpha), (g_\alpha+1,m'_\alpha))$ is a link in $\B{k_\alpha}$. With these definitions, an $\alpha$-link participates in $k_{\alpha'} k_{\alpha''}$ cubes. Similarly a cube face with a ``normal'' in the $\alpha$ direction (this face has four links that define it, two the $\alpha'$-direction and two in the $\alpha''$-direction) is shared by $k_\alpha$ cubes. Finally, every site of the arena participates in $k_1 k_2 k_3$ distinct cubes.

To define the \Xcube~model on $\Bd{3}{k_1,k_2,k_3}$, we introduce a qubit on every link of this three dimensional arboreal arena. For every cube $c$, the magnetic term $B_c$ is introduced as
\beq\label{eqn:XcubeB}
B_c = \prod_{I/c} Z_{I/c}
\eeq
where $I/c$ are twelve links that  make up the cube $c$. Next, for each ``direction'' $\alpha$, we can define the star operator defined at every site 
\beq\label{eqn:XcubeA}
A_{s \alpha} = \prod_{I/s} X_I
\eeq
where $I/s$ are $(k_{\alpha'} + k_{\alpha''})$ links that touch the site $s$ in ``orthogonal'' directions to $\alpha$. The X-cube model is defined as
\beq\label{eqn:XcubeH}
H_{\textup{X-cube}} = -J \sum_{c} B_c -K \sum_{s,\alpha} A_{s \alpha}
\eeq
where $J,K >0$ are the energy scales. 

It is easily verified that the operators $B_c$ in \eqnref{eqn:XcubeB} and $A_{s \alpha}$ in \eqnref{eqn:XcubeA} commute with each other. A ground state of the model, which has $B_c = 1$ and $A_s =1$ for all $c$ and $s$, is
\beq\label{eqn:XCGS}
\ket{\GS_{\textup{X-cube}}} = \prod_{c} (1 + B_c) \ket{\Rightarrow}
\eeq
where $\ket{\Rightarrow} = \prod_I \left(\frac{\ket{\uparrow}_I + \ket{\downarrow}_I}{2} \right)$. 

While the ground state of the X-cube model defined on the arboreal arena has very similar features as the X-cube model defined on the cubic lattice, the nature of excitations are different and interesting. Consider first the ``electric charge'' excitation at a site where two of the $A$ operators have a value of $-1$. Such excitations can be created from the ground state by the application of the $Z_I$ operator on a link $I$. If $I$ is a 1-link, this will result in dipoles of $A_{s 2}=-1$ and $A_{s 3}=-1$ electric charge excitations (each of which cost an energy of $2K$) where $s=s_1,s_2$ are the two sites that define the chosen  1-link $I$. Pick another (any one of $k_1-1$ possibilities) 1-link $I'$ emanating from the site $s_2$, and apply the operator $Z_{I'}$ to the state obtained after the application of $Z_I$. We see that the charges $A_{s_2 (2,3)} = -1$ are transported to a new site $s_{3}$ as $A_{s_3 (2,3)} = -1$, where $s_3$ is the other site of $I'$, without any additional energy cost. More generally, for any given $\alpha$, a charges $A_{s \alpha'}=-1$ and $A_{s \alpha''}=-1$ located at $s=((g^o_1,m^o_1),(g^o_2,m^o_2),(g^o_3,m^o_3))$ can be transported to any point on the tree $\B{k_\alpha}$. For example, if $\alpha=1$, then the charges can be transported to any other point $s'= ((g_1,m_1),(g^o_2,m^o_2),(g^o_3,m^o_3))$ where $(g_1,m_1)$ is any other point on the tree $\B{k_1}$. We thus see that linenonic electric charges of the X-cube model defined on a cubic lattice, generalize to ``{\em treeonic}'' charges -- charges with mobility restrictions constrained to a tree!

Consider now the monopole excitations where some cubes obtain $B_c = -1$. Such excitations are produced by application of the $X_I$ operator on the ground state at link $I$. When this link $I$ in the $\alpha$ direction, this process  produces an excited state that is a bound state of $k_{\alpha'} k_{\alpha''}$ monopoles each with $B_c=-1$. For  the X-cube model defined on a cubic lattice, this process will produce a bound state of four monopoles. However, in the cubic lattice, the quadrupole of monopoles can be `split' into two dipoles, and these dipoles can move freely in a plane. The situation is quite different in the arboreal arena. Consider $\alpha=1$, i.~e., the link $I$ is an $1$-link. Application of $X_I$ on the ground state will produce $k_2 k_3$ monopoles. Now consider the application of a second $X$ operator on a  $1$-link $I'$ connected to a $2$-link $I_2$ which in turn is connected to the original $1$-link $I$. This whole process will  produce a total of $2 (k_2-1) k_3$ monopoles. In other words, the application of the second spin flip operator (in an attempt to move a subset  of monopoles) will result in the creation of $(k_2-2)k_3$ {\em additional} monopoles. We thus see that in an arboreal three dimensional lattice ($k_\alpha >2$), there are no multipoles of $B_c$ excitations that are mobile.

\subsection{The gauge theory}

% \begin{itemize}
%     \item Definition of the X-cube gauge theory, Gauss' law imposed.
%     \item Confined and deconfined phases
%     \item What is known about this theory in the cubic lattice.
%     \item We study the system on a finite arboreal arena. Several different possibilities present themselves. We will focus on two cases with all rough boundaries and some smooth boundaries. Additionally,  important special cases involve here $k_3$ is 2
%     \item Variational formulation.
%     \item Main results for rough boundaries. Fully rough boundaries, smooth rough
%     \item Mention dualities.
% \end{itemize}

We here study the X-cube gauge theory defined on an three dimensional arboreal arena. We will consider arena with boundaries focusing, among the variety of possibilities,  on $\RC{k_1}{M_1} \cart \RC{k_2}{M_2} \cart \RC{k_3}{M_3}$ with all rough boundaries, and $\SC{k_1}{M_1} \cart \RC{k_2}{M_2} \cart \RC{k_3}{M_3}$ with one smooth boundary. The Hamiltonian we consider is
\beq\label{eqn:XGT}
H_{XGT} = - J \sum_c B_c - h \sum_I X_I + H_\dou 
\eeq
where $J$ and $h$ are energy scales, $B_c$ is the cube term (see \eqnref{eqn:XcubeB} defined on all allowed cubes, $X_I$ is the operator that acts of the qubit placed at link $I$. The Hamiltonian is invariant under local transformations generated by all the allowed operators $A_{s\alpha}$ defined at each site $s$ (see \eqnref{eqn:XcubeA}). As in the case of the $\Integers_2$-gauge theory, additional boundary terms arise as these are invariant under the action of the local transformations generated by $A_{s\alpha}$. The theory is studied in the gauge-invariant sector of the Hilbert space which satisfies the generalized Gauss' law,
\beq
A_{s \alpha} = 1.
\eeq

When $h \ll J$, the theory \eqnref{eqn:XGT} reduces to the X-cube model  \eqnref{eqn:XcubeH} in an infinite three dimensional arboreal lattice. The ground state, in this regime, is in the deconfined phase of the theory. For large $h$, the ground state is the state $\ket{\Rightarrow}$ defined near \eqnref{eqn:XCGS}, and is in the confined phase of the theory. In a three dimensional cubic lattice, it is known that, upon increase of $h$ from $h \ll J$ to $h \gg J$,  a transition from the deconfined to confined phase occurs at a critical value of $h$ via a first order transition~\cite{MuhlhauserSchmidt2020}. The natural question to address is the nature of the transition on the arboreal arena, and \eqn{eqn:XGT} is introduced on a finite system to aid this analysis.

Before we discuss the phase transition anticipated above, we will briefly describe the ground state degeneracy $D_G$ of the system in the limit  $h \ll J$. An explicit calculation shows that
\beq
% \begin{split}
%     &\ln_{2}D_G = \\
%     & \sum_{\alpha} \left(k_\alpha \left((k_\alpha-1)^{M_\alpha -3} - 1 \right) \right) \left( \left[ \sum_{\gamma = \alpha',\alpha''}\left( 1+  \frac{k_{\gamma} ((k_{\gamma}-1)^{M_{\gamma}} -1}{k_{\gamma} - 2} \right)\right] - 1 \right)
%     \end{split}
\begin{split}
    &\ln_{2}D_G = \\
    & \left(k_1 \left((k_1-1)^{M_1 -6} - 1 \right) \right) \left( \left[ \sum_{\gamma = 2,3}\left( 1+  \frac{k_{\gamma} ((k_{\gamma}-1)^{M_{\gamma}} -1}{k_{\gamma} - 2} \right)\right] - 1 \right)
    \end{split}
\eeq
for the case of \Xcube~gauge theory (with $h \ll J$, which is effectively the \Xcube-model)  defined on $\RC{k_1}{M_1} \cart \SC{k_2}{M_2} \cart \SC{k_3}{M_3}$, taking into account the boundary operators allowed. Such a large degeneracy arises owing the large number of Wilson line operators that become possible in the three dimensional arboreal arena in a fashion similar to that illustrated in \figref{fig:BoundaryOperators} for the $\Ztwo$ gauge theory.

\begin{figure}
    \centering
    \includegraphics[width=\columnwidth]{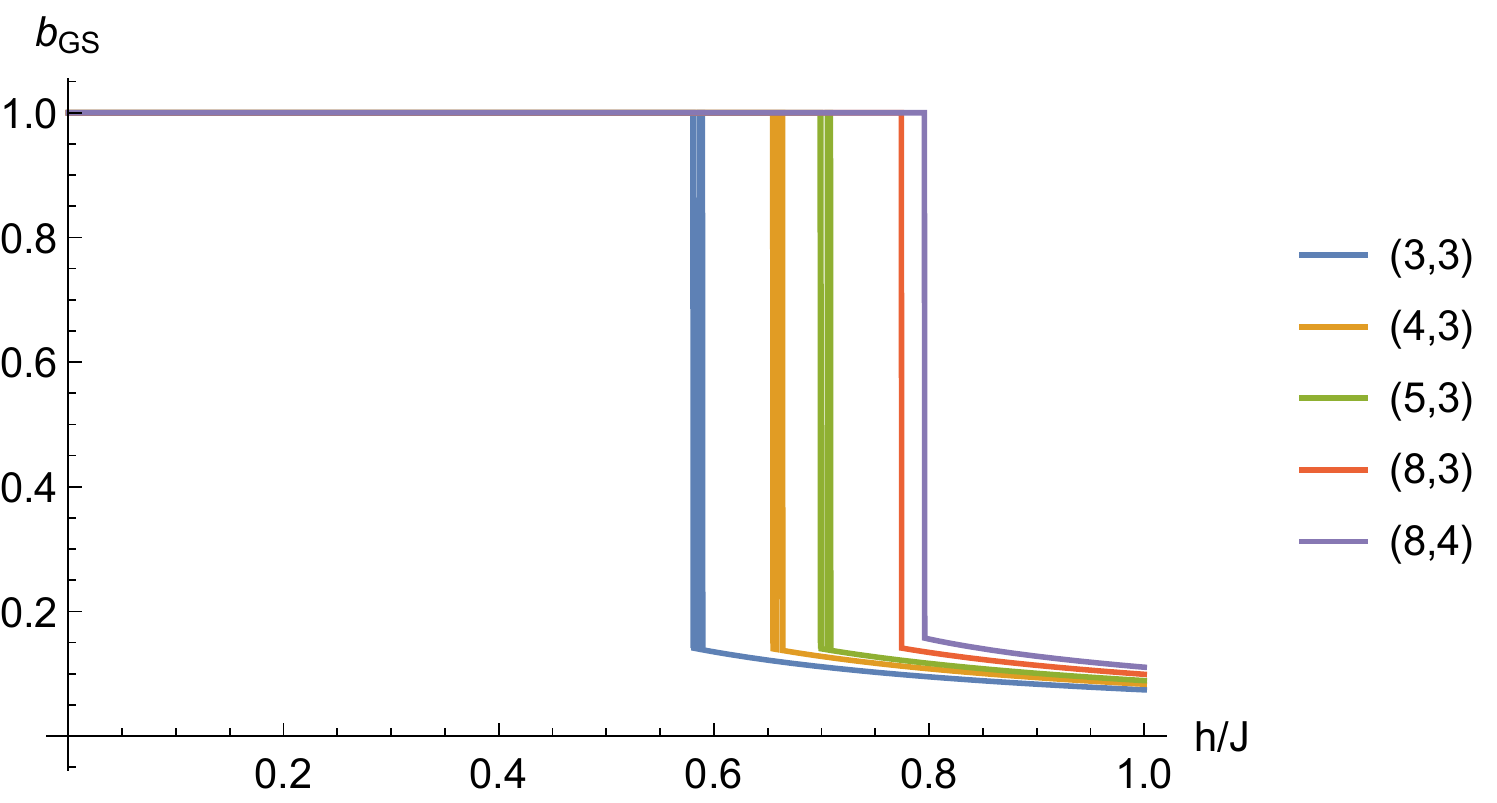}
    \caption{Dependence of $b_{\GS}$ on  $h/J$ for the X-cube gauge theory defined on $\SC{k_1}{M_1} \cart \RC{k_2}{M_2} \cart \RC{2}{M_3}$ for various $(k_1,k_2)$ indicated in the legend. For large $M_1,M_2,M_3$, the transition from the confined to deconfined phase is first order.}
    \label{fig:Var_Xcube_s2}
\end{figure}

\begin{figure}
    \centering
    \includegraphics[width=\columnwidth]{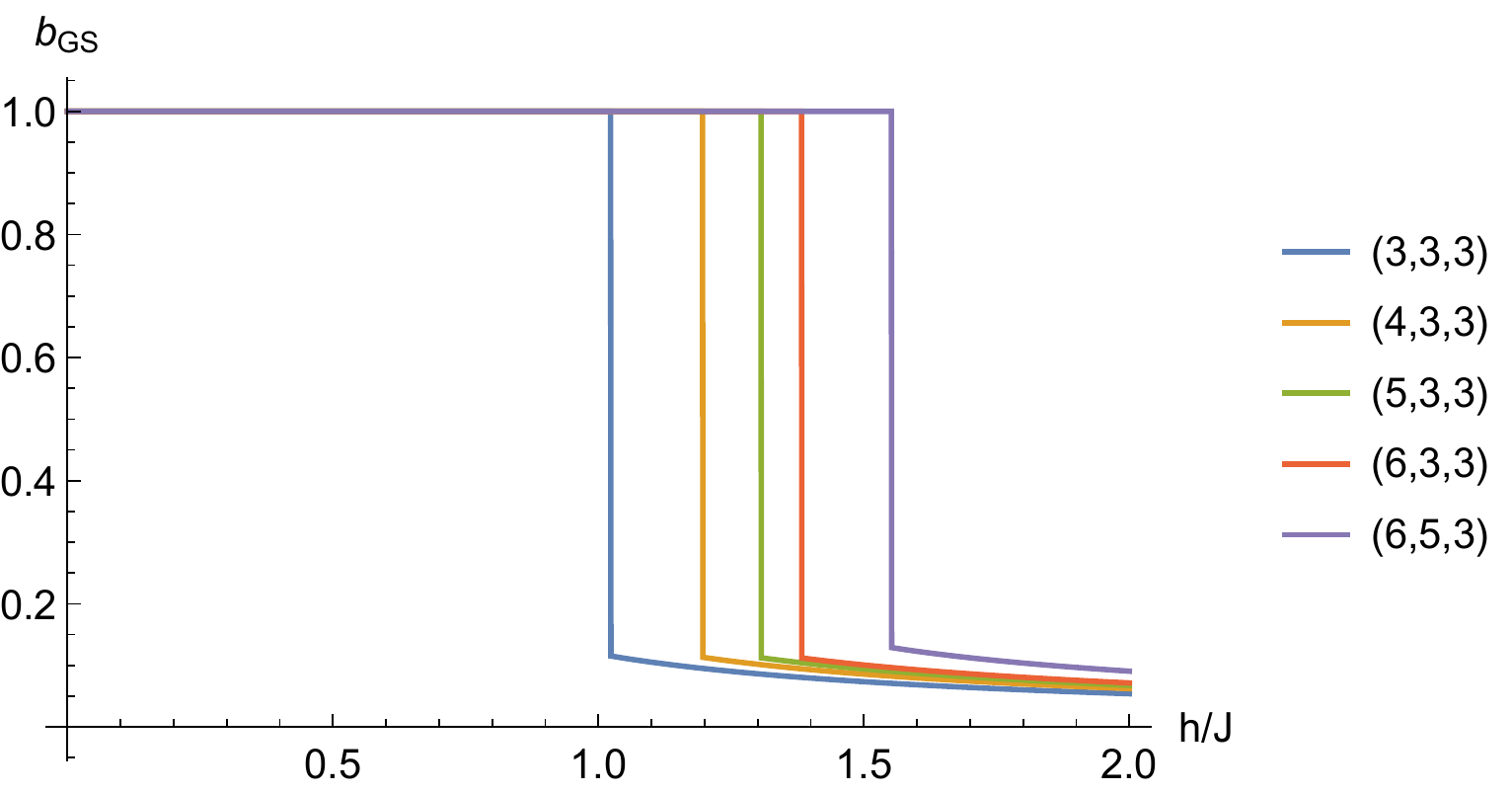}
    \caption{Dependence of $b_{\GS}$ on  $h/J$ for the X-cube gauge theory defined on $\SC{k_1}{M_1} \cart \RC{k_2}{M_2} \cart \RC{k_3}{M_3}$, for various $(k_1,k_2,k_3)$ indicated in the legend. For large $M_1,M_2,M_3$, a first order transition from the deconfined to confined case is obtained when the values of $k_1,k_2,k_3$ are ``small'' ($\lesssim 10$). For larger values of $k$s, a second order transition from the confined to deconfined phase is obtained (see appendix~\ref{sec:VCXcube}). 
    }
    \label{fig:Var_Xcube_s3}
\end{figure}

\begin{figure}
    \centering
    \includegraphics[width=\columnwidth]{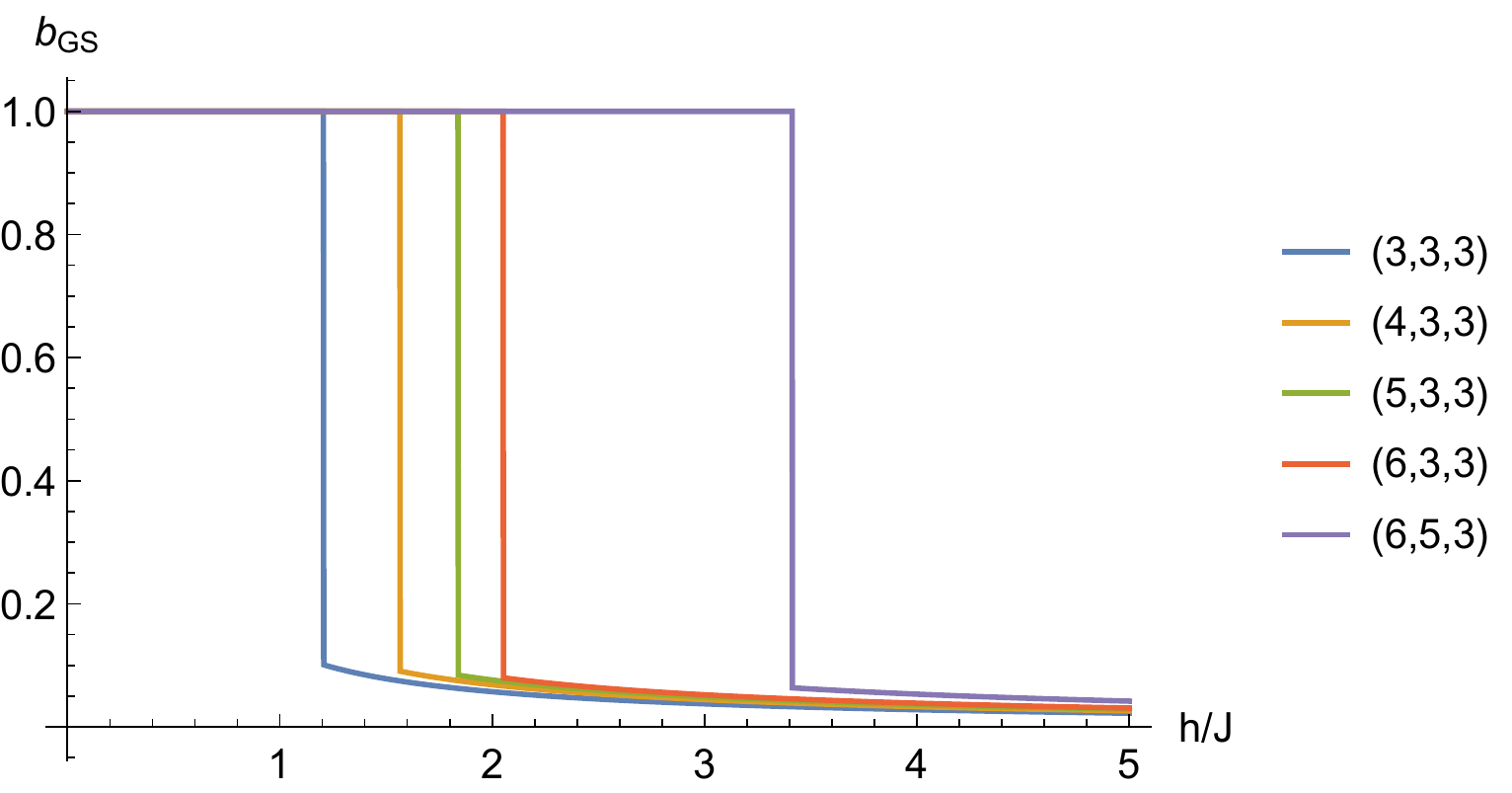}
    \caption{Dependence of $b_{\GS}$ on  $h/J$ for the X-cube gauge theory defined on $\RC{k_1}{M_1} \cart \RC{k_2}{M_2} \cart \RC{k_3}{M_3}$, for various $(k_1,k_2,k_3)$ indicated in the legend. A first order transition from the deconfined to confined case is obtained for large $M_1,M_2,M_3$.}
    \label{fig:Var_Xcube_r3}
\end{figure}

To study the evolution of the ground state as a function of $h/J$, we use the variational ansatz similar to \eqnref{eqn:GSb} as
\beq
\ket{\GS(b)} = \prod_c (1 + b B_c) \ket{\Rightarrow}
\eeq
where $c$ also runs over all boundary terms in addition to the usual cube terms. The variational ground state is obtained for that value of $b$ called $b_{\GS}$ at which the energy is minimized. The calculations are detailed in appendix~\ref{sec:VCXcube}.

 We first consider X-cube gauge theory defined on $\SC{k_1}{M_1} \cart \RC{k_2}{M_2} \cart \RC{2}{M_3}$ where smooth boundaries are combined with rough boundaries. In these finite arboreal lattices the third direction has $k_3=2$, i.~e., this as an extruded three dimensional arena. The main result, as seen from \figref{fig:Var_Xcube_s2} is that the transition from the deconfined to confined phase occurs via a first order transition  for values of $k_1,k_2 \lesssim 10$ (see below). We also note that the transition is first order in general if all the boundaries are rough (results not shown).

Turning now to more general finite three dimensional arboreal lattices with a smooth boundary $\SC{k_1}{M_1} \cart \RC{k_2}{M_2} \cart \RC{k_3}{M_3}$, we find that the transition is generically first order as illustrated in \figref{fig:Var_Xcube_s3}.   In fact, for some values of $k$, we find, as shown in appendix \ref{sec:VCXcube}, that there are {\em two} transitions, the first continuous one going from the deconfined to the confined phase, and a second first order transition in the confined phase, indicating that there are two types of confined phases. Moreover for very large values of $k$s, we find that the first order transition between the confined phases is no longer present.  The nature and physical underpinnings of these findings require further investigation which is a future direction to be pursued.

Finally, we note that dual model (see \tabref{tab:dualities}) to the \Xcube~gauge theory can be constructed (we do not elaborate this here) as generalized quantum face Ising models ``face'' represents the fact that interaction terms are determined not by hyperlinks, but by ``higher dimensional'' object such as faces and volumes formed by the hyperlinks). Such generalized quantum face Ising models will posses sub-dimensional symmetries, with the duality operating in the singlet sector of these symmetries along with the Gauss' law constraint on the gauge theory side. These ideas are natural generalizations of the dualities presented in \cite{Vijay2016} to arboreal arenas.

\section{Arboreal Topological and Fracton Orders}
\label{sec:ATFO}

\begin{figure}
    \centering
    \includegraphics[width=\columnwidth]{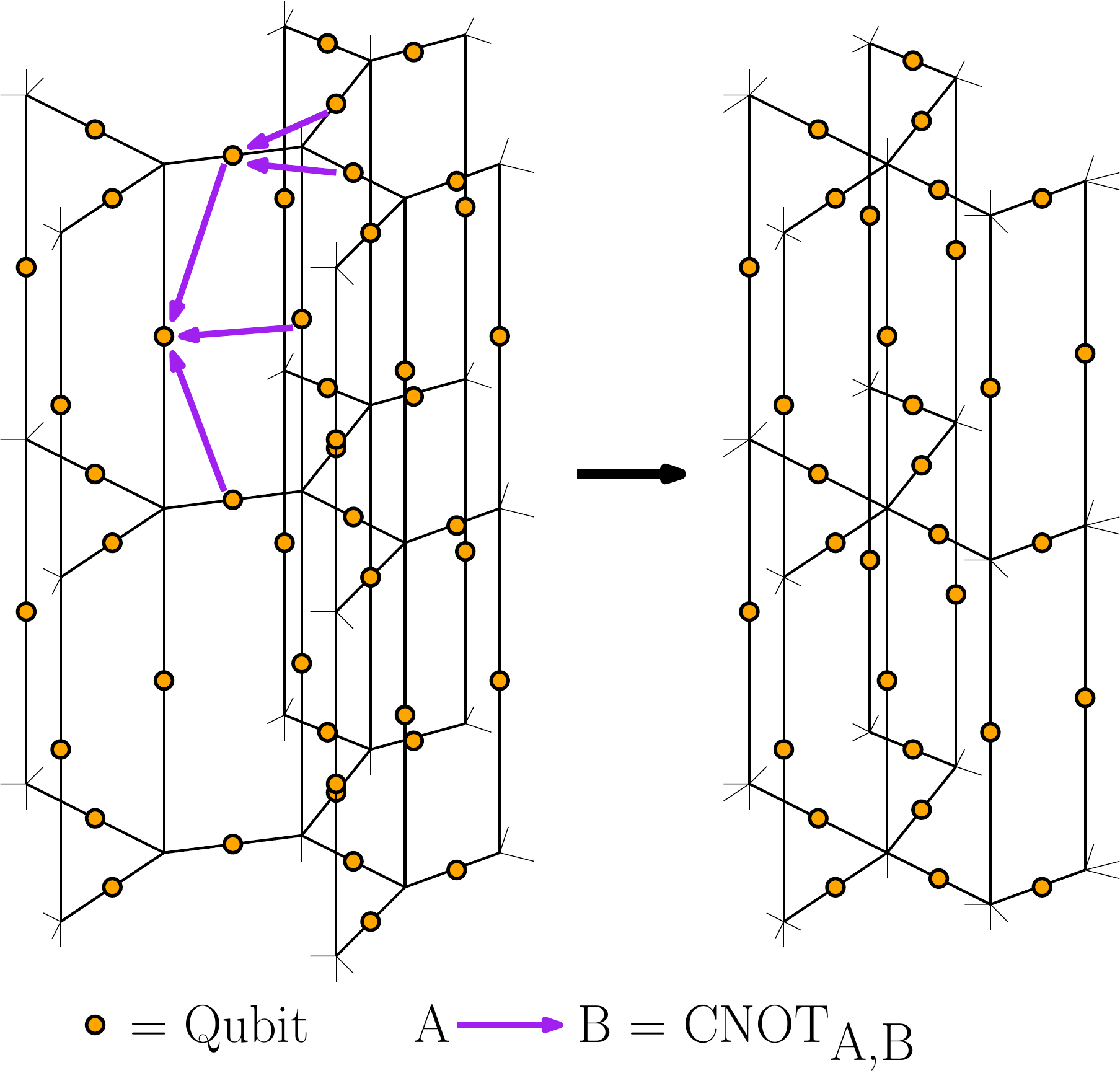}
    \caption{Entanglement renormalization procedure demonstrating that the arboreal topological order on a $\B{3} \cart \B{2}$ arena is equivalent to the one on $\B{4} \cart \B{2}$. The CNOT operations indicated have to be performed between all pairs equivalent qubits, a process that is equivalent to a finite depth quantum circuit. The ground state of $\B{3} \cart \B{2}$  is converted to that  $\B{4} \cart \B{2}$ producted with a set of qubits (not shown) in fixed states.  }
    \label{fig:er_tree}
\end{figure}

The results of the previous sections raise many interesting questions. For example, it is natural to enquire the relationship between the arboreal topological order found the $\Integers_2$ gauge theory on $\B{k} \cart \B{2}$ for different values of $k$. For example, are they ``different phases''? How are they related? A key idea to be exploited in addressing these questions is that two systems are considered to be equivalent (``same phase'') if, for example, the ground state of one can be transformed to that of the other by a finite depth unitary quantum circuit~\cite{Vidal2007,Chen2010} and a set of entangled degrees of freedom. A generalization of this idea to fracton phases is also available, and will be discussed below.

To address these questions, consider the toric code \eqnref{eqn:HToricCode} defined on $\B{3} \cart \B{2}$. We use the entanglement renormaliztion process~\cite{Vidal2007,Chen2010,Vidal2011} to convert the ground state \eqnref{eqn:TCGS} of the toric code on $\B{3} \cart \B{2}$ to that of the toric code on $\B{4} \cart \B{2}$ times a set of unentangled qubits in fixed states. The procedure, demonstrated in \figref{fig:er_tree}, uses a set of CNOT gates (for details, see \cite{Vidal2007}) to produce a finite depth quantum circuit that act on the toric code ground state  on $\B{3} \cart \B{2}$ to that of $\B{4} \cart \B{2}$ producted with unentangled qubit states. It is immediately evident that this process converts the toric code ground state on $\B{k} \cart \B{2}$ to that of $\B{k+1} \cart \B{2}$ whenever $k>2$, suggesting that the arboreal topological order on $\B{k} \cart \B{2}$ are equivalent. Note, however, that this order is distinct from the topological order on $\B{2} \cart \B{2}$, the square lattice, as there no finite depth unitary that will transform the toric code ground state on $\B{2} \cart \B{2}$ to that of $\B{3} \cart \B{2}$!

These observations become more interesting when we note that the same process can be used to show that the arboreal topological order encoded in the ground state of the toric code defined on $\B{k_1} \cart \B{k_2}$ when $k_1,k_2 > 2$ can be transform to the that of $\B{k_1 +1} \cart \B{k_2}$ (times unentangled qubits), or it can be transformed to that of $\B{k_1} \cart \B{k_2+1}$. We thus arrive at a remarkable conclusion that the arboreal toplogical orders encoded in the ground state of $\B{k_1} \cart \B{k_2}$ for all $k_1,k_2 >2$ are equivalent!

The above discussion allows us to ``classify'' arboreal orders of the toric code ground states. There are three types. First is the usual topological order of the toric code defined on the square lattice. The second is the arboreal topological order on extruded trees of the kind $\B{k} \cart \B{2}$. The third one is the arboreal topological order on general two dimensional arboreal lattices $\B{k_1} \cart \B{k_2}$ with $k_1,k_2 > 2$.

Moving to the fractonic models, we first observe that ideas from the notion of foliated fracton phases can be applied to understand and classify arboreal fracton orders. Using the notions introduced in \cite{Shirley2018,Shirley2019} two fracton states are considered to be equivalent ( ``in the same phase'' ) if one can be transformed to another times unentangled layers of topologically ordered states and unentangled qubits in fixed states. By application of a finite depth quantum circuit process similar to the one shown  in \figref{fig:er_tree} (see \cite{Shirley2018,Dua2020}), the X-cube ground state \eqnref{eqn:XCGS} on $\B{k}\cart \B{2}\cart \B{2}$, $k >2$ can be transformed to that of $\B{k+1} \cart \B{2} \cart \B{2}$ times unentangled toric code layers each of which carries the topological order of the  first kind discussed in the previous paragraph. This establishes the equivalence of the X-cube fracton order on $\B{k}\cart \B{2}\cart \B{2}$ for all $k >2$. Similar arguments show that  X-cube fracton order on $\B{k_1} \cart \B{k_2} \cart \B{2}$ for $k_1,k_2 > 2$ are all equivalent. The key point to note here is that the finite depth quantum circuit transforms the X-cube ground state on $\B{k_1} \cart \B{k_2} \cart \B{2}$ to that of $\B{k_1+1} \cart \B{k_2} \cart \B{2}$ times uncoupled layers of topological order of the kind $\B{k_2} \cart \B{2}$, i.~e., the arboreal topological order of the second kind discussed in the para above. Finally, X-cube orders on  $\B{k_1} \cart \B{k_2} \cart \B{k_3}$ with $k_1,k_2,k_3 > 2$ are all equivalent, in that X-cube ground state on  $\B{k_1} \cart \B{k_2} \cart \B{k_3}$ can be transformed to that of 
 $\B{k_1+1} \cart \B{k_2} \cart \B{k_3}$ times unentangled layers each of which carries an arboreal topological order of the third kind (see previous para) on $\B{k_2} \cart \B{k_3}$. This leads us to the conclusion that there are four types of X-cube fracton orders. The first one is the usual X-cube fracton order on a cubic lattice. The second is the X-cube order on $\B{k}\cart \B{2}\cart \B{2}$ for all $k >2$. The X-cube order on $\B{k_1} \cart \B{k_2} \cart \B{2}$ for $k_1,k_2 > 2$ form the third class. The fourth and final class is X-cube fracton order on $\B{k_1} \cart \B{k_2} \cart \B{k_3}$ for all $k_1,k_2,k_3 > 2$.
 
\section{Concluding Remarks}
\label{sec:CR}

We conclude the paper with a couple of remarks.  First, we note that our work points to interesting new possibilities that are offered by going beyond ``manifold-arenas'', as exemplified by the properties of the $\Ztwo$-gauge theories of the arboreal arenas. Will such systems offer a fresh direction that enables them to be utilized gainfully for quantum information processing? Naturally, this entails costs in the design and construction of qubit connectivities and controls that are admittedly more complex. The issue to be explored is that if such constructions are possible (which, almost definitely they will be at a future date), do they provide cost-effective and efficient platforms for quantum information storage and processing? Further work is required to address this question. Second, we note that we have  explored only limited types of topological and fracton ordered phases on the arboreal arena.  It will be an interesting, if obvious, direction to explore the physics of models~\cite{Levin2005,Walker2012,Burnell2013,Shirley2020} that produce other types of orders.

\noindent
{\bf Acknowledgements:} NM thanks the KVPY Programme, and VBS acknowledges DST, SERB for support.

\bibliographystyle{apsrev4-2}
\bibliography{fracton}

\clearpage

\appendix

\begin{widetext}

\centerline{{\large {\bf Appendices}}}

\section{Variational Calculation -- $\Integers_2$ Gauge Theory}
\label{sec:VarCalc}

We have
\beq
\ket{\GS(b)} = \underbrace{\left[ \prod_p (1 + b B_p )\right]}_{P(b)} \ket{\Rightarrow} 
\eeq
where $B_P$ include all terms (including those in $H_\dou$, if present) and
\beq
\braket{\GS(b)}{\GS(b)} = \bra{\Rightarrow} P(b)^2 \ket{\Rightarrow} 
\eeq
Now,
\beq
\begin{split}
P(b)^2 & = \prod_{p,q} (1 + b B_p) (1 + b B_q) \\ & 
=  \left[\prod_{p} \left(1 + b^2 + 2 b B_p \right) \right] 
\end{split}
\eeq
To compactify the equations we introduce $\zeta(b) = \frac{2b}{1+b^2}$.

% \subsection{Square Lattice with PBC}
% For a square lattice with periodic boundary conditions (PBC),
% \beq
% \begin{split}
% \braket{\GS(b)}{\GS(b)} & = \bra{\Rightarrow} \left[\prod_{p} \left(1 + b^2 + 2 b B_p \right) \right] \ket{\Rightarrow} \\
%   & = (1+ b^2)^{N_B} \bra{\Rightarrow} \left[\prod_{p} \left(1 + \frac{2 b}{1 + b^2} B_p \right) \right] \ket{\Rightarrow} \\
%   & =  
%   (1+ b^2)^{N_B} \left[ 1+ \left(\zeta(b) \right)^{N_B}  \right].
%   \end{split} 
% \eeq
% where $N_B$ is the number of plaquettes.
% Further, we obtain,
% \beq
% \begin{split}
% \bra{\GS(b)} \left( \sum_p B_p \right) \ket{\GS(b)} & = N_B (1 + b^2)^{N_B}  \left( \zeta(b) + (\zeta(b))^{N_B -1} \right) 
% \\
% \bra{\GS(b)} \left( \sum_I X_I \right) \ket{\GS(b)} & = N_L (1+b^2)^{N_B-2} (1-b^2)
% \end{split}
% \eeq
% This leads to an energy per site given by
% \beq
% \frac{{\cal E}}{J}= - \frac{2 b}{1+b^2} - \frac{h}{J} \frac{1-b^2}{(1+b^2)}
% \eeq

\subsection{$\SC{k_1}{M_1} \cart \SC{k_2}{M_2}$}
Taking $N_B$ as the number of plaquttes,
\beq
\begin{split}
\braket{\GS(b)}{\GS(b)} =& (1+b^2)^{N_B} \\
\bra{\GS(b)} \left( \sum_p B_p \right) \ket{\GS(b)} =& N_B (1 + b^2)^{N_B} \zeta(b) 
\end{split}
\eeq
There are two types of links. Links that are in the interior and links that are ``tangent'' to the boundary, which we will call \emph{surface links}. This is not to be confused with boundary links which are ``perpendicular'' to the boundary and is connected to only one site. A link in the 1-direction, a 1-link, in the interior will contribute to $k_2$ plaquettes, while it contributes only a single plaquette when it is on the surface (similarly for a 2-link).  Let there be $l_\alpha$ links, $v_\alpha$ vertices (of which $(\dou v)_\alpha$ are boundary vertices) in $\SC{k_\alpha}{M_\alpha}, \alpha=1,2$. Then 
\beq
\begin{split}
\bra{\GS(b)} \left( \sum_I X_I \right) \ket{\GS(b)} & = \\  l_1(v_2 - (\dou v)_2) (1+b^2)^{N_B - k_2} (1 - b^2)^{k_2} & +  l_1 (\dou v)_2 (1+b^2)^{N_B - 1} (1 - b^2)  \\ 
+ l_2(v_1 - (\dou v)_1) (1+b^2)^{N_B - k_1} (1 - b^2)^{k_1} & +  l_2 (\dou v)_1 (1+b^2)^{N_B - 1} (1 - b^2)
\end{split}
\eeq
This leads to an energy per site ${\cal E}$ given by
 \beq
 \frac{{\cal E}}{J}= -\frac{2
   b}{1+b^2} -\frac{h}{J}
   \left(\frac{1}{k_1-1}\left(\frac{1-b^2}{1+b^2}\right)^{k_1}+\frac{1}{k_2-1}\left(\frac{1-b^2}{1+b^2}\right)^{k_2}+ \left(\frac{1-b^2}{1+b^2}\right) 
   \left(\frac{k_1-2}{k_1-1}+\frac{k_2-2}{k_2-1}\right)\right)
 \eeq

\subsection{$\RC{k_1}{M_1} \cart \SC{k_2}{M_2}$}
In this case there are additional boundary operators (see \eqnref{eqn:HwithDou}), which we will denote by $W^\dou_q$, of which there are $N_\dou$ in number. We will assume that these have the same coupling constant $J$ as the bulk plaquette operators, i.e.,
\beq
H_\dou = -J \sum_q W^\dou_q
\eeq
Note that there are two types of boundary operators \figref{fig:BoundaryOperators}, first those that involving two links these are $N_{\dou 2}$ in number, and those that involve four links $N_{\dou 4}$, with $N_\dou = N_{\dou 2} + N_{\dou 4}$.
We now get,
\beq
\begin{split}
\braket{\GS(b)}{\GS(b)} =& (1+b^2)^{N_B+N_\dou} \\
\bra{\GS(b)} \left( \sum_p B_p \right) \ket{\GS(b)} =& N_B (1 + b^2)^{N_B + N_\dou} \zeta(b) \\
\bra{\GS(b)} \left( \sum_q B^\dou_q \right) \ket{\GS(b)} =& N_\dou (1 + b^2)^{N_B + N_\dou} \zeta(b) %\\
%\bra{\GS(b)} \left( \sum_I X_I \right) \ket{\GS(b)} =& N_{L_1} (1+b^2)^{N_B - k_2} (1 - b^2)^{k_2} \\ & + N_{L_2} (1+b^2)^{N_B - k_1} (1 - b^2)^{k_1}
\end{split}
\eeq

To compute the $\bra{\GS(b)} \left( \sum_I X_I \right) \ket{\GS(b)}$, we have to take care of a few things. First of all, the links in 2-direction are straightforward to deal with
\beq\label{eqn:RSXterm1}
\bra{\GS(b)} \left( \sum_{I_2} X_{I_2} \right) \ket{\GS(b)} = N_{L_2} (1+b^2)^{(N_B + N_\dou -k_1)} (1-b^2)^{k_1} 
\eeq
Now we note that there are eight different types of 1-links, i.~e., links in the 1-direction.  There are a total of $N_{L_1} = \underbrace{L(\RC{k_1}{M_1})}_{l_1} \times \underbrace{V(\SC{k_2}{M_2})}_{v_2}$. The first kind of 1-link is in the ``interior'' of the graph and appears in $k_2$ plaquettes. This contributes
\beq\label{eqn:RSXterm2}
\bra{\GS(b)} \left( \sum_{I_1^1} X_{I_1^1} \right) \ket{\GS(b)} = (l_1 (v_2 - (\dou v)_2)) (1+b^2)^{(N_B  + N_\dou -k_2)} (1-b^2)^{k_2}
\eeq
where $(\dou v)_2$ are the number of boundary vertices of $\SC{k_2}{M_2}$.
The second type of 1-link is ``tangent to the smooth boundary'' (surface link) and participates only in a single plaquette, there are $l_1$ such links are
\beq\label{eqn:RSXterm3}
\bra{\GS(b)} \left( \sum_{I_1^2} X_{I_2^1} \right) \ket{\GS(b)} = (l_1 ((\dou v)_2  -1)) (1+b^2)^{(N_B  + N_\dou -1)} (1-b^2)
\eeq
%\begin{widetext}
The remaining $l_1$ surface 1-links are of five types. There are $l_1 - k_1((k_1-1)^{M_1-1} + (k_1-1)^{M_1})$ boundary links each of which participates in a single plaquette giving
\beq
\bra{\GS(b)} \left( \sum_{I_1^3} X_{I_1^3} \right) \ket{\GS(b)} = (l_1 -  k_1((k_1-1)^{M_1-1} + (k_1-1)^{M_1})) (1+b^2)^{(N_B  + N_\dou -1)} (1-b^2)
\eeq
The next type of link participates in 1 boundary operators of the type $W^2$ and 1 plaquette operator, of which there are $k_1(k_1 -1)^{M_1 -2} (k_1-2)$ and give
\beq
\bra{\GS(b)} \left( \sum_{I_1^4} X_{I_1^4} \right) \ket{\GS(b)} = (k_1(k_1 -1)^{M_1 -1} (k_1-2)) (1+b^2)^{(N_B  + N_\dou -2)} (1-b^2)^2
\eeq
The next type of boundary link participates in 1 plaquette operator, $(k_1-2)$ boundary $W^2$ operators and $1$ boundary $W^4$ operator. The $k_1(k_1)^{M_1 -2} (k_1 -2)$ 1-link provide
\beq
\bra{\GS(b)} \left( \sum_{I_1^5} X_{I_1^5} \right) \ket{\GS(b)} = (k_1(k_1 -1)^{M_1 -2} (k_1-2)) (1+b^2)^{(N_B  + N_\dou - k_1)} (1-b^2)^{k_1}
\eeq
This is followed by 1-link which participates in 1 plaquette operator, $(k-2)$ operators of $W^2$ type, and $(k-2)$ operators of the $W^4$ type. There is only one such link, giving
\beq
\bra{\GS(b)}X_{I_1^6}\ket{\GS(b)} = (1+b^2)^{(N_B  + N_\dou - (2 k_1 -1))} (1-b^2)^{(2k_1 -1)}
\eeq
Yet another type of link relates to 1 plaquette operator, and one $W^4$ operator. There are $k_1(k_1 -1)^{M_1-2} (k-2)$ links of this type, and
\beq
\bra{\GS(b)} \left( \sum_{I_1^7} X_{I_1^7} \right)= (k_1(k_1 -1)^{M_1 -2} (k_1-2)) (1 + b^2)^{N_B + N_\dou - 2} (1-b^2)^2.
\eeq
Finally, there is one link that contributes to one plaquette and $(k_1 -2)$ boundary operators of type $W^4$. This gives
\beq
\bra{\GS(b)} X_{I_1^8} \ket{\GS{(b)}} = (1+b^2)^{(N_B  + N_\dou - (k_1 -1))} (1-b^2)^{(k_1 -1)}
\eeq
%\end{widetext}
One can see that only terms \eqnref{eqn:RSXterm1}, \eqnref{eqn:RSXterm2}, and \eqn{eqn:RSXterm3} make a finite contribution to the free energy density in the limit of $M_1,M_2 \to \infty$, leading to
\beq
\frac{\cal{E}}{J} = -\frac{2 b \left(k_1-1\right)}{1+b^2} - \frac{h}{J} \left(\left(\frac{1-b^2}{1+b^2}\right)^{k_1}+\left(\frac{k_1-1}{k_2-1}\right)
   \left(\frac{1-b^2}{1+b^2}\right)^{k_2}+\frac{
   \left(k_1-1\right) \left(k_2-2\right)}{
   \left(k_2-1\right)} \left(\frac{1-b^2}{1+b^2}\right)\right)
\eeq

\subsection{$\RC{k_1}{M_1} \cart \RC{k_2}{M_2}$}

The key point here is to handle the Wilson surface constraints. We get
\beq
\begin{split}
\braket{\GS(b)}{\GS(b)} & = (1+b^2)^{N_B} \bra{\Rightarrow} \left(\prod_p (1 + \zeta(b) B_p) \right)\ket{\Rightarrow} \\
&= (1+b^2)^{N_B} \left(1 + \sum_{j=1}^{N_W} \zeta(b)^{N_P(j)} \right)
\end{split}
\eeq
where $N_W$ is the number Wilson surfaces, $N_P(j)$ is the  number of plaquettes in the $j$-th Wilson surface.
%\begin{widetext}
Further,
\beq
\begin{split}
& \bra{\GS(b)} \left( \sum_p B_p \right) \ket{\GS(b)}  =  (1+b^2)^{N_B}\sum_p \bra{\Rightarrow} \left( \prod_{q} (1 + \zeta(b) B_q) \right) B_p \ket{\Rightarrow}  \\
& = (1+b^2)^{N_B} \sum_p \left(\zeta(b) + \sum_{j=1}^{N_W(p)} \zeta(b)^{N_P(j) - 1} \right) \\
& = N_B (1+b^2)^{N_B} \zeta(b) + (1+b^2)^{N_B} \sum_p \left(\sum_{j=1}^{N_W(p)} \zeta(b)^{N_P(j) - 1} \right) 
\end{split}
\eeq
where $N_W(p)$ is the number of Wilson surfaces containing the plaquette $p$, and $N_P(j)$ is the number of plaquettes in the $j$-th Wilson surface that contains the plaquette $p$.
%\end{widetext}
Finally, we have
\beq
\begin{split}
& \bra{\GS(b)} \left( \sum_I X_I \right) \ket{\GS(b)}  = \\
& l_1 v_2 \left( (1+b^2)^{N_B - k_2} (1-b^2)^{k_2} \right) \\  + &  l_2 v_1 \left( (1+b^2)^{N_B - k_1} (1-b^2)^{k_1}  \right)
\end{split}
\eeq

These considerations lead to an energy density (energy per site)
\beq
\frac{\cal E}{J} = -\frac{2 b \left(k_1-1\right) \left(k_2-1\right)}{1+b^2}-h
   \left(\left(k_2-1\right)
   \left(\frac{1-b^2}{1+b^2}\right)^{k_1}+\left(k_1-1\right)
   \left(\frac{1-b^2}{1+b^2}\right)^{k_2}\right)
\eeq

\end{widetext}

\begin{figure}
    \centering
    \includegraphics[width=\columnwidth]{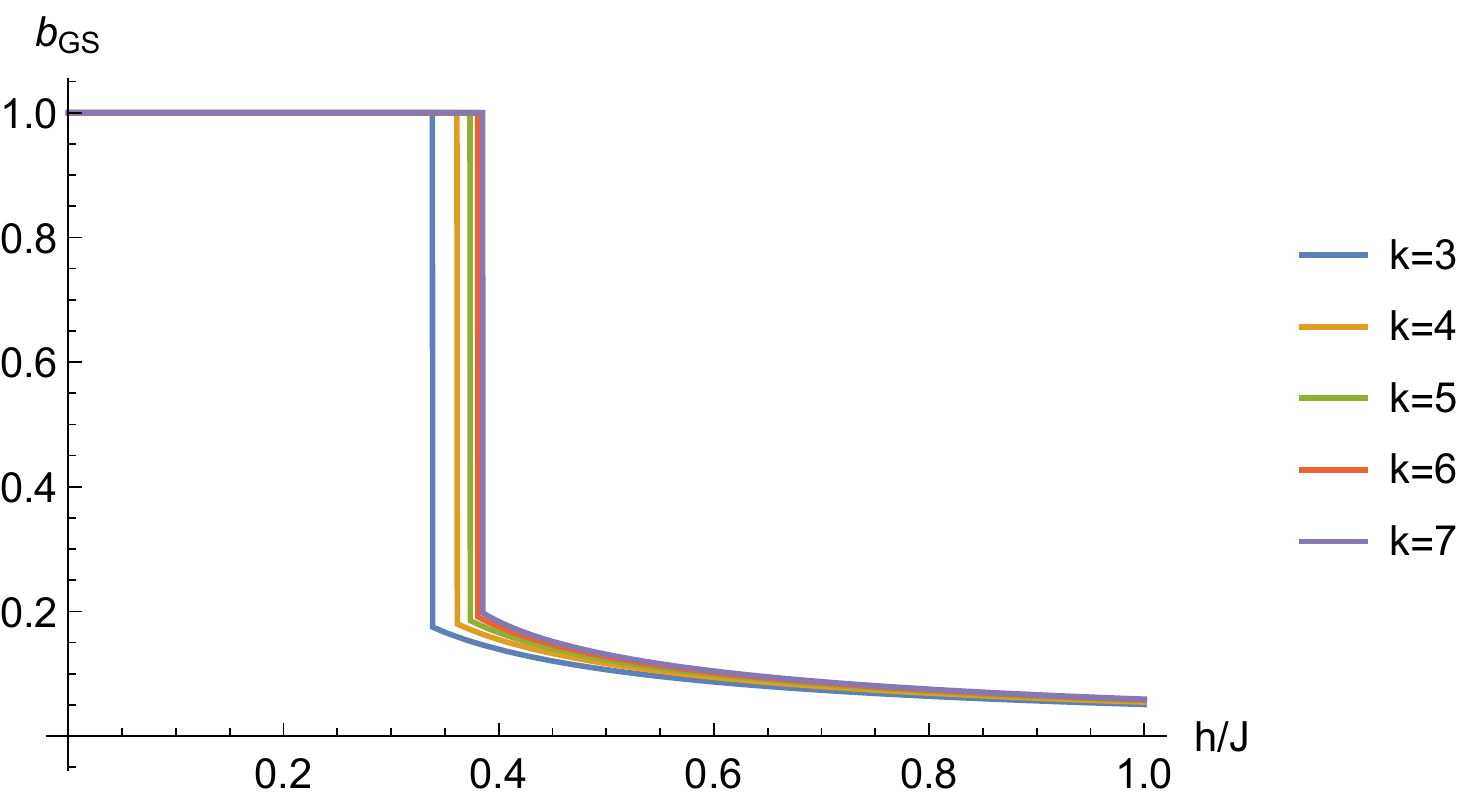}
    \caption{Plot of $b_{\GS}$ as a function of $h$ for the X-cube model on $\SC{k}{M1}\cart \RC{2}{M_2}\cart \RC{2}{M_3}$ for large values of $M_1,M_2,M_3$. A  first-order phase transition for all $k$.}
    \label{fig:Var_Xcube_s1}
\end{figure}
\begin{figure}
    \centering
    \includegraphics[width=\columnwidth]{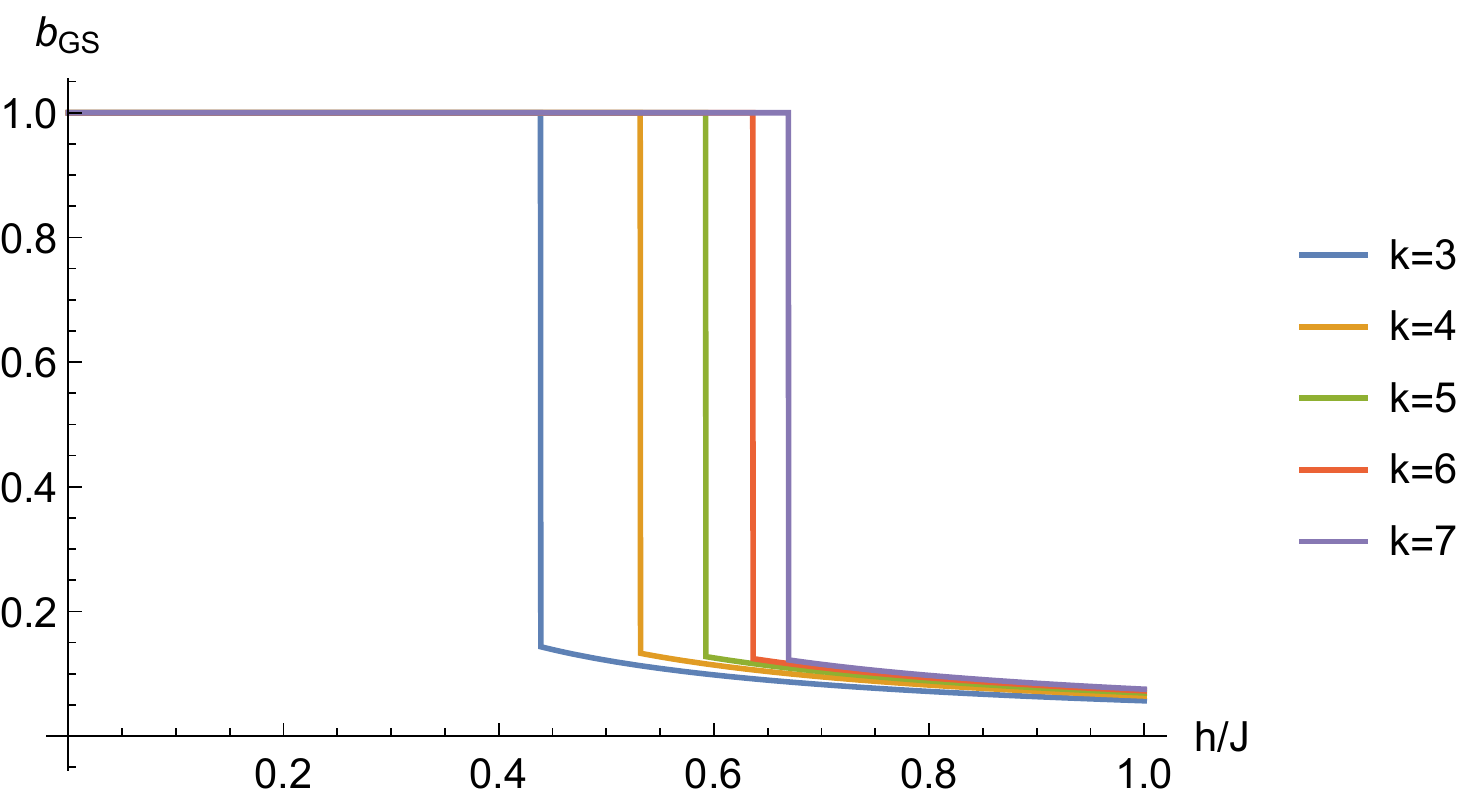}
    \caption{Plot of $b_{\GS}$ as a function of $h$ for the X-cube model on $\RC{k}{M1}\cart \RC{2}{M_2}\cart \RC{2}{M_3}$ for large values of $M_1,M_2,M_3$. A  first-order phase transition for all $k$.}
    \label{fig:Var_Xcube_r1}
\end{figure}
% \begin{figure}
%     \centering
%     \includegraphics[width=\columnwidth]{p_vs_hc_Xcube1.png}
%     \caption{Plot of the critical coupling $h_c$ as a function of $p$ for smooth boundaries (starting from 3).  }
%     \label{fig:p_vs_hc_Xcube1}
% \end{figure}
\begin{figure}
    \centering
    \includegraphics[width=\columnwidth]{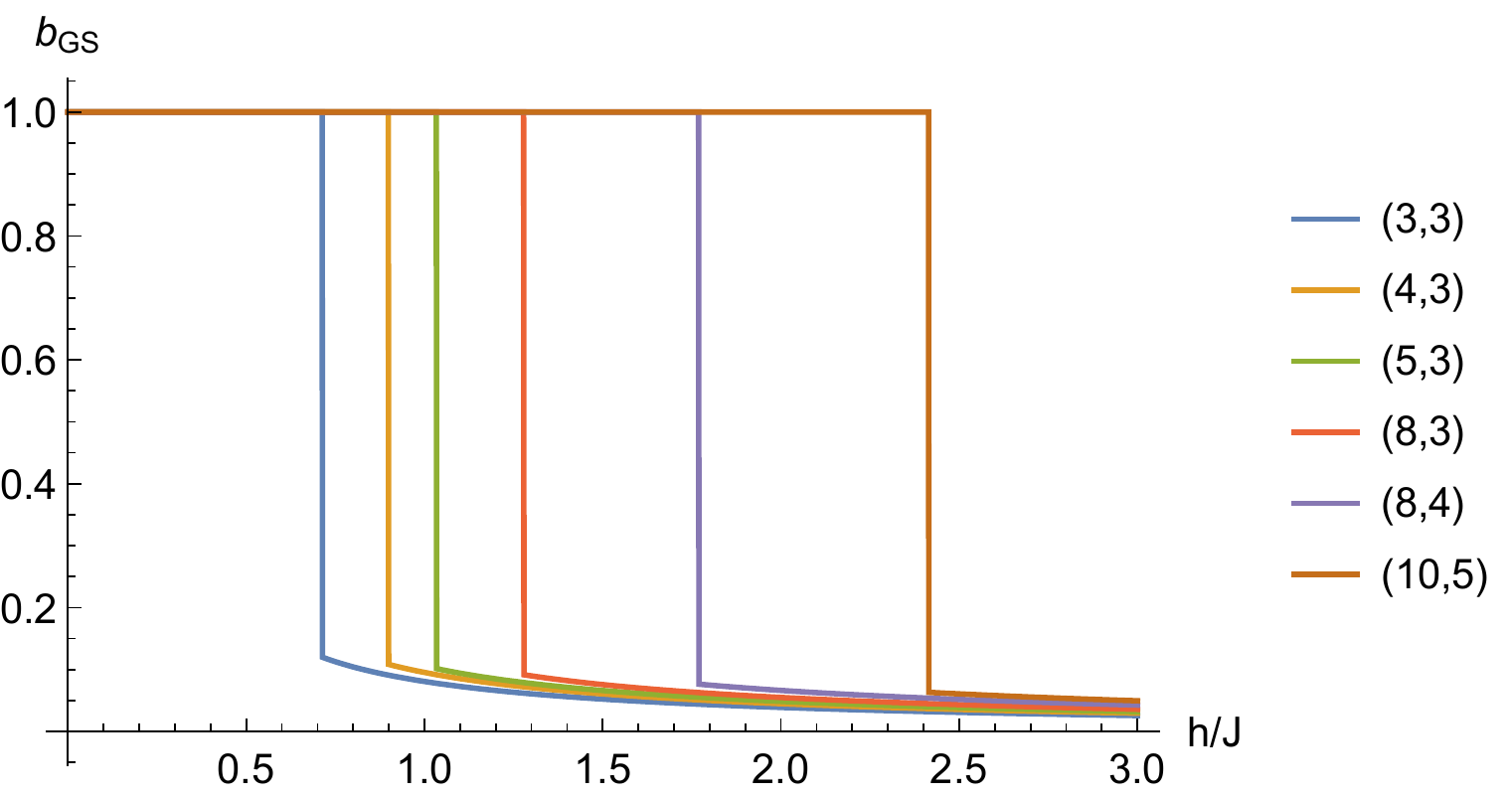}
    \caption{Plot of $b_{\GS}$ as a function of $h/J$ for the X-cube guage theory defined on $\RC{k_1}{M_1}\cart \RC{k_2}{M_2} \cart \RC{2}{M_3}$. The legend indicates $(k_1,k_2)$. }
    \label{fig:Var_Xcube_r2}
\end{figure}

\begin{figure}
    \centering
    \includegraphics[width=\columnwidth]{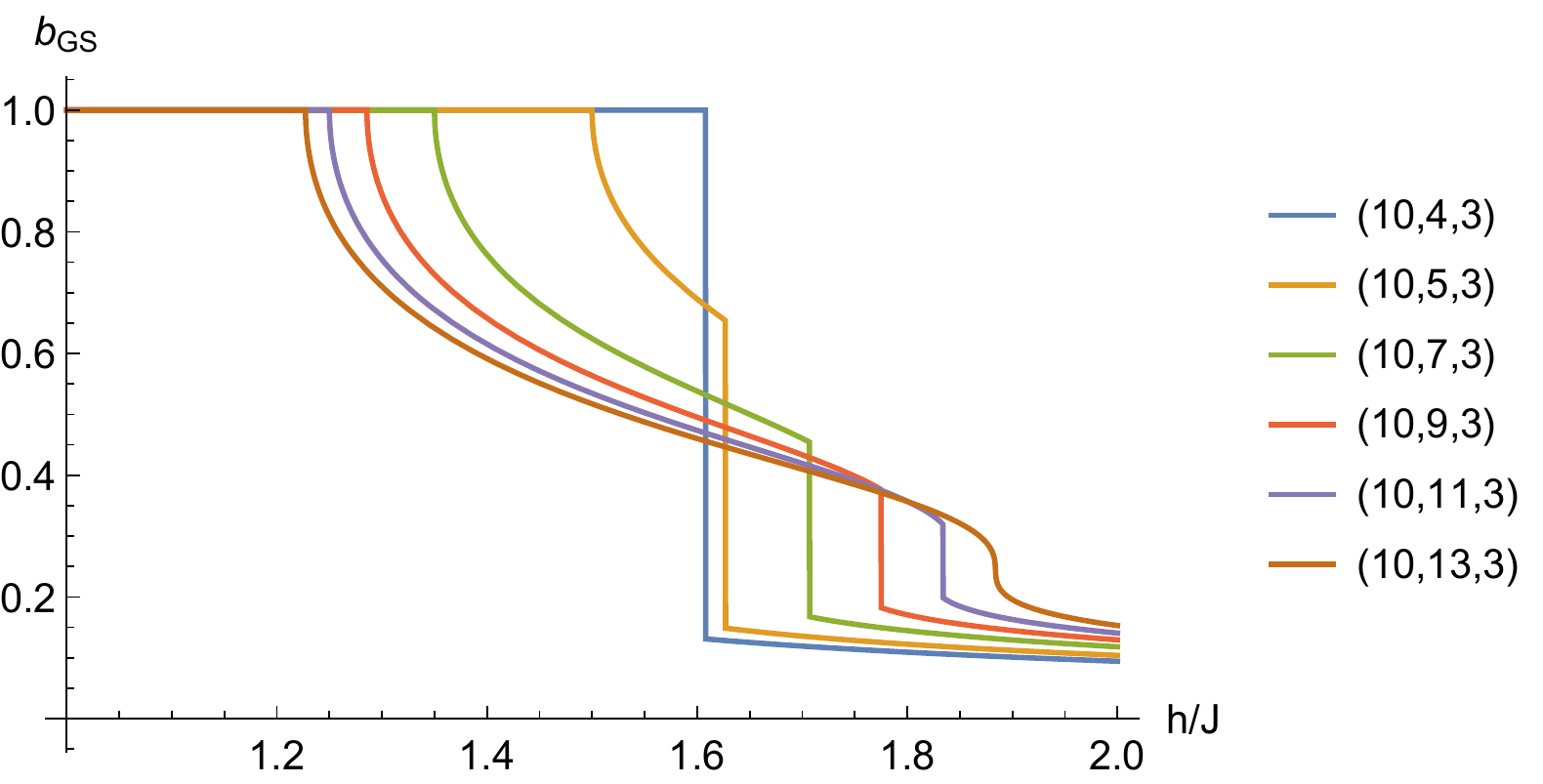}
    \caption{Two phase transitions for special values of $k_1,k_2,k_3$ for the X-cube model on $\RC{k_1}{M_1} \cart \RC{k_2}{M_2} \cart \RC{2}{M_3}$. $(k_1,k_2,k_3)$ are indicated in the legend.}
    \label{fig:Var_Xcube_s3Complex}
\end{figure}

\begin{widetext}

%\begin{widetext}
\section{Variational Calculations - X-cube Gauge Theory}
\label{sec:VCXcube}

The Hamiltonian for the X-cube gauge theory is given by
\beq
H = -J\sum_c B_c - J \sum_\dou B_\dou - h \sum_I X_I
\eeq
where $B_\dou$ are the boundary terms -- gauge invariant local products of $Z$ operators that occur at the boundary. We only include boundary terms that are independent of each other and of the $B_c$ terms -- these are the boundary terms that will split some of the degeneracy. In the language of the excitations, these boundary terms are operators that transport a treeonic charge excitation from a rough boundary to another rough boundary of the same tree. The number of such independent boundary operators -- denoted $N_\dou$ -- will be much smaller than $N_c \sim N_I$, the number of cubes and links respectively. This is crucial to simplify our calculation later in this section.

The variational ansatz for the ground state is parametrized by $b\in [0,1]$ and the state is
\beq
\ket{\text{GS}(b)} = \prod_{\ind={c},{\dou}}\term{1+b B_\alpha}\ket{\Rightarrow}
\eeq
where $B_\alpha$ now includes both cube terms and boundary terms. Note that this state is not normalized.

First, we consider the X-cube $\mathbb{Z}_2$ gauge theory on a $\RC{k}{M_1} \cart \RC{2}{M_2} \square \RC{2}{M_3}$ and $\SC{k}{M_1} \cart \RC{2}{M_2} \square \RC{2}{M_3}$. First, we calculate the norm-squared of the state,
\begin{align}
    \braket{\GS(b)}{\GS(b)} &= \bra{\Rightarrow}\prod_{\alpha = c,\dou} \term{1+b B_\alpha}\prod_\alpha \term{1+b B_\alpha}\ket{\Rightarrow} \\
    &= \bra{\Rightarrow}\prod_\alpha \term{(1+b^2)+2b B_\alpha}\ket{\Rightarrow} \\
    &= \term{1+b^2}^{N_c + N_\dou}\bra{\Rightarrow} \prod_\alpha \term{1+\frac{2b}{(1+b^2)} B_\alpha}\ket{\Rightarrow} \\
    &\approx \term{1+b^2}^{N_c+N_\dou}
\end{align}
$N_c$ and $N_\dou$ are the number of cube terms and boundary terms respectively.
Now we calculate the expectation value of energy
\begin{align}
        \bra{\GS(b)}H\ket{\GS(b)} &= \bra{\Rightarrow}\prod_\ind \term{1+b B_\ind}\term{-J\sum_\ind B_\ind -h\sum_I X_I}\prod_\ind \term{1+b B_\ind}\ket{\Rightarrow}
\end{align}
There are two different terms. The first 
\begin{align}
    \bra{\GS(b)}\sum_\ind B_\ind\ket{\GS(b)} &= \sum_{\ind'} \bra{\Rightarrow}\prod_\ind \term{1+b B_\ind} B_{\ind'} \prod_\ind \term{1+b B_\ind}\ket{\Rightarrow} \\
    &= \sum_{\ind'}\bra{\Rightarrow}\prod_\ind \term{(1+b^2)+2b B_\ind} B_{\ind'}\ket{\Rightarrow} \\
    &= \term{1+b^2}^{N_\ind}\sum_{\ind'}\bra{\Rightarrow} \prod_\ind \term{1+\frac{2b}{(1+b^2)} B_\ind} B_{\ind'}\ket{\Rightarrow} \\
    &= 2b \term{1+b^2}^{N_\ind-1} N_\ind
\end{align}
where $N_\ind \equiv N_c + N_\dou$. The second
\begin{align}
    \bra{\GS(b)}\sum_I X_I\ket{\GS(b)} &= \sum_{I} \bra{\Rightarrow}\prod_\ind \term{1+b B_\ind} X_{I} \prod_\ind \term{1+b B_\ind}\ket{\Rightarrow}
\end{align}
Let us divide the sum into ``$1$-links'', ``$2$-links'', ``$3$-links'' and ``surface links'' (not counted earlier in the $1,2,3$-links) which touch 4, $2k$, $2k$ and 2 cubes respectively. There are also ``edge links'' which touch only one cube, but the number of these will be negligible in a $\RC{k}{M_1} \cart \RC{2}{M_2} \cart \RC{2}{M_3}$. In fact, the presence of a thermodynamically large number of edge links will destroy the deconfined phase and the phase transition (in this variational treatment) analogous to how smooth boundaries in the toric code destroy the deconfined phase (\figref{fig:Var_Manyp_S}), so we avoid these cases. Like seen in Appendix \ref{sec:VarCalc}, there are also special links that touch the boundary operators $B_\dou$. But since the number of boundary operators are not thermodynamically large, the difference in contribution due to these terms will vanish in the limit of large $M_1,M_2,M_3$. This simplifies our calculation greatly. Writing down each type of link explicitly,
\begin{align}
    \bra{\GS(b)}\sum_I X_I\ket{\GS(b)} &= \sum_{I_1,I_2,I_3,I_{\text{surf}}} \bra{\Rightarrow}\prod_\ind \term{1+b B_\ind} X_{I} \prod_\ind \term{1+b B_\ind}\ket{\Rightarrow}
    \label{eqn:123}
\end{align}
where $I_{1,2,3,\textup{surf}}$ are 1-links, 2-links, 3-links and boundary links.
Now we move $X_I$ to the right such that the plaquette terms touching the link $I$ pick up a minus sign (as they anticommute with $X_I$) and $X_I$ gives eigenvalue 1 when acting on $\ket{\Rightarrow}$. 
%\begin{widetext}
\begin{multline}
    \bra{\GS(b)}\sum_I X_I\ket{\GS(b)} = N_{I_1} \term{1-b^2}^4 \term{1+b^2}^{N_\ind-4} + N_{I_2} \term{1-b^2}^{k+2} \term{1+b^2}^{N_\ind-k-2} \\ + N_{I_z} \term{1-b^2}^{k+2} \term{1+b^2}^{N_\ind-k-2} + N_{I_\text{surf}} \term{1-b^2}^2 \term{1+b^2}^{N_\ind-2}
\end{multline}
%\end{widetext}
Now, 
\begin{align}
    N_\ind \approx N_c &\approx \frac{4M_2M_3k(k-1)^{M_1}}{p-2} \\
    N_{I_1} &\approx \frac{4M_2M_3k(k-1)^{M_1}}{k-2} \\
    N_{I_2} &\approx \frac{4M_2M_3k(k-1)^{M_1-1}}{k-2} \\
    N_{I_3} &\approx \frac{4M_2M_3k(k-1)^{M_1-1}}{k-2} \\
    N_{I_{\text{surf}}} &\approx \left\{ \begin{array}{cc}
    {4M_2M_3k(k-1)^{M_1-1} }    & \SC{k}{M_1}\cart \RC{2}{M_2}\cart \RC{2}{M_3} \\
    0     & \RC{k}{M_1}\cart \RC{2}{M_2}\cart \RC{2}{M_3} 
    \end{array}\right. 
\end{align}
 where we have also considered the three dimensional arboreal lattice with smooth boundaries. The boundary condition, smooth or rough, on the 2 and 3 directions do not affect the result when the trees in those directions are lines (2-trees), and hence only rough boundaries are considered in those directions.
We obtain the energy density ${\cal E}$
\begin{align}
    \frac{{\cal E}}{J} \sim  \left\{ \begin{array}{cc}
    -\term{ \frac{2b}{1+b^2}} -\frac{(h/J)}{k-1}\term{(k-1) \frac{\term{1-b^2}^4}{\term{1+b^2}^4} + 2\frac{\term{1-b^2}^{k+2}}{\term{1+b^2}^{k+2}} + (k-2)\frac{\term{1-b^2}^2}{\term{1+b^2}^2}}     &  \text{for } \SC{k}{M_1}\cart \RC{2}{M_2}\cart \RC{2}{M_3} 
    \\
    -\term{ \frac{2b}{1+b^2}} -\frac{(h/J)}{k-1}\term{(k-1) \frac{\term{1-b^2}^4}{\term{1+b^2}^4} + 2\frac{\term{1-b^2}^{k+2}}{\term{1+b^2}^{k+2}}}     &  \text{for } \RC{k}{M_1}\cart \RC{2}{M_2}\cart \RC{2}{M_3} 
    \end{array}\right.
\end{align}

We the value of $b$ for the ground state, $b_\GS$ is plotted in \figref{fig:Var_Xcube_s1} and \figref{fig:Var_Xcube_r1} for the smooth an rough cases. We observe a first order transition for all $p$ and all boundaries.  %The smooth boundary case has an interesting dependence of $h_c$ vs $p$, which I have plotted in \figref{fig:p_vs_hc_Xcube1}. 
Next we study the case of $\RC{k_1}{M_1}\cart \RC{k_2}{M_2} \cart \RC{2}{M_3}$ and  $\SC{k_1}{M_1} \cart \RC{k_2}{M_2} \cart \RC{2}{M_3}$   with $k_1,k_2 > 2$. The case where all boundaries are smooth does not have any phase transition due to the presence of a large number of edge links, and hence not discussed. In this case, we have
%\begin{widetext}
\begin{align}
   \bra{\GS(b)}\sum_\ind B_\ind\ket{\GS(b)}  &= -\frac{2b}{1+b^2} N_c \\
\bra{\GS(b)}\sum_I X_I\ket{\GS(b)} &=\!\begin{multlined}[t][10.5cm]
  -N_{I_1} \term{1-b^2}^{k_2+2} \term{1+b^2}^{N_c-k_2-2} -N_{I_2} \term{1-b^2}^{k_1+2} \term{1+b^2}^{N_c-k_1-2} \\ - N_{I_3} \term{1-b^2}^{k_1+k_2} \term{1+b^2}^{N_c-k_1-k_2}-N_{I_\text{surf}} \term{1-b^2}^2 \term{1+b^2}^{N_c-2}.
 \end{multlined}
\end{align}
where all the $N$s have same meanings as in the previous paragraphs. We obtain
\begin{align}
    N_\ind \approx N_c &\approx \frac{2M_3k_1k_2(k_1-1)^{M_1}(k_2-1)^{M_2}}{(k_1-2)(k_2-2)} \\
    N_{I_1} &\approx \frac{2M_3k_1 k_2(k_1  -1)^{M_1}(k_2 -1)^{M_2-1}}{(k_1  -2)(k_2 -2)} \\
    N_{I_2} &\approx \frac{2M_3k_1 k_2(k_1  -1)^{M_1-1}(k_2 -1)^{M_2}}{(k_1  -2)(k_2 -2)} \\
    N_{I_3} &\approx \frac{2M_3k_1 k_2(k_1  -1)^{M_1-1}(k_2 -1)^{M_2-1}}{(k_1  -2)(k_2 -2)} \\
    N_{I_{\text{surf}}} &\approx \left\{ \begin{array}{cc}
    2M_3k_1 k_2(k_1  -1)^{M_1-1}(k_2 -1)^{M_2-1}    & \SC{k_1}{M_1} \cart \RC{k_2}{M_2} \cart \RC{2}{M_3}\\
    0     & \RC{k_1}{M_1} \cart \RC{k_2}{M_2} \cart \RC{2}{M_3}
    \end{array}\right. 
\end{align}

For rough boundaries, we obtain first order transitions as shown  \figref{fig:Var_Xcube_r2}, while for smooth boundaries, first-order transitions are obtained for small values of $k_1,k_2$ as discussed in the main text (see \figref{fig:Var_Xcube_s2}).

We obtain the energy density as
\begin{align}
    \frac{{\cal E}}{J} \sim  \left\{ \begin{array}{cc}
    -\term{ \frac{2b}{1+b^2}} -\frac{(h/J)}{(k_1  -1)(k_2 -1)}\term{(k_1  -1) \frac{\term{1-b^2}^{k_2+2}}{\term{1+b^2}^{k_2+2}} + (k_2 -1)\frac{\term{1-b^2}^{k_1+2}}{\term{1+b^2}^{k_1+2}} +\frac{\term{1-b^2}^{k_1+k_2}}{\term{1+b^2}^{k_1+k_2}} + (k_1  -2)(k_2 -2)\frac{\term{1-b^2}^2}{\term{1+b^2}^2}}      \\
    -\term{ \frac{2b}{1+b^2}} -\frac{(h/J)}{(k_1  -1)(k_2 -1)}\term{(k_1  -1) \frac{\term{1-b^2}^{k_2+2}}{\term{1+b^2}^{k_2+2}} + (k_2 -1)\frac{\term{1-b^2}^{k_1+2}}{\term{1+b^2}^{k_1+2}} +\frac{\term{1-b^2}^{k_1+k_2}}{\term{1+b^2}^{k_1+k_2}} }
    \end{array}\right.
\end{align}
for $\SC{k_1}{M_1} \cart \RC{k_2}{M_2} \cart \RC{2}{M_3}$ and $\RC{k_1}{M_1} \cart \RC{k_2}{M_2} \cart \RC{2}{M_3}$ respectively.
Finally, consider  $\RC{k_1}{M_1}\cart \RC{k_2}{M_2} \cart \RC{k_3}{M_3}$ and $\RC{k_1}{M_1}\cart \RC{k_2}{M_2} \cart \SC{k_3}{M_3}$ with $k_1,k_2,k_3 > 2$. We have
\begin{align}
    \bra{\GS(b)}\sum_\ind B_\ind\ket{\GS(b)} &= -\frac{2b}{1+b^2} N_\ind \\
    \bra{\GS(b)}\sum_I X_I\ket{\GS(b)}  &= \!\begin{multlined}[t][10.5cm]
  -N_{I_1} \term{1-b^2}^{k_2+k_3} \term{1+b^2}^{N_c-k_2-k_3} -N_{I_2} \term{1-b^2}^{k_1+k_3} \term{1+b^2}^{N_c-k_1-k_3} \\- N_{I_3} \term{1-b^2}^{k_1+k_2} \term{1+b^2}^{N_c-k_1-k_2}-N_{I_\text{surf}} \term{1-b^2}^2 \term{1+b^2}^{N_c-2}
 \end{multlined}
\end{align}
%\end{widetext}
where
\begin{align}
    N_\ind \approx N_c &\approx \frac{k_1k_2k_3(k_1  -1)^{M_1}(k_2 -1)^{M_2}(k_3 -1)^{M_3}}{(k_1  -2)(k_2 -2)(k_3 -2)} \\
    N_{I_1} &\approx \frac{k_1k_2k_3(k_1  -1)^{M_1}(k_2 -1)^{M_2-1}(k_3 -1)^{M_3-1}}{(k_1  -2)(k_2 -2)(k_3 -2)} \\
    N_{I_1} &\approx \frac{k_1k_2k_3(k_1  -1)^{M_1-1}(k_2 -1)^{M_2}(k_3 -1)^{M_3-1}}{(k_1  -2)(k_2 -2)(k_3 -2)} \\
    N_{I_1} &\approx \frac{k_1k_2k_3(k_1  -1)^{M_1-1}(k_2 -1)^{M_2-1}(k_3 -1)^{M_3}}{(k_1  -2)(k_2 -2)(k_3 -2)} \\
    N_{I_{\text{surf}}} &\approx \left\{ \begin{array}{cc}
    \frac{k_1k_2k_3(k_1  -1)^{M_1-1}(k_2 -1)^{M_2-1}(k_3 -1)^{M_3}}{(k_3 -2)}    &  \RC{k_1}{M_1}\cart \RC{k_2}{M_2} \cart \SC{k_3}{M_3}\\
    0     & \RC{k_1}{M_1}\cart \RC{k_2}{M_2} \cart \RC{k_3}{M_3}
    \end{array}\right. 
\end{align}
We obtain
\begin{multline}
    \frac{{\cal E}}{J} \sim  
    -\term{ \frac{2b}{1+b^2}} -\frac{(h/J)}{(k_1  -1)(k_2 -1)(k_3 -1)}\left((k_1  -1) \frac{\term{1-b^2}^{k_2+k_3}}{\term{1+b^2}^{k_2+k_3}} + (k_2 -1)\frac{\term{1-b^2}^{k_1+k_3}}{\term{1+b^2}^{k_1+k_3}}\right. \\ \left.+(k_3 -1) \frac{\term{1-b^2}^{k_1+k_2}}{\term{1+b^2}^{k_1+k_2}} + (k_1  -2)(k_2 -2)(k_3 -2)\frac{\term{1-b^2}^2}{\term{1+b^2}^2}     \right)
\end{multline}
for $\RC{k_1}{M_1}\cart \RC{k_2}{M_2} \cart \SC{k_3}{M_3}$, and
\begin{align}
    \frac{{\cal E}}{J} \sim 
    -\term{ \frac{2b}{1+b^2}} -\frac{(h/J)}{(k_1  -1)(k_2 -1)(k_3 -1)}\term{(k_1-1) \frac{\term{1-b^2}^{k_2+k_3}}{\term{1+b^2}^{k_2+k_3}} + (k_2 -1)\frac{\term{1-b^2}^{k_1+k_3}}{\term{1+b^2}^{k_1+k_3}} +(k_3 -1) \frac{\term{1-b^2}^{k_1+k_2}}{\term{1+b^2}^{k_1+k_2}}}      
\end{align}
for $\RC{k_1}{M_1}\cart \RC{k_2}{M_2} \cart \RC{k_3}{M_3}$. The resulting phases obtained for different boundary conditions are discussed in the main text (see \figref{fig:Var_Xcube_s3} and \figref{fig:Var_Xcube_r3}). Fig.~\ref{fig:Var_Xcube_s3Complex} shows the two phase transitions seen for particular values of $(k_1,k_2,k_3)$ for $\SC{k_1}{M_1}\cart \RC{k_2}{M_2} \cart \RC{k_3}{M_3}$. The physics underlying these transitions needs further investigation to be taken up in a future work.

\end{widetext}

\end{document}